\documentclass[11pt,a4paper]{article}

\usepackage[utf8]{inputenc}
\def\papertitle{\bf Study of $CP$ violation in $\Lambda_b^0/\Xi^-_b\rightarrow \Lambda(1520)M$ decays with the final-state rescattering mechanism} 

\usepackage{authblk}
\title{\papertitle}
\author[1]{Tian-Liang Feng, \footnote{Email: fengtl2018@lzu.edu.cn}}
\author[2]{Hui-Qiang Shang,\footnote{Email: she@hust.edu.cn}}
\author[3]{Jing Gao,\footnote{Email: gao@hiskp.uni-bonn.de}}
\author[2]{Qin Qin\footnote{Email: qqin@hust.edu.cn}}
\author[1]{Fu-Sheng Yu\footnote{Email: yufsh@lzu.edu.cn}}

\affil[1]{Frontiers Science Center for Rare Isotopes, and School of Nuclear Science and Technology, Lanzhou University, Lanzhou 730000, China}
\affil[2]{School of Physics, Huazhong University of Science and Technology, Wuhan 430074, China}
\affil[3]{Helmholtz-Institut f\"{u}r Strahlen- und Kernphysik and Bethe Center for Theoretical Physics, Universit\"{a}t Bonn, D-53115 Bonn, Germany}

\usepackage[final]{changes}
\usepackage[top=1in, bottom=1.25in, left=1in, right=1in]{geometry}
\usepackage{microtype}
\usepackage{lineno}
\usepackage{xspace}
\usepackage{booktabs}
\usepackage{comment}
\usepackage{slashed}
\usepackage{physics}

\usepackage{caption} 

\usepackage{graphicx}  

\usepackage{color}
\usepackage{colortbl}
\DeclareGraphicsExtensions{.pdf,.PDF,png,.PNG}

\usepackage{amsmath} 
\usepackage{amsfonts,amssymb,amsthm,bm,braket,yhmath}
\usepackage{upgreek} 

\usepackage{hyperref}
\usepackage{hyperxmp}
\usepackage[figuresright]{rotating}
\usepackage{float} 
\usepackage{hypcap} 
\hypersetup{hidelinks} 
\usepackage{txfonts}
\usepackage{amssymb}
\usepackage{cases}
\usepackage{algorithm}
\usepackage{algpseudocode}
\usepackage{mathrsfs}
\usepackage{color}
\usepackage{amscd,amssymb,amsthm,amsmath,bm,graphicx,psfrag}
\usepackage{epsfig,mathrsfs}
\usepackage[bf,SL,BF]{subfigure}
\usepackage{amsmath}
\usepackage{pict2e,keyval,fp}
\usepackage[bf,SL,BF]{subfigure}
\usepackage{tikz}
\usepackage[compat=1.1.0]{tikz-feynman}

\theoremstyle{definition}

\numberwithin{equation}{section}

\def\kaon   {{\ensuremath{K}}\xspace}
\def\Kbar    {{\kern 0.2em\overline{\kern -0.2em \kaon}{}}\xspace}

\def\KorKbar    {\kern 0.18em\optbar{\kern -0.18em K}{}\xspace}

\def\Dbar    {{\kern 0.2em\overline{\kern -0.2em D}{}}\xspace}

\def\DorDbar    {\kern 0.18em\optbar{\kern -0.18em D}{}\xspace}

\def\Lbar        {{\ensuremath{\kern 0.1em\overline{\kern -0.1em\Lambda}}}\xspace}
\def\LorLbar    {\kern 0.18em\optbar{\kern -0.18em \PLambda}{}\xspace}

\newcommand{\tev}{\ensuremath{\mathrm{\,Te\kern -0.1em V}}\xspace}
\newcommand{\gev}{\ensuremath{\mathrm{\,Ge\kern -0.1em V}}\xspace}
\newcommand{\mev}{\ensuremath{\mathrm{\,Me\kern -0.1em V}}\xspace}
\newcommand{\kev}{\ensuremath{\mathrm{\,ke\kern -0.1em V}}\xspace}
\newcommand{\gevc}{\ensuremath{{\mathrm{\,Ge\kern -0.1em V\!/}c}}\xspace}
\newcommand{\mevc}{\ensuremath{{\mathrm{\,Me\kern -0.1em V\!/}c}}\xspace}
\newcommand{\gevcc}{\ensuremath{{\mathrm{\,Ge\kern -0.1em V\!/}c^2}}\xspace}
\newcommand{\gevgevcccc}{\ensuremath{{\mathrm{\,Ge\kern -0.1em V^2\!/}c^4}}\xspace}
\newcommand{\mevcc}{\ensuremath{{\mathrm{\,Me\kern -0.1em V\!/}c^2}}\xspace}

\usepackage{cite} 
\usepackage{mciteplus}

\begin{document}
\maketitle
\begin{abstract}
Recently, the LHCb collaboration reported the first observation of $CP$ violation in baryon decays, with a significance of more than $5\sigma$. 
This strongly motivates us to investigate the $CP$ violation in more baryon decay processes. 
In this work, we employ the final-state rescattering mechanism with introducing two model parameters, $\Lambda_{\rm{charm}}$ and $\Lambda_{\rm{charmless}}$, and calculate two-body non-leptonic baryon decays $\Lambda^0_b \rightarrow \Lambda(1520)\,\pi^0/\kappa(700)/f_0(500, 980)/\rho^0/K^{*0}/\phi$ and $\Xi^-_b \rightarrow \Lambda(1520)\,K^-$. 
Consequently, we evaluate the corresponding branching ratios, $CP$ asymmetries, and interference effects between different decay amplitudes. 
Our theoretical predictions for certain decay channels are in good agreement with current experimental measurements, while the remaining processes—particularly the remarkably large $CP$ violation observable revealed by the kinematic analysis are expected to be tested in future experiments.
\newpage
\end{abstract} 
\tableofcontents
\section{Introduction}

The observed matter-antimatter asymmetry of the universe provides a strong motivation for searching
new sources of $CP$ violation, which can manifest as observable effects in baryonic processes \cite{Sakharov:1967dj}. In the Standard Model (SM), the $CP$ violation originates exclusively from the Cabibbo-Kobayashi-Maskawa (CKM) quark-mixing matrix~\cite{Cabibbo:1963yz,Kobayashi:1973fv}. 
Therefore, studying the $CP$ violation in baryonic sector under the CKM mechanism is essential, as it directly tests a necessary condition for cosmological asymmetry using the established mechanism of the SM.

The $CP$ violation has been extensively investigated, both theoretically and experimentally, in meson systems. 
It was first discovered in $K$ mesons~\cite{Muller:1960ph, Christenson:1964fg, KTeV:1999kad}, and then observed in $B$ and $D$ mesons~\cite{BaBar:2001ags, Belle:2001zzw, BaBar:2004gyj, Belle:2004nch, LHCb:2013syl, LHCb:2019hro}, collectively confirming the validity of the CKM mechanism in the meson sector. 
Recently, the LHCb collaboration reported a measurement of $CP$ asymmetry $\mathcal{A}_{CP}  = (2.45 \pm 0.46 \pm 0.10)\% $ for the process $\Lambda_b^0 \rightarrow pK^-\pi^+\pi^-$, marking the first observation of the $CP$ violation in baryonic sector with a significance of $5\sigma$~\cite{LHCb:2025ray}. 
This breakthrough opens a new chapter in the study of matter-antimatter asymmetry in the universe. From a theoretical standpoint, our focus is shifting gradually from “predicting discoveries” towards “precision computation and detailed analysis”.

In this work, we calculate the $CP$ violation of bottom baryon decays to $\Lambda(1520)$ within the final-state rescattering framework. This theoretical approach was initially successful in calculations of $B$ mesons decays~\cite{Cheng:2004ru}. After subsequent refinements, this method has been effectively applied to processes involving both charmed~\cite{Jia:2024pyb} and bottom baryons~\cite{Duan:2024zjv}, with results in good agreement with experimental data. 
The core physical picture describes the non-factorizable long-distance contributions in the weak decay via single-hadron exchange. This connects one weak vertex to two strong vertices through the exchanged hadron, resulting in a triangle diagram. After introducing two model-dependent form factors, performing the loop integral over the triangle diagram yields the complete decay amplitude, which incorporates both the strong and weak phases. 
We specifically compute the branching ratios and $CP$ asymmetries for processes $\Lambda^0_b \rightarrow \Lambda(1520)\,\pi^0/\kappa(700)/f_0(500, 980)/\rho^0/K^{*0}/\phi$ and $\Xi^-_b \rightarrow \Lambda(1520)\,K^-$. By decomposing the helicity amplitudes and extracting the corresponding CKM factors, we analyze the origins of the numerical magnitudes in different channels. 
Furthermore, we construct the total amplitude including interference effects between channels with identical final states, and perform a detailed kinematic analysis of the results using multiple methodologies.
Notably, all employed methods consistently indicate a highly significant $CP$ violation observable, which is of considerable interest for future experimental studies.

The paper is organized as follows. In Section 2, we provide a detailed introduction to the theoretical framework, including the topological diagrams in the weak effective theory and the final-state rescattering. In Section 3, we list all input parameters and present the contributions of different CKM factors to the helicity amplitudes. In Section 4, we display the branching ratios, $CP$ asymmetries, and model-parameter dependencies for each decay channel, with a detailed discussion of these results. 
In Section 5, we take into account interference between certain channels, make an estimation of the total amplitude, and analyze the kinematic behavior. In Section 6, we give a brief summary. The appendices contain all triangle diagrams, the Feynman rules of the corresponding vertex as well as the effective Lagrangians, the Wilson coefficient expansions for the weak vertices, and the relevant model specifics employed in this work.
\section{Theoretical framework}\label{framework}

In this section, we first summarize the theoretical framework for heavy-flavor quark weak decays established in previous studies. Subsequently, we introduce the naive factorization approach to estimate the short-distance contributions to the decay amplitudes. Finally, we provide a detailed description of the method for calculating the non-factorizable long-distance contributions with the final-state interaction.

\subsection{From topological diagrams to rescattering mechanism}
In order to study the charmless two-body non-leptonic $\Lambda_b^0$ and $\Xi^-_b$ weak decays, we employ the effective field theory for bottom baryon weak decays at the quark level. For the decay of a bottom quark into light quarks, the $b\rightarrow u$ transition at tree level and the $b\rightarrow q$ transition (with $q=d,s$) at loop level are described by the effective Hamiltonian ~\cite{Buchalla:1995vs}
\begin{equation}
	\begin{aligned}
    \mathcal{H}_{eff}= & \frac{G_F}{\sqrt{2}}\left\{V_{u b} V_{u q}^*\left[C_1(\mu) \mathcal{O}_1^u(\mu)+C_2(\mu) \mathcal{O}_2^u(\mu)\right]-V_{t b} V_{t q}^*\left[\sum_{i=3}^{10} C_i(\mu) \mathcal{O}_i(\mu)\right]\right\}+ \text { h.c. }\ ,
	\end{aligned}
\end{equation}
where $C_{i}~(i=1,...,10)$ are the Wilson coefficients evaluated at the renormalization scale $\mu=m_b$, and $\mathcal{O}_{i}~(i=1,...,10)$ are the four quark operators. 
Topological diagrams provide a clear and straightforward method for visualizing the matrix elements of these local effective operators. These diagrams encode all strong interaction dynamics, namely both perturbative and non-perturbative contributions~\cite{Cheng:2004ru}, and are categorized according to the structure of the weak vertex. 
The bottom quark decays studied in this work involve topological diagrams such as $T$-diagram (color-allowed diagram with external $W$ emission), $C$-diagram (color-suppressed internal $W$-emission diagram), $E$-diagram ($W$-exchange diagram) and $P$-diagram (diagram with penguin operators)~\cite{Duan:2024zjv}. 
For the $T$-diagram, it can be factorized with the QCD-based method, such as  naive factorization approach~\cite{Leibovich:2003tw}, QCD factorization (QCDF)~\cite{} and soft collinear effective theory (SCET)~\cite{}. 
However, non-factorizable long-distance contribution plays a dominant role in the $C$- and $E$-diagram~\cite{Cheng:2004ru, Beneke:2001ev}. Although their branching ratios can be estimated with SCET~\cite{Leibovich:2003tw, Mantry:2003uz}, a reliable theoretical determination of $CP$ violation remains a significant challenge. Therefore, we take into account the non-factorizable contributions with the framework of final-state interaction.

\begin{figure*}[h]
    \centering
    \begin{tikzpicture}[scale=0.8, baseline={([yshift=-0.1cm]current bounding box.center)}]
        \begin{feynman}
            \vertex (a) {\Large\(\mathcal{B}_b\)};
            \vertex [right=1.8cm of a] (b);
            \vertex [above right=1.5cm of b] (c);
            \vertex [below right=1.5cm of b] (d);
            
            \diagram* {
                (a) -- [fermion] (b) -- [fermion, edge label={\Large\(f_1\)}] (c),
                (b) -- [fermion, edge label'={\Large \(f_2\)}] (d) 
            };
        \end{feynman}
    \end{tikzpicture}
    \hspace{0.2cm}
    \Large $+$
    \hspace{0.2cm}
    $\sum_{i_1i_2}$
    $($
    \begin{tikzpicture}[scale=0.8, baseline={([yshift=-0.1cm]current bounding box.center)}]
        \begin{feynman}
            \vertex (a) {\(\mathcal{B}_b\)};
            \vertex [right=1.8cm of a] (b);
            \vertex [above right=1.5cm of b] (c);
            \vertex [right=1.5cm of c] (f1) {\(f_1\)};
            \vertex [below right=1.5cm of b] (d);
            \vertex [right=1.5cm of d] (f2) {\(f_2\)}; 
            
            \diagram* {
                (a) -- [fermion] (b) -- [fermion, edge label = \(i_1\)] (c) -- [fermion] (f1),
                (b) -- [fermion, edge label'= \(i_2\)] (d) -- [fermion] (f2), 
                (c) -- [fermion, edge label= \(k\)] (d)  
            };
        \end{feynman}
    \end{tikzpicture}
     $)$
    \caption{Feynman diagrams for $\mathcal{B}_b\to f_1 +f_2$ process with the final-state interaction~\cite{Wolfenstein:1990ks, Suzuki:1999uc, Cheng:2004ru, Jia:2024pyb}.}
    \label{fig:FSI}
\end{figure*}
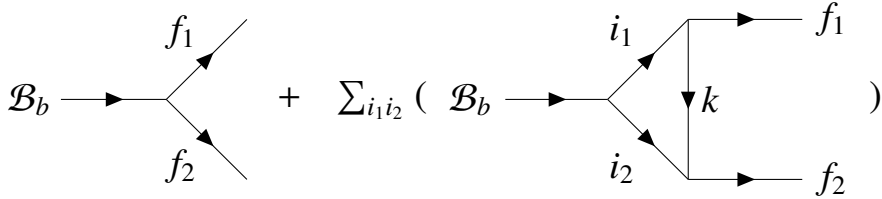

The final-state interaction provides a natural physical picture of the long-distance contribution in heavy flavor hadron decays, as it accounts for the non-factorizable contributions in $C$- and $E$-diagram, and generates the strong phase essential for $CP$ violation. In the time-evolution picture of scattering, the full amplitude can be expressed as~\cite{Cheng:2020ipp}
\begin{equation}\label{re-scattering}
	\begin{aligned}
\mathcal{A}(\mathcal{B}_b\to f)=\bra{f}\mathcal{H}_{eff}\ket{\mathcal{B}_b}+\sum_{i}\bra{f}U(+\infty,\tau)\ket{i}\bra{i}\mathcal{H}_{eff}\ket{\mathcal{B}_b} ,
	\end{aligned}
\end{equation}
where $\mathcal{B}_b$ represents the baryon either $\Lambda_b^0$ or $\Xi^-_b$. The Eq. (~\ref{re-scattering}) indicates that the decay process includes not only the weak matrix element of four-quark operators from the effective Hamiltonian, but also the rescattering contribution with the single particle exchange approximation at hadron level. These rescattering processes are described by triangle diagrams, which involve one weak vertex and two strong vertices, as illustrated in Fig.~\ref{fig:FSI}. A detailed explanation is provided in the following subsection.

\subsection{Short-distance contributions under the factorization hypothesis}

As shown in Eq.~\eqref{re-scattering}, the full amplitude contains two types of short-distance weak contributions: the direct short-distance weak process $\bra{f}\mathcal{H}_{eff}\ket{\mathcal{B}_b}$, and the short-distance weak vertex $\bra{i}\mathcal{H}_{eff}\ket{\mathcal{B}_b}$ occurring before rescattering process. Both of them describe $\mathcal{B}_b$ decays into a baryon and a meson, and can be expressed as:
 \begin{equation}\label{SD}
	\begin{aligned}
\bra{\mathcal{B}M}\mathcal{H}_{eff}\ket{\mathcal{B}_b}=\frac{G_{F}}{\sqrt{2}}V_{CKM}\sum_{j}C_{j}\bra{\mathcal{B}M}\mathcal{O}_{j}\ket{\mathcal{B}_b} .
	\end{aligned}
\end{equation}
Within the naive factorization approach, the matrix element on the right-hand side of the Eq. (~\ref{SD}) can be factorized into two parts : the first part is meson decay constant, the second part is the baryon transition matrix element. Employing Fiertz identities and the equation of motion, we simplify the matrix elements with different Wilson coefficients to the process-dependent coefficients multiplied by matrix elements of $V–A$ current. Then they can be further parameterized as heavy-to-light form factors.

For the direct short-distance weak process, we adopt the helicity-based approach to describe the baryonic matrix elements, since the final-state baryon $\Lambda(1520)$ is high spin. The matrix elements of the vector and axial vector currents are given by~\cite{Descotes-Genon:2019dbw,Das:2020cpv}
\begin{equation}
\begin{aligned}
	& \langle\Lambda(1520)(p_f)|\bar{s}\gamma_{\mu}b|\mathcal{B}_b(p_i)\rangle \\
    &= \bar{u}^{\alpha}(p_f)\Bigg\{(p_i)_{\alpha}\Bigg[f_{t}^{V}(q^{2})(m_{\mathcal{B}_b}-m_{\Lambda(1520)})\frac{q_{\mu}}{q^{2}} \\
	& +f_{0}^{V}(q^{2})\frac{m_{\mathcal{B}_b}+m_{\Lambda(1520)}}{s_{+}}\left((p_i)_{\mu}+(p_f)_\mu-(m_{\mathcal{B}_b}^{2}-m_{\Lambda(1520)}^{2})\frac{q_{\mu}}{q^{2}}\right) \\
	& + f_{\perp}^{V}(q^{2})\left(\gamma_{\mu}-\frac{2m_{\Lambda(1520)}}{s_{+}}(p_i)_{\mu}-\frac{2m_{\mathcal{B}_b}}{s_{+}}(p_f)_\mu\right) \bigg] \\
	& +f_{g}^{V}(q^{2})\biggl[g_{\alpha\mu}+m_{\Lambda(1520)}\frac{(p_i)_{\alpha}}{s_{-}}\biggl(\gamma_{\mu}-\frac{2p_{\mu}}{m_{\Lambda(1520)}}+\frac{2(m_{\Lambda(1520)}(p_i)_{\mu}+m_{\mathcal{B}_b}(p_f)_\mu)}{s_{+}}\biggr)\biggr]\biggr\}u_{\mathcal{B}_b}(p_i),
\end{aligned}
\end{equation}
\begin{equation}
\begin{aligned}
	& \langle\Lambda(1520)(p_f)|\bar{s}\gamma_{\mu}\gamma_{5}b|\mathcal{B}_b(p_i)\rangle \\
	& =- \bar{u}^{\alpha}(p_f)\gamma_{5}\Bigg\{(p_i)_{\alpha}\Bigg[g_{t}^{A}(q^{2})(m_{\mathcal{B}_b}+m_{\Lambda(1520)})\frac{q_{\mu}}{q^{2}}\Bigg] \\
	& +g_{0}^{A}(q^{2})\frac{m_{\mathcal{B}_b}-m_{\Lambda(1520)}}{s_{-}}\left((p_i)_{\mu}+(p_f)_\mu-(m_{\mathcal{B}_b}^{2}-m_{\Lambda(1520)}^{2})\frac{q_{\mu}}{q^{2}}\right) \\
	& +g_{\perp}^{A}(q^{2})\left(\gamma_{\mu}+\frac{2m_{\Lambda(1520)}}{s_{-}}(p_i)_{\mu}-\frac{2m_{\mathcal{B}_b}}{s_{-}}(p_f)_\mu\right) \bigg] \\
	& +g_g^A(q^2)\biggl[g_{\alpha\mu}-m_{\Lambda(1520)}\frac{(p_i)_\alpha}{s_+}\biggl(\gamma_\mu+\frac{2p_\mu}{m_{\Lambda(1520)}}-\frac{2(m_{\Lambda(1520)}(p_i)_\mu-m_{\Lambda_b}p_\mu)}{s_-}\biggr)\biggr]\biggr\}u_{\mathcal{B}_b}(p_i),
\end{aligned}
\end{equation}
where$s_{\pm}=(m_{\mathcal{B}_b}\pm m_\Lambda(1520))-q^2$. $u_{\mathcal{B}_b}(p_i)$ and $u^{\alpha}(p_f)$ are the Dirac spinors of the initial baryon $\mathcal{B}_{b}(p_{i})$ and the final baryon $\Lambda(1520)(p_f)$, respectively, and $q=p_i-p_f$ denotes the momentum transfer. $f^V$ and $g^A$ with different subscripts are the baryon transition form factors with different helicities, which are taken from the Ref.~\cite{Huang:2024oik}.
After replacing the meson decay constant, we obtain the full amplitude:

\begin{equation}
	\begin{aligned}
      \mathcal{A}(\mathcal{B}_{b}(p_{i})\to \Lambda(1520)(p_{f})P)=i\lambda f_pq^{\mu}\langle\Lambda(1520)(p_f)|\bar{s}\gamma_{\mu}(1-\gamma_5)b|\mathcal{B}_b(p_i)\rangle,
	\end{aligned}\label{Pesudo}
\end{equation}
\begin{equation}
	\begin{aligned}
       \mathcal{A}(\mathcal{B}_{b}(p_{i})\to  \Lambda(1520)(p_{f})V)=\lambda m_{\rm{v}} f_{\rm{v}} \epsilon^{*\mu}\langle\Lambda(1520)(p_f)|\bar{s}\gamma_{\mu}(1-\gamma_5)b|\mathcal{B}_b(p_i)\rangle,
	\end{aligned}\label{Vector}
\end{equation}
where $P$ and $V$ are the pseudoscalar and vector meson, respectively. $f_P$ and $f_V$ are correspond to the decay constants of pseudoscalar and vector mesons. $\epsilon^{*\mu}$ is the polarization of the vector meson $V$.  $\lambda$ is the process-dependent coefficient mentioned earlier, which we refer to as the $\lambda$ function. It represents the residual coefficients after simplifying the different Wilson coefficients into the matrix elements of $V–A$ current. The $\lambda$ functions for different processes are listed in Appendix~\ref{app.E}.

For short-distance weak vertex in triangle diagrams,  we employ a general parameterization scheme because of no high-spin particles involving, and provide the full amplitude as follows~\cite{Pakvasa:1990if}
\begin{equation}
	\begin{aligned}
       \mathcal{A}(\mathcal{B}_{b}(p_{i})\to \mathcal{B}(p_{f})P)=i\bar{u}(p_{f})\left[A+B\gamma_{5}\right]u(p_{i}),
	\end{aligned}\label{Pesudo}
\end{equation}
\begin{equation}
	\begin{aligned}
       \mathcal{A}(\mathcal{B}_{b}(p_{i})\to \mathcal{B}(p_{f})V)=\bar{u}(p_{f})\left[A_{1}\gamma_{\mu}\gamma_{5}+A_{2}\frac{(p_f)_\mu}{M_{i}}\gamma_{5}+B_{1}\gamma_{\mu}+B_{2}\frac{(p_f)_\mu}{M_{i}}\right]\epsilon^{*\mu}u(p_{i}),
	\end{aligned}\label{Vector}
\end{equation}
the parameters $A$, $B$, $A_{1},A_{2}$, $B_{1}$ and $B_{2}$ are derived within naive factorization
\begin{equation}
	\begin{aligned}
      A&=\lambda_A f_{P}(M_{i}-M_{f})f_{1}(q^{2}),\\
      B&=\lambda_B f_{P}(M_{i}+M_{f})g_{1}(q^{2}),\\
     A_{1}&=-\lambda m_{V}f_{V}\left[g_{1}(q^{2})+\frac{M_{i}-M_{f}}{M_{i}}g_{2}(q^{2})\right],\\
    B_{1}&=\lambda m_{V}f_{V}\left[f_{1}(q^{2})-\frac{M_{i}+M_{f}}{M_{i}}f_{2}(q^{2})\right],\\
         A_{2}&=-2\lambda m_{V}f_{V}f_{2}(q^{2}),\\
    B_{2}&=2\lambda m_{V}f_{V}g_{2}(q^{2}),
	\end{aligned}
\end{equation}
where $f_1,f_2,g_1$ and $g_2$ denote the heavy-to-light transition form factors in $\mathcal{B}_b$ decays. The numerical values for the form factors are summarized in the subsection~\ref{input}.

\subsection{Long-distance contributions with the rescattering mechanism}
For a given channel $\mathcal{B}_b \to f_1 + f_2$, there is only one direct short-distance weak process. In contrast, the rescattering process exists plenty of possibilities, since excited states can be inserted as intermediate particles in the triangle diagram. However, for these highly excited states, there is neither relevant Lagrangians nor information on strong coupling vertex. This will lead to large uncertainties in our calculations or even make the predictions impossible. Therefore, we require a limited set of particles involved in the rescattering process. In this work, we only take into account several possible types of triangle diagrams, as shown in Fig.~\ref{fig:six_feynman}. With a limited set, we have listed all possible triangle diagrams for each process in the Appendix~\ref{app.A}.

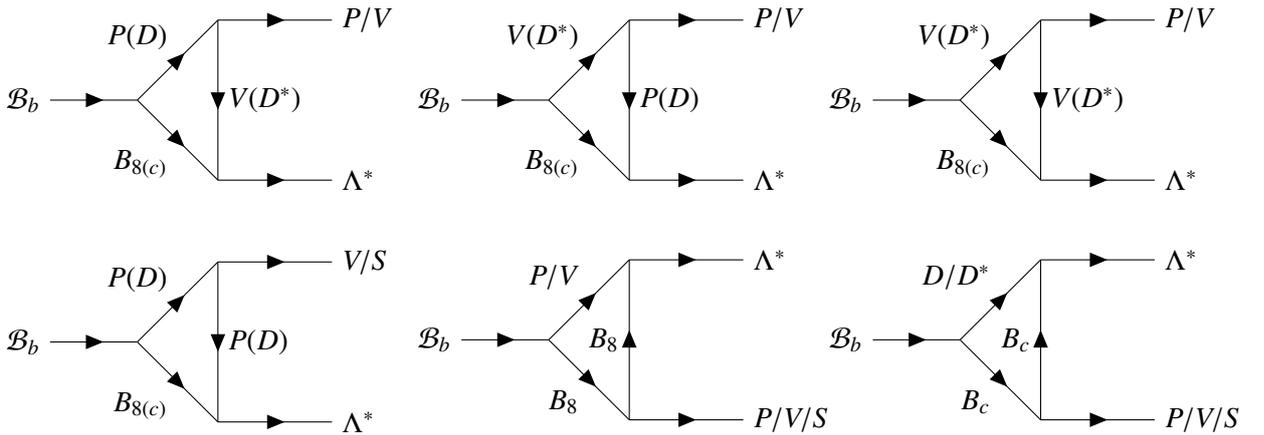
\begin{figure*}[h]
    \centering
    \begin{minipage}[b]{0.32\textwidth}
        \centering
        \begin{tikzpicture}[scale=0.4]
            \begin{feynman}
                \vertex (a) {\(\mathcal{B}_b\)};
                \vertex [right=of a] (b);
                \vertex [above right=of b] (c);
                \vertex [right=of c] (f1) {\(P/V\)};
                \vertex [below right=of b] (d);
                \vertex [right=of d] (f2) {\(\Lambda^*\)}; 
                
                \diagram* {
                    (a) -- [fermion] (b) -- [fermion, edge label = \(P(D)\)] (c) -- [fermion] (f1),
                    (b) -- [fermion, edge label'= \(B_{8(c)}\)] (d) -- [fermion] (f2), 
                    (c) -- [fermion, edge label= \(V(D^*)\)] (d)  
                };
            \end{feynman}
        \end{tikzpicture}
    \end{minipage}
    \hfill
    \begin{minipage}[b]{0.32\textwidth}
        \centering
        \begin{tikzpicture}[scale=0.4]
            \begin{feynman}
                \vertex (a) {\(\mathcal{B}_b\)};
                \vertex [right=of a] (b);
                \vertex [above right=of b] (c);
                \vertex [right=of c] (f1) {\(P/V\)};
                \vertex [below right=of b] (d);
                \vertex [right=of d] (f2) {\(\Lambda^*\)}; 
                
                \diagram* {
                    (a) -- [fermion] (b) -- [fermion, edge label = \(V(D^*)\)] (c) -- [fermion] (f1),
                    (b) -- [fermion, edge label'= \(B_{8(c)}\)] (d) -- [fermion] (f2), 
                    (c) -- [fermion, edge label= \(P(D)\)] (d)  
                };
            \end{feynman}
        \end{tikzpicture}
    \end{minipage}
    \hfill
    \begin{minipage}[b]{0.32\textwidth}
        \centering
        \begin{tikzpicture}[scale=0.4]
            \begin{feynman}
                \vertex (a) {\(\mathcal{B}_b\)};
                \vertex [right=of a] (b);
                \vertex [above right=of b] (c);
                \vertex [right=of c] (f1) {\(P/V\)};
                \vertex [below right=of b] (d);
                \vertex [right=of d] (f2) {\(\Lambda^*\)}; 
                
                \diagram* {
                    (a) -- [fermion] (b) -- [fermion, edge label = \(V(D^*)\)] (c) -- [fermion] (f1),
                    (b) -- [fermion, edge label'= \(B_{8(c)}\)] (d) -- [fermion] (f2), 
                    (c) -- [fermion, edge label= \(V(D^*)\)] (d)  
                };
            \end{feynman}
        \end{tikzpicture}
    \end{minipage}
    
    \vspace{0.5cm} 
    
    \begin{minipage}[b]{0.32\textwidth}
        \centering
        \begin{tikzpicture}[scale=0.4]
            \begin{feynman}
                \vertex (a) {\(\mathcal{B}_b\)};
                \vertex [right=of a] (b);
                \vertex [above right=of b] (c);
                \vertex [right=of c] (f1) {\(V/S\)};
                \vertex [below right=of b] (d);
                \vertex [right=of d] (f2) {\(\Lambda^*\)}; 
                
                \diagram* {
                    (a) -- [fermion] (b) -- [fermion, edge label = \(P(D)\)] (c) -- [fermion] (f1),
                    (b) -- [fermion, edge label'= \(B_{8(c)}\)] (d) -- [fermion] (f2), 
                    (c) -- [fermion, edge label= \(P(D)\)] (d)  
                };
            \end{feynman}
        \end{tikzpicture}
    \end{minipage}
    \hfill
    \begin{minipage}[b]{0.32\textwidth}
        \centering
        \begin{tikzpicture}[scale=0.4]
            \begin{feynman}
                \vertex (a) {\(\mathcal{B}_b\)};
                \vertex [right=of a] (b);
                \vertex [above right=of b] (c);
                \vertex [right=of c] (f1) {\(\Lambda^*\)};
                \vertex [below right=of b] (d);
                \vertex [right=of d] (f2) {\(P/V/S\)}; 
                
                \diagram* {
                    (a) -- [fermion] (b) -- [fermion, edge label = \(P/V\)] (c) -- [fermion] (f1),
                    (b) -- [fermion, edge label'= \(B_{8}\)] (d) -- [fermion] (f2), 
                    (d) -- [fermion, edge label= \(B_{8}\)] (c)  
                };
            \end{feynman}
        \end{tikzpicture}
    \end{minipage}
    \hfill
    \begin{minipage}[b]{0.32\textwidth}
        \centering
        \begin{tikzpicture}[scale=0.4]
            \begin{feynman}
                \vertex (a) {\(\mathcal{B}_b\)};
                \vertex [right=of a] (b);
                \vertex [above right=of b] (c);
                \vertex [right=of c] (f1) {\(\Lambda^*\)};
                \vertex [below right=of b] (d);
                \vertex [right=of d] (f2) {\(P/V/S\)}; 
                
                \diagram* {
                    (a) -- [fermion] (b) -- [fermion, edge label = \(D/D^*\)] (c) -- [fermion] (f1),
                    (b) -- [fermion, edge label'= \(B_{c}\)] (d) -- [fermion] (f2), 
                    (d) -- [fermion, edge label= \(B_{c}\)] (c)  
                };
            \end{feynman}
        \end{tikzpicture}
    \end{minipage}
    
    \caption{The long-distance rescattering contributions to $\mathcal{B}_{b}$ decays at hadron level in single-particle exchange approximation, where the $\mathcal{B}_{b},B_{c},B_{8},\Lambda^*$ denote bottom, charm ,octet baryons and $\Lambda(1520)$, and $P,V,S,D,D^*$ are pseudoscalar octet, vector octet, scalar mesons, $D$ and $D^{*}$ mesons, respectively.}
    \label{fig:six_feynman}
\end{figure*}

To perform a full calculation of the triangle diagram, we need combine the second type of short-distance vertex derived previously with the hadronic rescattering process. The hadronic rescattering process is governed by effective Lagrangians and all the Lagrangians used in this work are compiled in the Appendix~\ref{app.B}. 
By inserting the effective Lagrangians into the matrix elements for specific initial and final states, we obtain the Feynman rules for the two strong interaction vertices, which are listed in Appendix~\ref{app.C}. 
By performing loop integration, we can obtain the full analytical amplitude of the triangle diagrams shown in Fig.~\ref{fig:six_feynman}. According to the classification, all triangle diagrams of a given type follow identical Feynman rules, since they originate from the same effective Lagrangians. We list the Feynman rules for all types of triangle diagrams in Appendix~\ref{app.D}. 
Taking $\mathcal{B}_b \to [P,B_8;V] \to \Lambda^* P$ decay process as an example, $\mathcal{B}_b$ first decays into $B_8$ and $P$ through a weak vertex, then $B_8$ and $P$ exchange the particle $V$ via strong vertices, finally producing the final state $\Lambda^*$ and $P$. 
We present its full analytical amplitude with the method discussed above
\begin{equation}
	\begin{aligned}
\mathcal{M}[P,B_8;V]&=\displaystyle\int\frac{d^4p}{(2\pi)^4}(+i)\frac{g_{VB\Lambda}}{m_k^2}\frac{g_{PPV}}{\sqrt{2}}\bar{u^{\sigma}}(p_4, s_4)\sigma_{\mu\nu}k^{\nu}k_{\sigma}(\slashed{p_{2}} + m_{2})(A + B\gamma_{5}) u(p,s)(-g^{\mu\alpha} + \frac{k^\mu k^\alpha}{m_{k}^{2}})\\
&(p_1+p_3)_\alpha \mathcal{P}^3\mathcal{F},
    \end{aligned}
\end{equation}
in the expression, $p$ denotes the momentum of the initial state particle, $p_1$ and $p_2$ represent the momentum of the meson and baryon produced at the weak vertex, respectively, while $p_3$ and $p_4$ correspond to the momentum of the meson and baryon after the rescattering process. The momentum of the exchanged particle is denoted by $k$, besides, $\mathcal{P}^3$ denotes the three propagators, and $\mathcal{F}$ denotes the form factor, which is applied to regularize the UV divergence in the loop integral and adjust the strength of the two strong coupling constants. It is necessary to carry out adjustment because these coupling constants in the triangle diagrams are calculated based on the on-shell assumption, rather than the off-shell condition.
Following our previous work, we continue to use the following model for the form factor~\cite{Yue:2024paz,Duan:2024zjv}
\begin{equation}\label{FF}
\mathcal{F}(\Lambda, m_k) = \frac{\Lambda^4}{(k^2 - m_k^2)^2 + \Lambda^4},
\end{equation}
where $\Lambda$ is a model parameter. This model does not use the Pauli-Villars regularization scheme to handle UV divergence, so no additional ghost fields and imaginary parts are introduced and it is beneficial for us to calculate $CP$ violation. Depending on whether the triangle diagram contains charm quark, we define two model parameters, $\Lambda_{\rm{charm}}=1$ and $\Lambda_{\rm{charmless}}=0.5$. Because decay modes are same at the quark-level, these two model parameters are determined by the fits to experimental measurements of channels $\Lambda_b^0 \to pK^-$ and $p\pi^-$ in previous work~\cite{Duan:2024zjv}. It effectively distinguishes the mass difference between the processes with and without charm quark, which significantly affects the off-shell behavior of the intermediate particles, an effect that is not occurred in charmed baryon decays. This implies that all our subsequent calculations have no additional model freedom, as all parameters are taken from our earlier studies directly. This also presents a challenge to both the FSI framework and the model for the form factor.

\section{Helicity amplitudes }\label{hamplitudes}

In this section, we first present all the input parameters used in the numerical analysis. We then calculate the helicity amplitudes for the eight decay channels $\Lambda^0_b\to \Lambda(1520)+(\pi^0, \kappa(700), f_0(500, 980), \rho^0, K^{*0}, \phi)$ and $\Xi^-_b \to \Lambda(1520)+K^-$. Finally, we analyze and discuss the short-distance contribution, as well as the rescattering contributions involving charmed and charmless hadrons of each helicity amplitudes.

\subsection{Input parameters}\label{input}

We first summarize the mass parameters involved in all the decay channels in Table~\ref{tab:masses_all}, including the masses of baryons, mesons, and quarks~\cite{ParticleDataGroup:2022pth}. It should be noted that $\kappa(700)$ and $f_0(500, 980)$ are both scalar mesons with spin $J=0$. $\Lambda(1520)$ is an excited baryon with spin $J=\tfrac{3}{2}$ in comparison with the nucleons (proton and neutron) with spin $J=\tfrac{1}{2}$. The light quark masses correspond to the $\overline{\text{MS}}$ masses evaluated at the renormalization scale $\mu = 2~\text{GeV}$, and the heavy quark masses $m_b$ and $m_c$ are given in the $\overline{\text{MS}}$ scheme at the scale of their respective $\overline{\text{MS}}$ masses.
Subsequently, we list all the decay constants of the mesons in Table~\ref{tab:decay_constant}. 
The definition of $f_{\text{meson}}^{q\bar{q}}$ refers to the decay constants induced by the pseudoscalar current $\bar{q}\gamma_\mu\gamma_5 q$.

\begin{table}[htbp]
\centering
\caption{Masses of baryons, mesons, and quarks used in this work~\cite{ParticleDataGroup:2022pth}.}
\label{tab:masses_all}
\renewcommand{\arraystretch}{1.4}
\setlength{\tabcolsep}{4pt}
\begin{tabular}{c|c|c|c|c|c|c|c} 
\hline\hline
Parameters & $\Lambda_b^0$ & $\Xi^-_b$& $\Lambda_c^+$ & $p$ & $n$ & $\Lambda(1520)$  & $\Lambda^0$ \\
\hline
Mass [MeV] & 5.619 & 5.797& 2.286 & 0.938 & 0.940 & 1.515  & 1.116 \\
\hline
\hline
Parameters & $D^{\pm}$ & $D^0$ & $D^*$ & $D_s$ & $D_s^*$ & $K$ & $K^*$ \\
\hline
Mass [MeV] & 1.869 & 1.865 & 2.007 & 1.968 & 2.106 & 0.493 & 0.892 \\
\hline
\hline
Parameters & $\rho$ & $\pi$ & $\omega$ & $\phi$ & $f_0(500)$ & $f_0(980)$ & $\kappa(700)$ \\
\hline
Mass [MeV] & 0.770 & 0.140 & 0.783 & 1.020 & 0.513 & 0.980 &0.838  \\
\hline
\hline
Parameters & $u$ & $d$ & $s$ & $c$ & $b$ &  &  \\
\hline
Mass [MeV] & 2.16 & 4.70 & 93.5 & 1270 & 4180 &  &  \\
\hline\hline
\end{tabular}
\end{table}
\begin{table}[htbp]
\renewcommand{\arraystretch}{1.5}
\addtolength{\arraycolsep}{3pt}
	\centering
\caption{The decay constants of mesons used in this work~\cite{Zhu:2018jet,Qi:2024ddp,Liu:2019ymi}.} \label{tab:decay_constant}
	\begin{tabular}{c|c|c|c|c|c|c|c|c|c|c} 
		\hline
		\hline
Decay constant &  $f_\pi$ & $f_K$  &  $f_\rho$  & $f_\omega$  & $f_\phi$ &  $f_{K^*}$  & $f_{D}$ & $f_{D^{*}}$ & $f_{D_{s}}$ &  $f_{D^{*}_{s}}$\\
\hline
Value [MeV] &  $ 131 $ & $ 160 $ &  $ 216$ & $  195$ & $ 233 $ & $210 $ & $210$ & $220$ & $230$&  $ 271$\\
\hline
\hline
Decay constant  & $f_{J/\psi}$    & $f_{\eta}^{u\bar{u}}$  & $f_{\eta}^{d\bar{d}}$ &  $f_{\eta}^{s\bar{s}} $ & $f_{\eta^{\prime}}^{u\bar{u}}$ & $f_{\eta^{\prime}}^{d\bar{d}}$ & $f_{\eta^{\prime}}^{s\bar{s}}$ &$f_{f_0(500)}^{s\bar{s}}$&$f_{f_0(980)}^{s\bar{s}}$&$f_{\kappa(700)}$\\
\hline
Value [MeV]   & $ 395 $  & $  54$ & $ 54 $ & $-111 $ & $44$ & $44$ & $136$ & $113$ & $310$ & $420$\\ 
		\hline
		\hline
	\end{tabular} 
\end{table}

In the Standard Model, the CKM matrix describes the mixing among different quark flavors in the weak decays. The CKM matrix is expanded in the Wolfenstein parameterization as~\cite{Kobayashi:1973fv, Cabibbo:1963yz}
\begin{align}
V_{\text{CKM}} =
\begin{pmatrix}
V_{ud} & V_{us} & V_{ub} \\
V_{cd} & V_{cs} & V_{cb} \\
V_{td} & V_{ts} & V_{tb}
\end{pmatrix} = \begin{pmatrix}
    1 - \lambda_W^2/2 & \lambda_W & A\lambda_W^3(\rho - i\eta) \\ 
    -\lambda_W & 1 - \lambda_W^2/2 & A\lambda_W^2 \\
    A\lambda_W^3(1 - \rho -i\eta) & -A\lambda_W^2 & 1
\end{pmatrix} + \mathcal{O}(\lambda_W^4),
\end{align}
where the Wolfenstein parameters are taken to be $A=0.823, \rho=0.141,\eta=0.349$ and $\lambda_{W}=0.225$~\cite{ParticleDataGroup:2022pth}.

In the short-distance contributions, the heavy-to-light transition form factors  serve as the most fundamental and universal hadronic parameters 
describing the QCD effects in weak decays. 
They have been extensively investigated in various theoretical methods, such as perturbative QCD approach~\cite{Lu:2009cm}, (covariant) light-front quark model~\cite{Ke:2007tg,Wei:2009np}, QCD factorization~\cite{Zhu:2016bra}, soft-collinear effective theory~\cite{Feldmann:2011xf} and light cone sum rules~\cite{Khodjamirian:2011jp}. In this work, we adopt the theoretical results for the heavy-to-light baryonic form factors for $\Lambda_b \to p, n, \Lambda^0, \Lambda_c$ processes from Ref.~\cite{Zhu:2018jet}, and the heavy-to-light baryonic form factors for $\Xi_b \to \Xi_c, \Xi, \Sigma, \Lambda$ processes from Ref.~\cite{Faustov:2018ahb,Azizi:2011mw,Rui:2025iwa}  which are summarized in Table~\ref{tab:form factors}. It should be noted that the form factor values presented in the table correspond to $q^2 = 0$. The Non-zero $q^2$-dependence needed for the computation follows the analytical expressions from the original literature.

\begin{table}[htbp]
    \centering
    \caption{The values of the $\Lambda_b \to p, n, \Lambda, \Lambda_c$ and $\Xi_b \to \Xi_c, \Xi, \Sigma, \Lambda$ transition form factors from covariant light-front quark model and light cone sum rules~\cite{Zhu:2018jet,Faustov:2018ahb,Azizi:2011mw,Rui:2025iwa}.}
    \label{tab:form factors}
    \renewcommand{\arraystretch}{1.2} 
    \setlength{\tabcolsep}{28pt}      
    \begin{tabular}{c|c|c|c|c} 
    \hline \hline
       Decay & $f_1(0)$ & $f_2(0)$ & $g_1(0)$ & $g_2(0)$ \\ \hline
       $\Lambda_b \to \Lambda_c$ & $0.500$ & $-0.098$ & $0.509$ & $-0.015$ \\ \hline 
       $\Lambda_b \to p$ & $0.128$ & $-0.056$ & $0.129$ & $-0.033$ \\ \hline 
       $\Lambda_b \to \Lambda$ & $0.131$ & $-0.048$ & $0.132$ & $-0.023$ \\ \hline 
       $\Lambda_b \to n$ & $0.128$ & $-0.056$ & $0.129$ & $-0.033$ \\ \hline 
       $\Xi_b \to \Xi_c$ & $ 0.554$ & $ 0.012$ & $0.552$ & $-0.012$ \\ \hline 
       $\Xi_b \to \Xi$ & $0.142$ & $-0.020$ & $0.160$ & $-0.009$ \\ \hline 
       $\Xi_b \to \Sigma$ & $0.046$ & $-0.033$ & $0.067$ & $-0.024$ \\ \hline 
       $\Xi_b \to \Lambda$ & $0.092$ & $0.029$ & $0.077$ & $0.007$ \\ \hline \hline 
    \end{tabular}
\end{table}

In addition, the strong coupling constants are the essential input parameters for final-state rescattering processes, as they play a crucial role in determining the relative magnitudes of the same type triangle diagrams. We give priority to using the results obtained from experiments, in the absence of such results, we derive them from the existing results for vertices governed by the same Lagrangians with $SU(3)$ symmetry. For couplings that do not involve scalar mesons or $\Lambda(1520)$, no new parameters are introduced in this work and all the results are taken from previous studies~\cite{Duan:2024zjv, Cheng:2004ru, Aliev:2006xr, Aliev:2009ei, Janssen:1996kx, Meissner:1987ge, Yu:2017zst}. The new types of vertices that appeared in the work include $PB\Lambda(1520)$, $VB\Lambda(1520)$, $SPP$, and $SBB$, where the $P, S, V, B$ correspond to the pseudoscalar, scalar, vector mesons and baryons.

Coupling constants of the type \( PB\Lambda(1520) \) and \( VB\Lambda(1520) \) are derived from \( g_{N\bar{K}\Lambda(1520)} = 10.5 \) and \( g_{\rho NN(1520)} = 4.5 \), respectively, through their corresponding symmetry transformations~\cite{Nam:2005uq, Riska:2000gd}.
For the \( SPP \) vertices, the coupling constants are determined from 
\( g_{f_0(980)KK} = 4.4 \), \( g_{f_0(980)\pi\pi} = 1.6 \), 
\( g_{f_0(500)KK} = 1.4 \), and \( g_{f_0(500)\pi\pi} = 2.9 \)
following the same procedure~\cite{Ahmed:2020kmp}. The coupling constants of the \( SBB \) type are taken from Ref.~\cite{Erkol:2006eq}. 
For the vertices that do not involve charmless particles, we apply \( SU(3) \) the symmetry transformation, while
for those involving charmed particles, such as \( g_{D_s^-\Lambda_c^+\Lambda(1520)} \), 
we employ the \( SU(4) \) symmetry transformation. 
Although \( SU(4) \) symmetry is significantly broken due to the large mass of the charm quark, 
which renders the resulting coupling constants subject to large uncertainties, the model parameter $\Lambda_{\rm{charm}}$ introduced for charmed triangle diagrams compensates for this problem effectively. The form factor $\mathcal{F}(\Lambda, m_k)$ acts as a regulator of off-shell effects and simultaneously absorbs the uncertainties arising from \( SU(4) \) symmetry breaking. The detailed procedure of \( SU(4) \) symmetry method is given in Appendix~\ref{app.F}. 

\subsection{Numerical results}
We first compute the helicity amplitudes of 
the following eight decay channels: $\Lambda^0_b\to \Lambda(1520)\pi^0$, $\Lambda(1520)\kappa(700)$, $\Lambda(1520)f_0(500, 980)$, $\Lambda(1520)\rho^0$,    $\Lambda(1520)K^{*0}$, $\Lambda(1520)\phi$ and $\Xi^-_b \to \Lambda(1520)K^-$. The helicity amplitudes not only provide a foundation for further theoretical analysis but can also be tested experimentally through partial-wave analysis.
By factoring out the CKM matrix elements, the results make it possible to identify the specific contribution of each CKM element to the various helicity amplitudes, thereby providing a basis for subsequent studies of $CP$ violation in each channel.
The helicity amplitude results for the eight decay channels, corresponding to different CKM factors, are presented in Table~\ref{Tab: Helicity Lb to Lpi}- Table~\ref{Tab: Helicity Xb to Lk}. Here, $\mathcal{S}$, $\mathcal{NC}$, and $\mathcal{C}$ denote the short-distance amplitude, the charmless rescattering amplitude and the charmed rescattering amplitude, respectively.

\begin{sidewaystable}
\begin{table}[H]
\caption{Helicity amplitudes of $\Lambda_b^0 \rightarrow  \Lambda(1520)+\pi^0\, (10^{-7})$ with different CKM factors}
\centering
\begin{tabular}{ccccccc}
\toprule
\toprule
 decay modes & $H_{-\frac{1}{2}} ~(V_{ub}V_{us}^*) $ &$H_{-\frac{1}{2}} ~(V_{cb}V_{cs}^*)$ & $H_{-\frac{1}{2}} ~(V_{tb}V_{ts}^*)$ & $H_{\frac{1}{2}} ~(V_{ub}V_{us}^*) $ &$H_{\frac{1}{2}} ~(V_{cb}V_{cs}^*)$ & $H_{\frac{1}{2}} ~(V_{tb}V_{ts}^*)$ \\
\midrule
$\mathcal{S}(\Lambda_b^0 \rightarrow \Lambda(1520)+\pi^0)$& 0.20 $i$ & --- & 0.03 $i$  & 1.16 $i$ & --- & 0.16 $i$  \\
$\mathcal{NC}(\Lambda_b^0 \rightarrow \Lambda(1520)+\pi^0)$& 1.26+10.07 $i$ &--- & -0.003+0.07 $i$& 1.85+10.40 $i$&---&-0.008+0.03 $i$\\
\bottomrule
\bottomrule
\end{tabular}\label{Tab: Helicity Lb to Lpi}
\end{table}

\begin{table}[H]
\caption{Helicity amplitudes of $\Lambda_b^0\rightarrow \Lambda(1520)+\rho\, (10^{-7})$ with different CKM factors}
\centering
\begin{tabular}{cccccccc}
\toprule
\toprule
 decay modes & $H_{\frac{3}{2},1} ~(V_{ub}V_{us}^*) $ &$H_{\frac{3}{2},1} ~(V_{cb}V_{cs}^*)$&$H_{\frac{3}{2},1} ~(V_{tb}V_{ts}^*)$ & $H_{-\frac{3}{2},-1} ~(V_{ub}V_{us}^*) $ &$H_{-\frac{3}{2},-1} ~(V_{cb}V_{cs}^*)$&$H_{-\frac{3}{2},-1} ~(V_{tb}V_{ts}^*)$ \\
\midrule
$\mathcal{S}(\Lambda_b^0\rightarrow \Lambda(1520) +\rho )$&0.00&---& 0.00 &0.00&--- 
 &0.00 \\
$\mathcal{NC}(\Lambda_b^0\rightarrow \Lambda(1520) +\rho )$&0.36-3.56 $i$&---& 0.01-0.02$i$&-20.61-4.58 $i$ &---&-0.12-0.07 $i$\\
\midrule
 decay modes& $H_{\frac{1}{2},0} ~(V_{ub}V_{us}^*) $ &$H_{\frac{1}{2},0}  ~(V_{cb}V_{cs}^*)$&$H_{\frac{1}{2},0}  ~(V_{tb}V_{ts}^*)$ & $H_{-\frac{1}{2},0} ~(V_{ub}V_{us}^*) $ &$H_{-\frac{1}{2},0}  ~(V_{cb}V_{cs}^*)$&$H_{-\frac{1}{2},0} ~(V_{tb}V_{ts}^*)$ \\
\midrule
$\mathcal{S}(\Lambda_b^0\rightarrow \Lambda(1520) +\rho )$& 1.87 & --- & 0.19&-0.15&---& -0.02\\
$\mathcal{NC}(\Lambda_b^0\rightarrow \Lambda(1520) +\rho )$&130.12+31.69 $i$&---&0.66+0.48 $i$&-20.91+27.56 $i$&---&-0.09+0.10 $i$\\
\midrule
 decay modes& $H_{-\frac{1}{2},-1} ~(V_{ub}V_{us}^*) $ &$H_{-\frac{1}{2},-1} ~(V_{cb}V_{cs}^*)$&$H_{-\frac{1}{2},-1} ~(V_{tb}V_{ts}^*)$ & $H_{\frac{1}{2},1}~(V_{ub}V_{us}^*) $ &$H_{\frac{1}{2},1} ~(V_{cb}V_{cs}^*)$&$H_{\frac{1}{2},1}~(V_{tb}V_{ts}^*)$ \\
\midrule
$\mathcal{S}(\Lambda_b^0\rightarrow \Lambda(1520) +\rho)$& -4.54 & --- & -0.46&0.41&---& 0.04\\
$\mathcal{NC}(\Lambda_b^0\rightarrow \Lambda(1520) +\rho)$&-47.92-12.12$i$&---&-0.29-0.19 $i$&15.65-3.63 $i$&---&0.11+0.02 $i$\\
\bottomrule
\bottomrule
\end{tabular}\label{Tab: Lb to Lphi}
\end{table}

\end{sidewaystable}

\begin{sidewaystable}

\begin{table}[H]
\caption{Helicity amplitudes of $\Lambda_b^0\rightarrow\Lambda(1520)+f_0(980)\, (10^{-7})$ with different CKM factors}
\centering
\begin{tabular}{ccccccc}
\toprule
\toprule
 decay modes & $H_{-\frac{1}{2}} ~(V_{ub}V_{us}^*) $ &$H_{-\frac{1}{2}} ~(V_{cb}V_{cs}^*)$ & $H_{-\frac{1}{2}} ~(V_{tb}V_{ts}^*)$ & $H_{\frac{1}{2}} ~(V_{ub}V_{us}^*) $ &$H_{\frac{1}{2}} ~(V_{cb}V_{cs}^*)$ & $H_{\frac{1}{2}} ~(V_{tb}V_{ts}^*)$ \\
\midrule
$\mathcal{S}(\Lambda_b^0\rightarrow\Lambda(1520)+f_0(980))$&  ---  & --- & -0.60   &  --- & --- & -0.31   \\
$\mathcal{NC}(\Lambda_b^0\rightarrow\Lambda(1520)+f_0(980))$& -3.63-2.57 $i$ &--- & 0.35+0.11$i$& -7.70-3.37 $i$&---&0.03+0.02 $i$\\
$\mathcal{C}(\Lambda_b^0\rightarrow\Lambda(1520)+f_0(980))$& --- & -2.63-1.36$i$ & 0.20-0.29$i$ & ---&-2.79+9.99 $i$&-0.18+0.67 $i$\\
\bottomrule
\bottomrule
\end{tabular}\label{Tab: Helicity Lb to L980}
\end{table}

\begin{table}[H]
\caption{Helicity amplitudes of $\Lambda_b^0\rightarrow \Lambda(1520) +\phi\, (10^{-7})$ with different CKM factors}
\centering
\begin{tabular}{cccccccc}
\toprule
\toprule
 decay modes & $H_{\frac{3}{2},1} ~(V_{ub}V_{us}^*) $ &$H_{\frac{3}{2},1} ~(V_{cb}V_{cs}^*)$&$H_{\frac{3}{2},1} ~(V_{tb}V_{ts}^*)$ & $H_{-\frac{3}{2},-1} ~(V_{ub}V_{us}^*) $ &$H_{-\frac{3}{2},-1} ~(V_{cb}V_{cs}^*)$&$H_{-\frac{3}{2},-1} ~(V_{tb}V_{ts}^*)$ \\
\midrule
$\mathcal{S}(\Lambda_b^0\rightarrow \Lambda(1520) +\phi )$&---&---& 0.00 &---&--- 
 &0.00 \\
$\mathcal{NC}(\Lambda_b^0\rightarrow \Lambda(1520) +\phi )$&5.71 - 0.98 $i$&---& 0.14 + 0.03$i$&-1.17 + 0.61 $i$ &---&-0.07 - 0.03 $i$\\
$\mathcal{C}(\Lambda_b^0\rightarrow \Lambda(1520) +\phi )$&---& 7.38 + 4.55 $i$&0.07 + 0.04$i$&---&2.42 + 0.92 $ i$& -0.25 - 0.27 $i$\\
\midrule
 decay modes& $H_{\frac{1}{2},0} ~(V_{ub}V_{us}^*) $ &$H_{\frac{1}{2},0}  ~(V_{cb}V_{cs}^*)$&$H_{\frac{1}{2},0}  ~(V_{tb}V_{ts}^*)$ & $H_{-\frac{1}{2},0} ~(V_{ub}V_{us}^*) $ &$H_{-\frac{1}{2},0}  ~(V_{cb}V_{cs}^*)$&$H_{-\frac{1}{2},0} ~(V_{tb}V_{ts}^*)$ \\
\midrule
$\mathcal{S}(\Lambda_b^0\rightarrow \Lambda(1520) +\phi )$& --- & --- & -0.48&---&---& 0.12\\
$\mathcal{NC}(\Lambda_b^0\rightarrow \Lambda(1520) +\phi )$&14.84 - 9.15 $i$&---&-0.15 + 0.07 $i$&-7.50 + 2.61 $i$&---&0.34 + 0.01 $i$\\
$\mathcal{C}(\Lambda_b^0\rightarrow \Lambda(1520) +\phi)$&---&4.04 - 5.28 $i$&1.44 + 1.12 $i$&---& -21.70 - 0.05 $i$&0.08 - 1.54 $i$\\
\midrule
 decay modes& $H_{-\frac{1}{2},-1} ~(V_{ub}V_{us}^*) $ &$H_{-\frac{1}{2},-1} ~(V_{cb}V_{cs}^*)$&$H_{-\frac{1}{2},-1} ~(V_{tb}V_{ts}^*)$ & $H_{\frac{1}{2},1}~(V_{ub}V_{us}^*) $ &$H_{\frac{1}{2},1} ~(V_{cb}V_{cs}^*)$&$H_{\frac{1}{2},1}~(V_{tb}V_{ts}^*)$ \\
\midrule
$\mathcal{S}(\Lambda_b^0\rightarrow \Lambda(1520) +\phi)$& --- & --- & 1.51&---&---& -0.14\\
$\mathcal{NC}(\Lambda_b^0\rightarrow \Lambda(1520) +\phi)$&-8.56 + 0.66$i$&---&-0.22 - 0.01 $i$&14.90 - 3.32 $i$&---&0.04 + 0.02 $i$\\
$\mathcal{C}(\Lambda_b^0\rightarrow \Lambda(1520) +\phi )$&---&17.16 + 9.68$i$&0.11 - 1.43$i$&---& 15.30 + 18.29$i$& 0.30 + 0.13 $i$\\
\bottomrule
\bottomrule
\end{tabular}\label{Tab: Lb to Lphi}
\end{table}

\end{sidewaystable}

\begin{sidewaystable}
\begin{table}[H]
\caption{Helicity amplitudes of $\Lambda_b^0\rightarrow\Lambda(1520)+f_0(500)\, (10^{-7})$ with different CKM factors}
\centering
\begin{tabular}{ccccccc}
\toprule
\toprule
 decay modes & $H_{-\frac{1}{2}} ~(V_{ub}V_{us}^*) $ &$H_{-\frac{1}{2}} ~(V_{cb}V_{cs}^*)$ & $H_{-\frac{1}{2}} ~(V_{tb}V_{ts}^*)$ & $H_{\frac{1}{2}} ~(V_{ub}V_{us}^*) $ &$H_{\frac{1}{2}} ~(V_{cb}V_{cs}^*)$ & $H_{\frac{1}{2}} ~(V_{tb}V_{ts}^*)$ \\
\midrule
$\mathcal{S}(\Lambda_b^0\rightarrow\Lambda(1520)+f_0(500))$&  ---  & --- & 0.10   &  --- & --- & 0.03   \\
$\mathcal{NC}(\Lambda_b^0\rightarrow\Lambda(1520)+f_0(500))$& 8.64-0.63 $i$ &--- & 0.29+0.09$i$& 2.08-1.95 $i$&---&-0.33-0.03 $i$\\
$\mathcal{C}(\Lambda_b^0\rightarrow\Lambda(1520)+f_0(500))$& --- & -0.87-0.44$i$ & 0.06-0.12 $i$ & ---&-0.71+3.95 $i$&-0.06+0.26 $i$\\
\bottomrule
\bottomrule
\end{tabular}\label{Tab: Helicity Lb to L500}
\end{table}

\begin{table}[H]
\caption{Helicity amplitudes of $\Lambda_b^0\rightarrow \Lambda(1520) +K^{*0}\, (10^{-7})$ with different CKM factors}
\centering
\begin{tabular}{cccccccc}
\toprule
\toprule
 decay modes & $H_{\frac{3}{2},1} ~(V_{ub}V_{ud}^*) $ &$H_{\frac{3}{2},1} ~(V_{cb}V_{cd}^*)$&$H_{\frac{3}{2},1} ~(V_{tb}V_{td}^*)$ & $H_{-\frac{3}{2},-1} ~(V_{ub}V_{ud}^*) $ &$H_{-\frac{3}{2},-1} ~(V_{cb}V_{cd}^*)$&$H_{-\frac{3}{2},-1} ~(V_{tb}V_{td}^*)$ \\
\midrule
$\mathcal{S}(\Lambda_b^0\rightarrow \Lambda(1520) +K^{*0} )$&---&---& 0.00 &---&--- 
 &0.00 \\
$\mathcal{NC}(\Lambda_b^0\rightarrow \Lambda(1520) +K^{*0} )$&5.10 + 1.63 $i$&---& -0.97 - 0.36$i$&23.83+12.42 $i$ &---&0.70+0.06 $i$\\
$\mathcal{C}(\Lambda_b^0\rightarrow \Lambda(1520) +K^{*0} )$&---& 5.55+3.47 $i$&0.06+0.04$i$&---&1.27-0.08$ i$& -0.20-0.21 $i$\\
\midrule
 decay modes& $H_{\frac{1}{2},0} ~(V_{ub}V_{ud}^*) $ &$H_{\frac{1}{2},0}  ~(V_{cb}V_{cd}^*)$&$H_{\frac{1}{2},0}  ~(V_{tb}V_{td}^*)$ & $H_{-\frac{1}{2},0} ~(V_{ub}V_{ud}^*) $ &$H_{-\frac{1}{2},0}  ~(V_{cb}V_{cd}^*)$&$H_{-\frac{1}{2},0} ~(V_{tb}V_{td}^*)$ \\
\midrule
$\mathcal{S}(\Lambda_b^0\rightarrow \Lambda(1520) +K^{*0} )$& --- & --- & -0.56&---&---& 0.09\\
$\mathcal{NC}(\Lambda_b^0\rightarrow \Lambda(1520) +K^{*0} )$&-88.26 - 68.21 $i$&---&-3.76 + 0.03 $i$&11.91-18.78$i$&---&5.20+1.07 $i$\\
$\mathcal{C}(\Lambda_b^0\rightarrow \Lambda(1520) +K^{*0})$&---&15.64+15.18$i$&1.88+1.09 $i$&---& -26.34+14.14 $i$&0.29-1.86 $i$\\
\midrule
 decay modes& $H_{-\frac{1}{2},-1} ~(V_{ub}V_{ud}^*) $ &$H_{-\frac{1}{2},-1} ~(V_{cb}V_{cd}^*)$&$H_{-\frac{1}{2},-1} ~(V_{tb}V_{td}^*)$ & $H_{\frac{1}{2},1}~(V_{ub}V_{ud}^*) $ &$H_{\frac{1}{2},1} ~(V_{cb}V_{cd}^*)$&$H_{\frac{1}{2},1}~(V_{tb}V_{td}^*)$ \\
\midrule
$\mathcal{S}(\Lambda_b^0\rightarrow \Lambda(1520) +K^{*0})$& --- & --- & 1.55&---&---& -0.14\\
$\mathcal{NC}(\Lambda_b^0\rightarrow \Lambda(1520) +K^{*0})$&44.59 + 33.59$i$&---&1.79 + 0.43 $i$&8.73 - 1.48 $i$&---&-2.53 - 1.13 $i$\\
$\mathcal{C}(\Lambda_b^0\rightarrow \Lambda(1520) +K^{*0} )$&---&19.03+6.61$i$&0.29-1.42$i$&---& 14.29+26.01$i$&0.37+0.31 $i$\\
\bottomrule
\bottomrule
\end{tabular}\label{Tab: Lb to Lks}
\end{table}

\end{sidewaystable}

\begin{sidewaystable}

\begin{table}[H]
\caption{Helicity amplitudes of $\Lambda_b^0\rightarrow\Lambda(1520)+\kappa(700)\, (10^{-7})$ with different CKM factors}
\centering
\begin{tabular}{ccccccc}
\toprule
\toprule
 decay modes & $H_{-\frac{1}{2}} ~(V_{ub}V_{ud}^*) $ &$H_{-\frac{1}{2}} ~(V_{cb}V_{cd}^*)$ & $H_{-\frac{1}{2}} ~(V_{tb}V_{td}^*)$ & $H_{\frac{1}{2}} ~(V_{ub}V_{ud}^*) $ &$H_{\frac{1}{2}} ~(V_{cb}V_{cd}^*)$ & $H_{\frac{1}{2}} ~(V_{tb}V_{td}^*)$ \\
\midrule
$\mathcal{S}(\Lambda_b^0\rightarrow\Lambda(1520)+\kappa(700))$&  ---  & --- & -0.29  &  --- & --- & -0.68   \\
$\mathcal{NC}(\Lambda_b^0\rightarrow\Lambda(1520)+\kappa(700))$& -5.30-5.88 $i$ &--- & 0.03-0.04$i$& -4.97-2.76 $i$&---&0.04+0.09 $i$\\
$\mathcal{C}(\Lambda_b^0\rightarrow\Lambda(1520)+\kappa(700))$& --- & -1.30-0.79$i$ & 0.04+0.13$i$ & ---&-3.23-5.41$i$&0.06-0.23 $i$\\
\bottomrule
\bottomrule
\end{tabular}\label{Helicity Lb to L700}
\end{table}

\begin{table}[H]
\caption{Helicity amplitudes of $\Xi_b^- \rightarrow \Lambda(1520)+K^-\, (10^{-7})$ with different CKM factors}
\centering
\begin{tabular}{ccccccc}
\toprule
\toprule
 decay modes & $H_{-\frac{1}{2}} ~(V_{ub}V_{us}^*) $ &$H_{-\frac{1}{2}} ~(V_{cb}V_{cs}^*)$ & $H_{-\frac{1}{2}} ~(V_{tb}V_{ts}^*)$ & $H_{\frac{1}{2}} ~(V_{ub}V_{us}^*) $ &$H_{\frac{1}{2}} ~(V_{cb}V_{cs}^*)$ & $H_{\frac{1}{2}} ~(V_{tb}V_{ts}^*)$ \\
\midrule
$\mathcal{S}(\Xi_b^- \rightarrow \Lambda(1520)+K^-)$& -5.61 $i$ & --- & -0.86 $i$  & -21.14 $i$ & --- & -0.68 $i$  \\
$\mathcal{NC}(\Xi_b^- \rightarrow \Lambda(1520)+K^-)$& -1.06-1.30 $i$ &--- & -0.007+0.51 $i$& -1.64+12.51 $i$&---&-0.04+0.54 $i$\\
$\mathcal{C}(\Xi_b^- \rightarrow \Lambda(1520)+K^-)$& --- & -0.61-0.21$i$ & 0.19+0.09 $i$ & ---&4.37+2.99 $i$&0.16+0.04 $i$\\
\bottomrule
\bottomrule
\end{tabular}\label{Tab: Helicity Xb to Lk}
\end{table}

\end{sidewaystable}

\subsection{Discussions}

Based on the numerical results presented above, we now discuss several important points as follows:

\begin{itemize}

    \item From the analysis of short-distance contributions for all the channels, we observe that $\mathcal{S}(\Lambda^0_b \to \Lambda(1520)+\pi^0)$ and $\mathcal{S}(\Xi^-_b \to \Lambda(1520)+K^-)$ are purely imaginary, while $\mathcal{S}(\Lambda^0_b \to \Lambda(1520)+\rho^0)$, $\mathcal{S}(\Lambda^0_b \to \Lambda(1520+)f_0(980))$, $\mathcal{S}(\Lambda^0_b \to \Lambda(1520)+\phi)$, $\mathcal{S}(\Lambda^0_b \to \Lambda(1520)+f_0(500))$, $\mathcal{S}(\Lambda^0_b \to \Lambda(1520)+K^{*0})$ and $\mathcal{S}(\Lambda^0_b \to \Lambda(1520)+\kappa(700))$ are purely real. This can be attributed to the fact that the only source of phase in the short-distance calculation is the weak phase of the CKM matrix, which implies that $CP$ violation cannot arise from the short-distance contribution alone. Therefore, we take into account the rescattering contribution in each channel, which can induce a sizable strong phase, consistent with our previous studies.
    Furthermore, the short-distance contributions of all amplitudes are found to be much smaller than the rescattering contributions, which differs from our earlier work. This is mainly because the short-distance parts do not include the numerically important $T$-diagrams and the transition form factors $\Lambda_b^0 \to \Lambda(1520)$ are relatively small.
    
    \item It is found that the CKM factors  differ among these decay amplitudes, thereby influencing the branching fractions and $CP$ violation in the eight decay channels.
    First, $\Lambda^0_b \to \Lambda(1520)+\pi^0$ and $\Lambda^0_b \to \Lambda(1520)+\rho^0$ are not contributed by charmed triangle diagrams, because isospin-1 particles in the final state cannot form an allowed vertex with the $D_s$ particles produced at the weak vertex or exchanged at the strong vertices. Thus, these processes only involve $V_{ub}V_{us}^*$ and $V_{tb}V_{ts}^*$, without involving any charmed CKM matrix elements. 
    Second, the presence or absence of a strange quark in the final-state meson determines the type of CKM factors entering the decay amplitudes. This results in the decay modes $\Lambda^0_b \to \Lambda(1520)+\pi^0$, $\Lambda^0_b \to \Lambda(1520)+\rho^0$, $\Lambda^0_b \to \Lambda(1520)+f_0(980)$, $\Lambda^0_b \to \Lambda(1520)+f_0(500)$, $\Lambda^0_b \to \Lambda(1520)+\phi$ and $\Xi^-_b \to \Lambda(1520)+K^-$, which are governed by the CKM matrix elements $V_{qb}V_{qs}^*$; whereas the channels
 $\Lambda^0_b \to \Lambda(1520)+K^{*0}$ and $\Lambda^0_b \to \Lambda(1520)+\kappa(700)$ involve $V_{qb}V_{qd}^*$, with $q=u,c,d$. 
    For convenience in the subsequent analysis, the numerical values of the relevant CKM matrix element combinations are summarized as follows: $|V_{ub} V_{ud}^{*}| = 0.00360, |V_{ub} V_{us}^{*}| = 0.00083, |V_{cb} V_{cd}^{*}| = 0.00940, |V_{cb} V_{cs}^{*}| = 0.04071, |V_{tb} V_{td}^{*}| = 0.00856, |V_{tb} V_{ts}^{*}| = 0.04107$. It is worth noting that although $|V_{tb} V_{ts}^{*}|$ is large, the decay amplitudes are suppressed by a small Wilson coefficient, its contribution is essentially equivalent to $|V_{ub} V_{us}^{*}|$ in its final effect. 
    In comparison, $|V_{cb} V_{cs}^{*}|$ is not only large but also comparable to $ |V_{ub} V_{us}^{*}|$, accompanied by a large Wilson coefficient. c, the charmed triangle diagrams provide the dominant contribution to these decay processes and offer valuable insight into the branching ratios and $CP$ violation of each channel.
    
    \item In addition to variations in the CKM factors, the helicity amplitudes exhibit distinct relative magnitudes across the various decay channels. In our previous studies, the short-distance contributions were dominated by $T$-diagrams. The relatively simple chiral properties allow us to infer the relative magnitudes of the helicity amplitudes and estimate their approximate ratios using the SCET topological diagram hierarchy.
    However, in this work, we are not able to predetermine the relative magnitudes of the helicity amplitudes, since the short-distance amplitudes do not include the T-type diagrams and the short-distance vertices differ significantly. Nevertheless, a quantitative analysis allows us to identify certain patterns exhibited in the vector meson helicity amplitudes. For instance, helicity amplitudes $H_{\frac{3}{2},1}$ and $H_{-\frac{3}{2},-1}$ are both quiet small in all decay channels. Furthermore, $H_{\frac{1}{2},0}$ and $H_{-\frac{1}{2},0}$ in $\Lambda^0_b \to \Lambda(1520)\rho^0$ and $\Lambda^0_b \to \Lambda(1520)K^{*0}$ processes are much larger than others, whereas such an enhancement is not observed in  $\Lambda^0_b \to \Lambda(1520)\phi$ process. 
\end{itemize}
\section{Branching ratio and CP violation}\label{BRCPV}

In this section, we first introduce the definitions of the branching ratio and $CP$ asymmetry, and then analyze these results with reference to previous calculations and experimental measurements.

\subsection{Numerical result}
\begin{figure}[H]
\begin{center}
\includegraphics[width=\columnwidth, trim={0 225pt 0 225pt},clip]{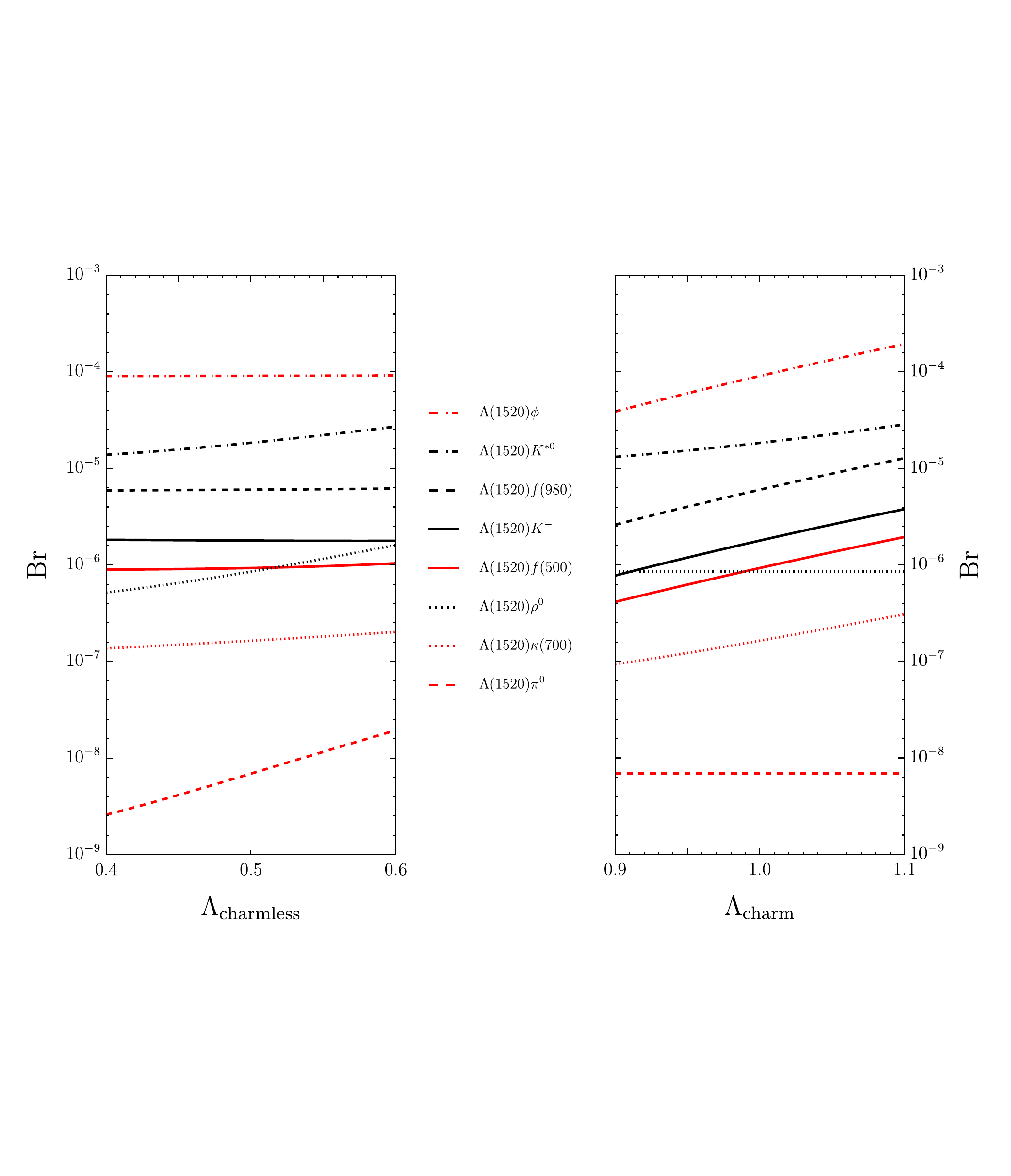}
\end{center}
\vspace{-20pt}
\caption{The dependence of the branching ratios of all decay channels on the model parameters $\Lambda_{\rm charmless}$ and $\Lambda_{\rm charm}$.}
\label{fig:Br}
\end{figure}

\begin{figure}[H]
\begin{center}
\includegraphics[width=\columnwidth]{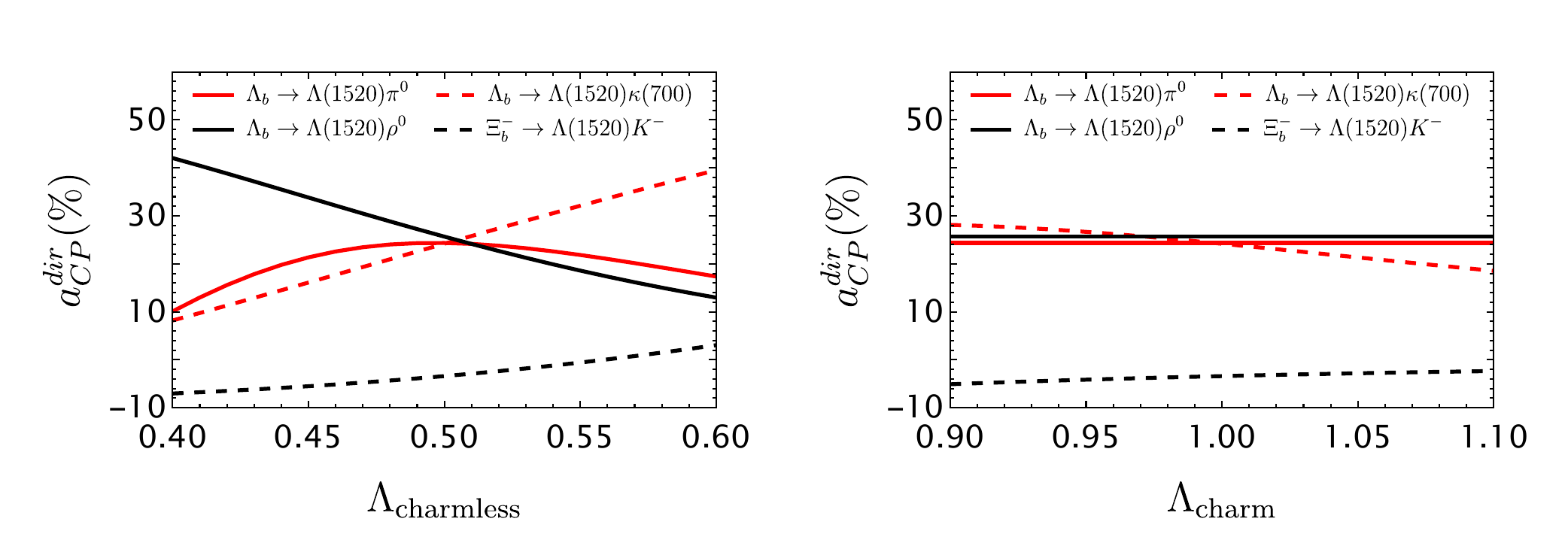}
\end{center}
\vspace{-20pt}
\caption{The dependence of the $CP$ violations of $\Lambda_b^0 \to \Lambda(1520)+ \pi^0/\rho^0/\kappa(700)$and $\Xi^-_b \to \Lambda(1520)+K-$ on the model parameters $\Lambda_{\rm charmless}$ and $\Lambda_{\rm charm}$.}
\label{fig:cp1}
\end{figure}

\begin{figure}[H]
\begin{center}
\includegraphics[width=\columnwidth]{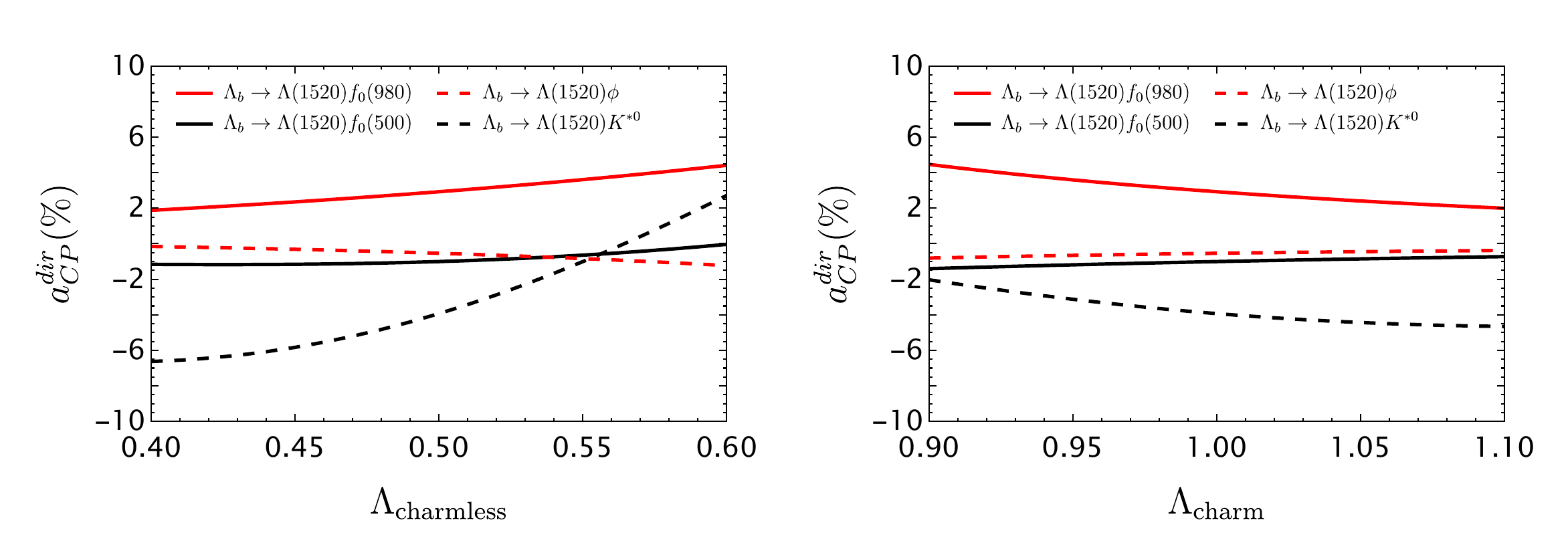}
\end{center}
\vspace{-20pt}
\caption{The dependence of the $CP$ violations of $\Lambda_b \to f_0(980,500)/\phi/K^{*0}$ on the model parameters $\Lambda_{\rm charmless}$ and $\Lambda_{\rm charm}$.}
\label{fig:cp2}
\end{figure}

The branching ratio and direct $CP$ violation for each channel are given in the following formula:
\begin{equation}
    \begin{aligned}
    &BR\left[\mathcal{B}_b\to \Lambda(1520)+P(S)\right]=\frac{\Gamma\left[\mathcal{B}_b\to \Lambda(1520)+P(S)\right]}{\Gamma_{\mathcal{B}_b}}=\frac{|p_{c}|}{8\pi M^{2}_{\mathcal{B}_b}\Gamma_{\mathcal{B}_b}}\frac{1}{2}\left(\left|H_{\frac12}\right|^{2}+\left|H_{-\frac12}\right|^{2}\right), 
\\ \nonumber
&BR\left[\mathcal{B}_b\to \Lambda(1520)+V\right]=\frac{\Gamma\left[\mathcal{B}_b\to \Lambda(1520)+V\right]}{\Gamma_{\mathcal{B}_b}}
\\ \nonumber
&=\frac{|p_{c}|}{8\pi M^{2}_{\mathcal{B}_b}\Gamma_{\mathcal{B}_b}}\frac{1}{2}\left(\left|H_{\frac32,1}\right|^{2}+\left|H_{-\frac32,-1}\right|^{2}+\left|H_{\frac12,0}\right|^{2}+\left|H_{-\frac12,0}\right|^{2}+\left|H_{-\frac12,1}\right|^{2}+\left|H_{-\frac12,-1}\right|^{2}\right),
\\ \nonumber
&a^{dir}_{CP}=\frac{\Gamma-\bar{\Gamma}}{\Gamma+\bar{\Gamma}},
\end{aligned}
\end{equation}
where $\mathcal{B}_b, P, S, V$ denote the bottom baryon in the initial state, and the pseudoscalar, scalar, and vector meson in the final state, respectively. $H$ denotes the helicity amplitudes appearing in these decay channels, and $p_c$ represents the momentum of the final-state baryon in the rest frame of $\mathcal{B}_b$.
Due to the large masses of charmed hadrons compared to light hadrons, we introduce two model parameters $\Lambda_{\rm{charm}}$ and $\Lambda_{\rm{charmless}}$ to describe the off-shell effects of the intermediate particles and to regularize the divergence of the triangle diagrams that contribute to the long-distance contribution.
We calculate the branching ratios and direct $CP$ asymmetries for all the decay channels with $\Lambda_{\rm{charm}}=1\pm 0.1$ and $\Lambda_{\rm{charmless}}=0.5\pm 0.1$ \cite{Duan:2024zjv}.
The numerical results are summarized in Table~\ref{brcpv}, categorized according to the experimentally observed final states indicated in parentheses, for example, both $\Lambda(1520)K^{*0}$ and $\Lambda(1520)\kappa(700)$ decay into $pK^-K^+\pi^-$. And the uncertainties of the two observables stem from the variations in model parameters. To further examine the dependence of the branching ratios and $CP$ asymmetries on the two model parameters, the variations of the branching ratios are shown in Fig.~\ref{fig:Br}, while those of the $CP$ asymmetries are displayed in Figs.~\ref{fig:cp1} and~\ref{fig:cp2}.

\begin{table}[H]
\caption{Branching ratios and direct $CP$ violations for the various decay channels}
\centering
\begin{tabular}{ccccc}
\toprule
\toprule
  & $BR$ & Direct $CP$ $(10^{-2})$ \\
\midrule
$ (\Lambda^0_b\rightarrow pK^-\pi^0) $ \\
$\Lambda(1520)\pi^0$  & $6.90_{-4.31}^{+12.47}\times 10^{-9} $ & $24.1^{+0.3}_{-14.0}  $  \\
\midrule
$ (\Lambda^0_b\rightarrow pK^-\pi^+\pi^-) $\\
$\Lambda(1520)f_0(980)$ & $ 6.01^{+6.98}_{-3.47}\times 10^{-6} $ & $ 2.9_{-1.6}^{+3.8}  $\\
$\Lambda(1520)\rho^0$  & $8.51^{+7.58}_{-3.33} \times 10^{-7} $ & $25.7_{-12.8}^{+16.3} $  \\
$\Lambda(1520)f_0(500)$ & $9.28^{+11.50}_{-5.42}\times 10^{-7} $ & $-1.0^{+1.1}_{-0.7} $  \\
\midrule
$ (\Lambda^0_b\rightarrow pK^-K^+K^-) $\\
$\Lambda(1520)\phi$ &  $9.07^{+10.73}_{-5.29} \times 10^{-5} $ & $-0.5_{-1.3}^{+0.4} $  \\
$\Lambda(1520)f_0(980)$ & $ 6.01^{+6.98}_{-3.47}\times 10^{-6} $ & $ 2.9_{-1.6}^{+3.8}  $\\
\midrule
$ (\Lambda^0_b\rightarrow pK^-K^+\pi^-) $\\
$\Lambda(1520)K^{*0}$& $1.79^{+1.98}_{-0.92} \times 10^{-5} $ &  $-3.9_{-1.7}^{+5.0} $ \\
$\Lambda(1520)\kappa(700)$ & $ 1.64^{+1.87}_{-0.93} \times 10^{-7} $ & $ 24.2_{-17.8}^{+17.3} $  \\
\midrule
$ (\Xi^-_b\rightarrow pK^-K^-) $\\
$\Lambda(1520)K^-$ &$ 1.79^{+2.03}_{-1.02} \times 10^{-6} $ & $ -3.4_{-6.7}^{+7.9}  $ \\
\bottomrule
\bottomrule
\end{tabular}\label{brcpv}
\end{table}

\subsection{Discussions}

Based on the above numerical results, we proceed to the following discussion and analysis:

\begin{itemize}

    \item Based on the previous quantitative analysis of the helicity amplitudes, we find that the branching ratios and $CP$ asymmetries of each channel are strongly related to the corresponding CKM matrix elements. According to the magnitudes of the involved CKM matrix elements and the corresponding Wilson coefficients, the processes proportional to $V_{cb}V_{cs}^*$ provide the dominant contributions, followed by those proportional to 
    $V_{cb}V_{cd}^*$, while the contributions from 
    $V_{ub}V_{ud}^*$ and $V_{ub}V_{us}^*$ are relatively small.
    This is also consistent with our numerical results, for instance, the branching ratio of the decay $\Lambda_b^0 \to \Lambda(1520) + \phi$ is the largest, while that of $\Lambda_b^0 \to \Lambda(1520) + \pi^0$ is the smallest.
    With regard to $CP$ violation, the dominant contribution from \( V_{cb}V_{cs}^* \) tends to obscure the weak-phase interference, resulting in a small \( CP \) violation, as exemplified by the decays \( \Lambda_b^0 \to \Lambda(1520)\phi \) and \( \Lambda_b^0 \to \Lambda(1520)f_0(500) \). In contrast, the processes without charm-loop diagrams exhibit significant interference effects, leading to a relatively large \( CP \) violation, such as the decays \( \Lambda_b^0 \to \Lambda(1520)\pi^0 \) and \( \Lambda_b^0 \to \Lambda(1520)\rho^0 \).
     These results reveal an interesting correlation between the $CP$ violation and the branching ratio: decay channels with significant $CP$ violations typically exhibit suppressed branching ratios, rendering them difficult to observe. Conversely, the processes with large branching ratios generally feature small $CP$ violations, making the experimental measurements of $CP$ violation challenging. Based on this, we can interpret the parameter dependence of the branching ratios for different processes shown in Fig.~\ref{fig:Br}: processes without charm-related CKM matrix elements depend strongly only on $\Lambda_{\text{charmless}}$, while the other processes depend almost exclusively on $\Lambda_{\text{charm}}$. Among these, processes containing the $V_{cs}V^*_{cb}$ matrix element are particularly prominent.

    \item It is found that both the branching ratios and $CP$ asymmetries exhibit considerable uncertainties, which originate from their dependence on the two model parameters $\Lambda_{\rm{charm}}$ and
    $\Lambda_{\rm{charmless}}$. The regulator form factor $\mathcal{F}(\Lambda, m_k)$ scales as $\mathcal{O}(\Lambda^{4})$, while the branching ratio depends on the squared amplitude, resulting in a scaling of order $\mathcal{O}(\Lambda^{8})$. Numerically, when $\Lambda_{\text{charm}}$ is set to its upper limit, the branching ratio increases by a factor of 2.14 relative to corresponding central value, whereas for the lower limit of $\Lambda_{\text{charm}}$, the branching ratio is $0.43$ times smaller than the corresponding central value. This strong parameter dependence explains the large numerical uncertainties observed in the branching ratio predictions.
    On the other hand, the uncertainties in $CP$ violations is affected by the opposite interplay between the two model parameters. The parameter $\Lambda_{\text{charmless}}$ controls the magnitude of the charmless triangle diagrams, while $\Lambda_{\text{charm}}$ governs the contribution of the charmed triangle diagrams. These two types of diagrams involve different CKM matrix elements and interfere with each other, thereby influencing the magnitude of $CP$ violation. In particular, when the two parameters deviate from each other, the interference becomes weaker, whereas it is enhanced when they are close. Since the interference effect depends on the fourth power of the model parameters, this leads to a relatively large uncertainty in $CP$ violation. Based on the CKM matrix elements involved in each reaction and the error behavior analyzed above, we can also interpret the dependence of the $CP$ violation for each process on the model parameters, as shown in Figs.~\ref{fig:cp1} and~\ref{fig:cp2}: any change in the model parameters that brings the two distinct weak phase sources—one from charm-containing and the other from charmless CKM matrix elements—closer in magnitude will enhance the $CP$ violation, whereas any change that drives them further apart will suppress it.

    \item For the decay processes investigated in this work, the LHCb collaboration has reported a measurement of $\Xi_b^{-} \to \Lambda(1520) K^{-}$, with a branching fraction 
$BR = (0.76 \pm 0.09 \pm 0.08 \pm 0.30) \times 10^{-6}$ 
and a direct $CP$ asymmetry 
$A_{CP} = (-5 \pm 9_{\text{(stat)}} \pm 8_{\text{(syst)}}) \times 10^{-2}$~\cite{LHCb:2025ray}. 
It can be clearly seen that both the experimental measurement and the theoretical prediction suffer from sizable uncertainties; nevertheless, the theoretical results are consistent with the experimental data within the same order, indicating that our calculation still provides a meaningful theoretical reference. 
For the remaining decay channels, we have so far calculated only the primary decays, while the experimental signals correspond to the final states originating from the subsequent secondary decays~\cite{LHCb:2025ray}. 
To enable a direct comparison with experimental results and facilitate the verification of our theoretical predictions, a complete treatment of the full decay chains is necessary. 
Therefore, in the next section, we will perform a comprehensive analysis of the full decay processes 
$\Lambda_b^{0} \to p K^{-} \pi^{+} \pi^{-}$, 
$\Lambda_b^{0} \to p K^{-} K^{+} K^{-}$, 
and 
$\Lambda_b^{0} \to p K^{-} K^{+} \pi^{-}$, 
taking into account a detailed assessment of interference effects among them.
\end{itemize}
\section{Interference effects}\label{interf}

In this section, we are now ready to explore the complete decay processes and evaluate the interference effects between the primary and secondary decay amplitudes. Then a detailed numerical analysis and a comprehensive discussion of the corresponding results are presented.

\subsection{The $CP$ violation with interference effects}
To compute the $CP$ violation in full decay chain, it is necessary to account for both the primary decay and the subsequent decay of the two final-state particles. The most straightforward approach is to extend the Feynman rules of the primary decay amplitude by attaching propagators to each final-state particle and incorporating the vertices for the secondary decays. However, this procedure is computationally prohibitive, as it requires evaluating hundreds of triangle diagrams involving multiple vertices, loop integrals, and complicated Lorentz structures arising from the $\Lambda(1520)$ resonance. 
Therefore, we adopt a more efficient strategy, regarding as an approximation to the full calculation. The computed amplitudes for the primary and secondary decays are treated as numerical inputs and connected via the Breit–Wigner propagators, followed by integration over the relevant momenta. Moreover, it is unnecessary to consider the secondary decay of the $\Lambda(1520)$ in the primary decay, since its decay amplitude is identical for all channels and the corresponding contribution can be canceled. 
Taking the decay channel $\Lambda_b^{0} \to p K^{-} K^{+} \pi^{-}$ as an example, the expression for a single helicity amplitude can be written as:
\begin{equation}
    \begin{aligned}
\mathcal{H}^\frac12_j&(\Lambda^0_b \to \Lambda(1520)K^+\pi^-)\\
=&\int[\mathcal{H}^\frac12_{j,j-\frac12}(\Lambda^0_b \to \Lambda(1520)+K^{*0})\frac{i}{q^2-m_{K^{*0}}^2+im_{K^{*0}}\Gamma_{K^{*0}}}d^1_{0,j-\frac12}(\theta)\mathcal{H}(K^{*0}\to K^+\pi^-)\\
+&\mathcal{H}^\frac12_{j,j-\frac12}(\Lambda^0_b \to \Lambda(1520)+\kappa(700))\frac{i}{q^2-m_{\kappa(700)}^2+im_{\kappa(700)}\Gamma_{\kappa(700)}}\mathcal{H}(\kappa(700)\to K^+\pi^-)]dq,
    \end{aligned}
\end{equation}
where $\mathcal{H}$ denotes the helicity amplitude for the specific decay channel, with the superscript and subscript representing the helicities of the initial- and final-state particles, respectively. The parameters $m$ and $\Gamma$ correspond to the mass and decay width of the intermediate particles. And $d$ refers to the Wigner-$d$ function. The calculation is performed in the rest frame of $\Lambda_b^{0}$ and the decay direction of the meson are defined as the positive $z$-axis. The angle between the $z$-axis and the direction of the secondary decay is denoted by $\theta$.

The conjugate amplitude can be obtained by taking the complex conjugate of the CKM matrix elements in the primary decay amplitude. The total decay rate and its conjugate are then calculated by summing over the squared of the amplitudes over all possible helicity configurations. Finally, we can estimate the total $CP$ violation by performing a weighted average with the branching ratios of the primary decays obtained from our calculations and those of the secondary decays obtained from the experimental data
\begin{equation}
A_{cp}^{total} = \frac{\sum_{M}\Gamma(\Lambda_b\rightarrow\Lambda(1520)+M)\times \Gamma(M\rightarrow \rm{final\, state})\times A_{CP}^M}{\sum_{M}\Gamma(\Lambda_b\rightarrow\Lambda(1520)+M)\times \Gamma(p\rightarrow \rm{final\, state})} .
\end{equation}
Due to the limited experimental data on the secondary decays, only the approximate ranges of the branching ratios are available. Although the weighted average method mainly serves as an estimation approach, the final results demonstrate that our theoretical predictions agree well with the experimentally measured ranges without introducing any significant uncertainties. All numerical results are summarized in Table~\ref{newres}.
\begin{table}[H]
\caption{Total $CP$ violations}
\centering
\begin{tabular}{lcc}
\toprule
\toprule
Decay Channel  & Direct $CP$ violations $(10^{-2})$ \\
\midrule
{$\Lambda^0_b\rightarrow \Lambda(1520)\pi^+\pi^-$} 
 & {$10.8^{+0.1}_{-4.7}$} \\

{$\Lambda^0_b\rightarrow \Lambda(1520)K^+K^-$} 
 & {$0.1^{+0.1}_{-0.1}$} \\

{$\Lambda^0_b\rightarrow \Lambda(1520)K^+\pi^-$} 
& {$-4.6^{+5.6}_{-0.4}$} \\

\bottomrule
\bottomrule
\end{tabular}
\label{newres}
\end{table}

\subsection{Kinematic expansion}

In this section, we systematically investigate the total $CP$ violation in each decay channel and explore its fundamental origin. 
The relative magnitudes of contributions from different channels are analyzed, allowing us to identify decay modes and observables that are particularly promising for measurement, thereby offering useful guidance for future experiments.
As established in the previous section, the decay amplitude for any primary process with a vector meson in the final state inherently includes a $\cos\theta$ term. As a result, the decay width can be expressed as a quadratic form of $a\cos^{2}\theta+b\cos\theta+c$.
Taking the decay channel $\Lambda_b^{0} \to p K^{-} K^{+} \pi^{-}$ as an example, the composition of the coefficients for each power of $\cos\theta$ is as follows:
\begin{equation}\label{Coe}
    \begin{aligned}
a=&-\chi_1(|\mathcal{H}(K^{*0})^\frac12_{\frac32,1}|^2+\mathcal{H}(K^{*0})^{-\frac12}_{-\frac32,-1}|^2
+\mathcal{H}(K^{*0})^{\frac12}_{-\frac12,-1}|^2+\mathcal{H}(K^{*0})^{-\frac12}_{\frac12,1}|^2)\\
&+2\chi_1(|\mathcal{H}(K^{*0})^\frac12_{\frac12,0}|^2+\mathcal{H}(K^{*0})^{-\frac12}_{-\frac12,0}|^2),\\
b=&-2\chi_2(\Re(\mathcal{H}(K^{*0})^\frac12_{\frac12,0})\Re(\mathcal{H}(\kappa(700))^\frac12_{\frac12,0})+\Im(\mathcal{H}(K^{*0})^\frac12_{\frac12,0})\Im(\mathcal{H}(\kappa(700))^\frac12_{\frac12,0})\\
&+\Re(\mathcal{H}(K^{*0})^{-\frac12}_{-\frac12,0})\Re(\mathcal{H}(\kappa(700))^{-\frac12}_{-\frac12,0})+\Im(\mathcal{H}(K^{*0})^{-\frac12}_{-\frac12,0})\Im(\mathcal{H}(\kappa(700))^{-\frac12}_{-\frac12,0}))\\
&+\chi_2(\Re(\mathcal{H}(K^{*0})^\frac12_{\frac12,0})\Im(\mathcal{H}(\kappa(700))^\frac12_{\frac12,0})-\Im(\mathcal{H}(K^{*0})^\frac12_{\frac12,0})\Re(\mathcal{H}(\kappa(700))^\frac12_{\frac12,0})\\
&+\Re(\mathcal{H}(K^{*0})^{-\frac12}_{-\frac12,0})\Im(\mathcal{H}(\kappa(700))^{-\frac12}_{-\frac12,0})-\Im(\mathcal{H}(K^{*0})^{-\frac12}_{-\frac12,0})\Re(\mathcal{H}(\kappa(700))^{—\frac12}_{-\frac12,0})),\\
c=&\chi_1(|\mathcal{H}(K^{*0})^\frac12_{\frac32,1}|^2+\mathcal{H}(K^{*0})^{-\frac12}_{-\frac32,-1}|^2
+\mathcal{H}(K^{*0})^{\frac12}_{-\frac12,-1}|^2+\mathcal{H}(K^{*0})^{-\frac12}_{\frac12,1}|^2)\\
&+\chi_3(|\mathcal{H}(\kappa(700))^\frac12_{\frac12,0}|^2+|\mathcal{H}(\kappa(700))^{-\frac12}_{-\frac12,0}|^2),
    \end{aligned}
\end{equation}
where $\mathcal{H}$ denotes the helicity amplitude of the primary decay involving different resonant states in the final state. The symbol $\chi$ represents positive real coefficients, whose values are determined by the propagators and the amplitudes of the secondary decays.
The coefficient $a$ originates from the squared modulus of the helicity amplitudes for the vector mesons. The coefficient $b$ arise from the cross terms between the helicity amplitudes of the vector and scalar mesons, thus characterizing the interference effects. The coefficient $c$, on the other hand, is composed of the sum of the modulus squares of certain helicity amplitudes from both vector and scalar mesons. 
Since these coefficients can be directly measured and have clear physical meaning, we proceed to compute them along with their charge-conjugate counterparts with the aim of defining and evaluating the $CP$ asymmetries associated with each coefficient. 
Our approach for defining $CP$ violation is consistent with the framework used for branching ratio calculations.
The expression is given as follows:
\begin{equation}
a^a_{CP}=\frac{a-\Bar{a}}{a+\Bar{a}},\quad
a^b_{CP}=\frac{b+\Bar{b}}{b-\Bar{b}},\quad
a^c_{CP}=\frac{c-\Bar{c}}{c+\Bar{c}},
\end{equation}
where $\bar{a}$, $\bar{b}$, and $\bar{c}$ represent the charge-conjugated counterparts of the coefficients $a$, $b$, and $c$, respectively. It is worth noting that because the coefficient $b$ possesses a $P_{odd}$ nature, an additional minus sign is introduced in its $CP$ violation, which is the same as the method used in the definition of $CP$ violation for the Lee-Yang parameters.
According to this definition, the complete results are presented in Table~\ref{orgres2} and Table~\ref{orgres2cp}.
\begin{table}[H]
\caption{The values of different coefficients}
\centering
\begin{tabular}{ccccccccc}
\toprule
\toprule
 &$a(10^{-15})$&$\Bar{a}(10^{-15})$&$b(10^{-15})$&$\Bar{b}(10^{-15})$& $c(10^{-15})$ & $\bar{c}(10^{-15})$ \\
\midrule
$\Lambda^0_b\rightarrow \Lambda(1520)\pi^+\pi^- $&$2.07^{+3.08}_{-1.33}$&$1.17^{+2.73}_{-1.02}$&$-3.29^{+0.91}_{-0.51}$&$3.77^{+0.47}_{-0.91}$&$3.13^{+2.55}_{-1.09}$&$2.77^{+2.51}_{-1.02}$ \\

\midrule
$ \Lambda^0_b\rightarrow \Lambda(1520)K^+K^-$ &$0.81^{+0.11}_{-0.11}$&$0.72^{+0.22}_{-0.19}$&$1.02^{+0.88}_{-0.49}$&$0.97^{+1.03}_{-0.66}$&$9.66^{+10.42}_{-5.56}$&$9.67^{10.40}_{-5.48}$\\

\midrule
$ \Lambda^0_b\rightarrow \Lambda(1520)K^+\pi^-$&$17.05^{+0.15}_{-1.95}$&$17.99^{+2.65}_{-1.99}$&$0.13^{+0.55}_{-0.30}$&$0.30^{+0.38}_{-0.38}$&$3.51^{+1.75}_{-0.50}$&$3.23^{+3.80}_{-1.21}$\\

\bottomrule
\bottomrule
\end{tabular}\label{orgres2}
\end{table}
\begin{table}[H]
\caption{The values of $CP$ asymmetries.}
\centering
\begin{tabular}{ccccccc}
\toprule
\toprule
 &$A_{cp}^a(10^{-2})$&$A_{cp}^b(10^{-2})$&$A_{cp}^c(10^{-2})$\\
\midrule
$\Lambda^0_b\rightarrow \Lambda(1520)\pi^+\pi^- $&$27.7^{+36.7}_{-13.9}$&$1457.7^{+385.5}_{-2556.9}$&$6.2^{+1.1}_{-2.5}$ \\

\midrule
$ \Lambda^0_b\rightarrow \Lambda(1520)K^+K^-$&$6.0^{+7.7}_{-7.0}$ &$2.6^{+22.6}_{-5.2}$&$-0.1^{+0.2}_{-0.1}$\\

\midrule
$ \Lambda^0_b\rightarrow \Lambda(1520)K^+\pi^-$&$-8.8^{+11.5}_{-0.2}$&$-39.5^{+88.3}_{-2.3}$&$4.1^{+28.8}_{-18.3}$\\

\bottomrule
\bottomrule
\end{tabular}\label{orgres2cp}
\end{table}
The results indicate that pronounced $CP$ violation effects in some of the coefficients, demonstrating the value of the kinematic analysis. Motivated by this, we aim to introduce a $CP$ violation observable that is more convenient for both experimental measurement and theoretical computation.
In the new scheme, these coefficients are normalized before calculating the $CP$ violation. The normalization procedure can be described as follows:
\begin{equation}
A=\frac{a}{c}, B=\frac{b}{c},C=1,
\end{equation}
thus we obtain a new expression for $CP$ violation, which is given by:
\begin{equation}
a^A_{CP}=\frac{A-\Bar{A}}{A+\Bar{A}},a^B_{CP}=\frac{B+\Bar{B}}{B-\Bar{B}}.
\label{noabs}
\end{equation}
This new definition essentially reflects the $CP$ violation in the ratios between coefficients, 
thereby partially reducing the systematic uncertainties arising in experimental measurements.
Theoretically, after phase-space integration, the total amplitude depends only on the two coefficients $a$ and $c$, and one of them can be expressed in terms of the other using the integration result. 
This implies that the three kinematic coefficients actually represent only two independent physical degrees of freedom; accordingly, the above normalization can be interpreted as the removal of redundant physical quantities.
All numerical results based on this definition are presented in Table~\ref{firstres2}.
\begin{table}[H]
\caption{The values of different coefficients and the corresponding $CP$ asymmetries.}
\centering
\begin{tabular}{ccccccccc}
\toprule
\toprule
 &$B$&$\Bar{B}$&$A$&$\Bar{A}$& $a^B_{CP}(10^{-2})$ & $a^A_{CP}(10^{-2})$ \\
\midrule
$\Lambda^0_b\rightarrow \Lambda(1520)\pi^+\pi^- $&$-1.05^{+0.63}_{-0.81}$&$1.36^{+1.05}_{-0.82}$&$0.66^{+1.86}_{-0.53}$&$0.42^{+1.79}_{-0.39}$&$-782.6^{+16.1}_{-0.2}$&$21.8^{+40.9}_{-15.3}$ \\

\midrule
$ \Lambda^0_b\rightarrow \Lambda(1520)K^+K^-$&$0.11^{+0.02}_{-0.02}$ &$0.10^{+0.00}_{-0.03}$&$0.08^{+0.09}_{-0.04}$&$0.07^{+0.06}_{-0.03}$&$2.7^{+23.5}_{-5.7}$&$6.1^{+8.6}_{-7.1}$\\

\midrule
$ \Lambda^0_b\rightarrow \Lambda(1520)K^+\pi^-$&$0.04^{+0.09}_{-0.10}$&$0.09^{+0.00}_{-0.13}$&$4.29^{+0.00}_{-1.07}$&$5.56^{+4.64}_{-3.27}$&$ -19.0^{+32.8}_{-24.0} $&$-12.9^{+29.8}_{-27.9}$\\

\bottomrule
\bottomrule
\end{tabular}\label{firstres2}
\end{table}

If we want to obtain a definition of $CP$ violation whose results are strictly confined to the range between $-100\%$ and $100\%$, we can modify Eq.~(\ref{noabs}) by taking the absolute value of each coefficient in the denominator, which is expressed as
\begin{equation}
a^A_{CP}=\frac{A-\Bar{A}}{\abs{A}+\abs{\Bar{A}}},a^B_{CP}=\frac{B+\Bar{B}}{\abs{B}-\abs{\Bar{B}}}.
\end{equation}
This definition keeps all results in Table.~\ref{firstres2} unchanged, except for the $CP$ violation of the $b$ coefficient in the first channel, which is updated to $-100.0^{+0.0}_{-0.0}$.
Under this definition, whenever a coefficient and its charge-conjugated counterpart differ only by a sign, a $-100\%$ $CP$ violation is obtained, which could lead to significant distortion when the actual value of that coefficient is close to zero.

All three schemes described above indicate that the $b$ coefficient in the process $\Lambda_b^0 \to \Lambda(1520) \pi^+ \pi^-$ exhibits sizable $CP$ violation. 
However, owing to the nature of the definitions, it is not possible to determine whether this apparently large value reflects the same scale of observability as other experimentally measurable quantities.
More directly, in evaluating whether a claimed large physical observable is experimentally accessible, it is important to compare it with already measured quantities, focusing on whether their statistical uncertainties are comparable, rather than considering the magnitude alone.
If we only multiplied the previous definitions by a factor of $100$, all results would appear highly significant, but that would not imply they are easier to observe; at the very least, they could not be fairly compared with the $CP$ violation in the branching ratio.
Fortunately, recent studies on partial-wave $CP$ violation provide a highly effective solution~\cite{Qi:2025zna}, which is based on the Legendre expansion of the squared amplitude modulus :
\begin{equation}
|\mathcal{M}|^{2}=\sum_l \omega_l P_l(cos\theta),
\end{equation}
the $CP$ violation for each coefficient $\omega_l$ is defined as:
\begin{equation}
A_{cp}^l \equiv \eta_l \frac{\omega_l-\bar{\omega_l}}{\omega_0+\bar{\omega_0}},
\end{equation}
where $\eta_l$ is a normalization factor to ensure the $CP$ violations for different coefficients are comparable with each other, and also directly comparable with the $CP$ violation of the branching ratio. 
The normalization factors are determined by matching the resulting $CP$ violations to a constructed, experiment-friendly $CP$ observable that depends solely on directly measurable quantities. Here we adopt an approximate calculation scheme provided in the previous work
\begin{equation}
\eta_l=\frac12 \int_{\pi}^{0} |P_l(cos\theta)|(-sin\theta)d\theta,
\end{equation}
when the Legendre expansion is truncated at the second order, the error induced by this approximation scheme is less than $0.2\%$, thus can be neglected.
Finally, we present the results obtained with this scheme in Table.~\ref{newres2} and Table.~\ref{newres2cp}.

\begin{table}[H]
\caption{The values of different coefficients.}
\centering
\begin{tabular}{ccccccccc}
\toprule
\toprule
 &$\omega_0(10^{-15})$&$\Bar{\omega_0}(10^{-15})$&$\omega_1(10^{-15})$&$\Bar{\omega_1}(10^{-15})$& $\omega_2(10^{-15})$ & $\bar{\omega_2}(10^{-15})$ \\
\midrule
$\Lambda^0_b\rightarrow\Lambda(1520)\pi^+\pi^- $&$3.83^{+2.09}_{-0.40}$&$3.16^{+2.16}_{-0.56}$&$-3.28^{+0.90}_{-0.52}$&$3.77^{+0.47}_{-0.90}$&$1.38^{+2.05}_{-0.89}$&$0.78^{+1.82}_{-0.68}$ \\

\midrule
$ \Lambda^0_b\rightarrow \Lambda(1520)K^+K^-$&$9.93^{+11.23}_{-5.59}$ &$9.91^{+11.19}_{-5.54}$&$1.02^{+0.88}_{-0.47}$&$0.97^{+1.03}_{-0.66}$&$0.54^{+0.07}_{-0.08}$&$0.48^{+0.14}_{-0.13}$\\

\midrule
$ \Lambda^0_b\rightarrow \Lambda(1520)K^+\pi^-$&$9.75^{+0.34}_{-1.18}$&$9.23^{+3.15}_{-0.33}$&$0.13^{+0.54}_{-0.36}$&$0.30^{+0.38}_{-0.31}$&$ 11.30^{+0.16}_{-1.25} $&$11.99^{+1.77}_{-1.28}$\\

\bottomrule
\bottomrule
\end{tabular}\label{newres2}
\end{table}

\begin{table}[H]
\caption{The values of $CP$ asymmetries.}
\centering
\begin{tabular}{ccccccc}
\toprule
\toprule
 &$A_{cp}^0(10^{-2})$&$A_{cp}^1(10^{-2})$&$A_{cp}^2(10^{-2})$\\
\midrule
$\Lambda^0_b\rightarrow \Lambda(1520)\pi^+\pi^- $&$9.6^{+0.7}_{-4.2}$&$-50.5^{+27.2}_{-8.5}$&$3.3^{+1.4}_{-2.0}$ \\

\midrule
$ \Lambda^0_b\rightarrow \Lambda(1520)K^+K^-$&$0.1^{+0.0}_{-0.2}$ &$0.1^{+0.2}_{-0.2}$&$0.1^{+0.3}_{-0.1}$\\

\midrule
$ \Lambda^0_b\rightarrow \Lambda(1520)K^+\pi^-$&$-3.9^{+6.6}_{-2.1}$&$-0.4^{+0.5}_{-0.1}$&$-4.2^{+0.33}_{-0.5}$\\

\bottomrule
\bottomrule
\end{tabular}\label{newres2cp}
\end{table}

\subsection{Discussions}
We proceed to analyze and discuss the numerical results obtained in this section.
\begin{itemize}

    \item First of all, we strongly recommend detailed experimental measurements of channel $\Lambda^0_b \to \Lambda(1520)\pi^+\pi^-$. All the three schemes for defining $CP$ violation consistently show that the coefficient corresponding to the $\cos\theta$ (or $P_l(\cos\theta)$) term in this channel exhibits significant $CP$ violation. We employ the expansion scheme based on $\cos\theta$ to analyze this phenomenon due to its clear physical meaning and its ability to reflect the characteristics of a Legendre polynomial expansion. The analysis shows that this phenomenon originates from the large $CP$ violation in channel $\Lambda^0_b \to \Lambda(1520)\rho^0$, which arises from a substantial phase difference. This phase difference lead to the coefficient $B$ and its conjugate to have opposite signs in the calculation. When we artificially adjust this phase difference, the sign reversal disappears. Therefore, an experimental measurement of coefficient $B$ in channel $\Lambda^0_b \to pK^-\pi^+\pi^-$ can serve as a probe for the $CP$ violation in channel $\Lambda^0_b \to \Lambda(1520)\rho^0$. It should be noted that these particular results carry a degree of uncertainty, which we will elaborate on in the following analysis.
    
    \item 
   The sources of uncertainties in the total $CP$ violations are consistent with the analysis presented in the previous sections. However, the uncertainties associated with the coefficients of different powers of $\cos\theta$ terms in the kinematical analysis require further clarification. These coefficients reflect the correlations among various decay channels and thus vary collectively. As an illustrative example, we consider the coefficient defined as $A = a/c$ in the decay channel $\Lambda_b^{0} \to p K^{-} \pi^{+} \pi^{-}$, in order to analyze the origin of its relatively large uncertainty. In this ratio, the numerator arises from the squared modulus of the helicity amplitudes for the vector meson, while the denominator comprises the sum of the modulus squares of certain helicity amplitudes from both vector and scalar mesons. The dominant contribution to this process comes from the primary decay $\Lambda_b^{0} \to \Lambda(1520) f_{0}(980)$, yielding a central value of $A$ smaller than 1. When adopting the model parameters $\Lambda_{\text{charm}} = 1.1$ and $\Lambda_{\text{charmless}} = 0.4$, the amplitude of the charmless vector meson decay channel $\Lambda_b^{0} \to \Lambda(1520)\rho^{0}$ is suppressed by a factor of 0.16. In contrast, the amplitude of the scalar meson decay channel $\Lambda_b^{0} \to \Lambda(1520) f_{0}(980)$, governed by charmed triangle diagrams, is enhanced by approximately a factor of $2.3$. As a consequence, the denominator of $A$ increases while the numerator decreases, giving rise to a substantial overall uncertainty in the coefficient $A$. The uncertainties in other schemes stem from the same source.

    \item 
    We find that the coefficient $A$ exhibits significant variation across different decay channels. To understand this behavior, it is necessary to perform a detailed analysis of these coefficients. According to Eq.~\eqref{Coe}, the numerator of $A$ is determined by the difference between specific helicity amplitudes in the vector meson processes, which characterizes the deviation between the $\rm{S}_{\rm{vector}}=0$ and $\rm{S}_{\rm{vector}}\neq0$  helicity components. As inferred from Sec.~\ref{hamplitudes}, this difference is relatively small in the decay channel $\Lambda_b^{0} \to \Lambda(1520)\,\phi$, but large in $\Lambda_b^{0} \to \Lambda(1520)\,K^{*0}$. This observation naturally accounts for the pronounced variation of the coefficient $A$ among different decay modes.
    For the coefficient $B$, its value in the decay process $\Lambda_b^{0} \to p K^{-} \pi^{+} \pi^{-}$ is found to be significantly larger than those obtained in $\Lambda_b^{0} \to p K^{-} K^{+} K^{-}$ and $\Lambda_b^{0} \to p K^{-} K^{+} \pi^{-}$. According to the definition of $B = b/c$, if the amplitude of either the vector or scalar meson decay strongly dominates over the other, the denominator $c$ becomes numerically dominant over the numerator $b$, naturally suppressing the coefficient $B$. As shown in Table.~\ref{brcpv}, the two decay channels with small coefficient $B$ indeed follow this pattern. The coefficient $B$ reliably reflects the interference effects only when the amplitudes of the multiple interfering processes are comparable. For instance, in the decay $\Lambda_b^{0} \to p K^{-} \pi^{+} \pi^{-}$, when we artificially adjust the decay widths of the intermediate mesons to enhance the interference, the value of coefficient $B$ increases markedly, supporting this interpretation.


    \item For the first channel, the experimental paper presents the composition of the $\pi^+\pi^-$ invariant mass distribution in the figure~\cite{LHCb:2025ray}. However, our inference based on the total $CP$ violation suggests that the fraction of channel $\Lambda^0_b \to \Lambda(1520)\rho^0$ is not as small relative to channel $\Lambda^0_b \to \Lambda(1520)f_0(980)$ as the experimental data indicates. We propose two potential explanations for this discrepancy. Firstly, in the $pK^-$ invariant mass region studied, we only considered the $\Lambda(1520)$ resonance, while significant contributions from other peaks are present in this region experimentally. From a FSI perspective, we should include all possible intermediate states and consider all allowed triangle diagrams. It is possible that channel $\Lambda^0_b \to \Lambda(1520)f_0(980)$ constitutes a larger fraction in processes involving other resonances, and the combined contributions might bring our predictions into agreement with the experimental results. Secondly, and in our view the most probable reason, is the incomplete understanding of the $f_0(980)$ particle. This lack of clarity affects our calculations in two distinct ways: On one hand, the strong coupling constant $g_{f_0(980)\pi\pi}$ is not precisely determined. Different studies report different values depending on the assumed internal structure and different experimental analyses. In this work, we used $g_{f_0(980)\pi\pi} = 1.6$; however, if we instead adopt a value of $2.1$~\cite{KLOE:2002deh,KLOE:2002kzf}, the total $CP$ violation changes from $10.8$ to $8.0$. The primary decay fractions inferred with this value are then largely consistent with experiment. This demonstrates the significant sensitivity of our results to the choice of this coupling constant. On the other hand, our calculation might only account for a part of the $\pi\pi$ contribution around $980$MeV. The strong coupling constant we used is based on the tetra-quark model, but other contributions, such as from two-quark states, quark-gluon hybrids, or molecular states, might also be present in this mass region. In essence, both aspects above stem from the same fundamental issue: the unclear nature of the $f_0(980)$ particle. We hope future studies based on a better understanding of this resonance will lead to results more consistent with experimental data. Furthermore, the computational uncertainties arising from the imperfect knowledge of scalar particles could potentially affect some conclusions mentioned earlier. For instance, the previously discussed $100\%$ $CP$ asymmetry for coefficient $B$ in channel $\Lambda^0_b \to pK^-\pi^+\pi^-$ might be altered if the actual fraction of the primary decay channel $\Lambda^0_b \to \Lambda(1520)\rho^0$ is smaller than currently estimated, thereby reducing its influence on $B$.

\end{itemize}
\section{Summary}\label{Summary}

In this work, we have investigated eight two-body non-leptonic decay channels of $\Lambda_b^0$ and $\Xi_b^-$ to $\Lambda^0(1520)$ with the framework of final-state interaction. 
Adopting the naive factorization approach, the hadronic matrix elements for these processes are factorized into meson decay constants, as well as $\Lambda_b^0$ and $\Xi_b^-$ to $\Lambda^0(1520)$ transition matrix elements. For the non-factorizable contributions, we evaluate all possible triangle diagrams for each decay mode using the final-state rescattering mechanism.
In calculating the long-distance contributions arising from triangle diagrams, we introduce two model parameters, $\Lambda_{\rm{charm}}$ and $\Lambda_{\rm{charmless}}$, to account for the off-shell effects of the exchanged particles and to regularize the UV divergence.

Subsequently, we calculated the helicity amplitude  for each decay process and extracted the  corresponding contributions of different CKM matrix elements, which clearly exhibit the differences of each decay amplitude and provide an explanation for the variations in their numerical behaviors. 
Then we predicted the branching ratios and $CP$ asymmetries for each channel and discussed the model dependence on the two parameters, $\Lambda_{\rm{charm}}$ and $\Lambda_{\rm{charmless}}$.
Furthermore, in order to compare with experimental observables directly, we incorporated the interference effects between different decay chains leading to the same final state and evaluated the total $CP$ violation. Accordingly, we defined several valuable observables based on a kinematic analysis of the final results. By exploring the origins of the differences in decay amplitudes, we identified the most promising channels for future experimental observations, it remains pronounced across all three computational schemes.

This work extends the theoretical framework established in our previous studies and updates the necessary input parameters, applying this approach  to the decay channels involving high-spin baryonic resonances for the first time. In light of the experimental observation of $CP$ violation in $\Lambda_b$ decays, our results offer an essential theoretical benchmark for upcoming high-precision experiments. Furthermore, this method shows the potential for broader applications to higher excited states and even multiquark states, such as $\Lambda_b^0 \to P_c(4312)^+ \pi^-$ and $\Lambda_b^0 \to P_c(4457)^+ \pi^-$ decays.

\section*{Acknowledgment}

The authors acknowledge Prof. Zhen-Hua Zhang for providing excellent solution and Jian-Peng Wang, Zhu-Ding Duan for valuable discussions. This work is supported by the National Science Foundation of China under Grant No. 12335003, No. 12375086 and No. 12522506, and by the Fundamental Research Funds for the Central Universities under No. lzujbky-2023-stlt01 and lzujbky-2024-oy02.
\appendix
\section{Full expressions of amplitudes}\label{app.A}
Here we give the full amplitude expressions of eight decay channels we consider in this work:

\begin{equation}
\begin{aligned}
\mathcal{A}(\Lambda_{b} \rightarrow \Lambda(1520) \pi^{0}) & =\mathcal{S}(\Lambda_{b} \rightarrow \Lambda(1520) \pi^{0})+\mathcal{M}(K^{-},p;K^{*-})+\mathcal{M}(K^{*-},p;K^{-})+\mathcal{M}(K^{*-},p;K^{*-})\\
&+\mathcal{M}(K^{-},p;p)+\mathcal{M}(K^{*-},p;p)+\mathcal{M}(\bar{K^{0}},n;\bar{K^{*0}})+\mathcal{M}(\bar{K^{*0}},n;\bar{K^{0}})+\mathcal{M}(\bar{K^{*0}},n;\bar{K^{*0}})\\
&+\mathcal{M}(\pi^{0},\Lambda;\Sigma^{0})+\mathcal{M}(\rho^{0},\Lambda;\Sigma^{0})+\mathcal{M}(\bar{K^{0}},n;n)+\mathcal{M}(\bar{K^{*0}},n;n),
\end{aligned}
\end{equation}

\begin{equation}
\hspace{-5em} 
\begin{aligned}
\mathcal{A}(\Lambda_{b} \rightarrow \Lambda(1520) \phi) & =\mathcal{S}(\Lambda_{b} \rightarrow \Lambda(1520) \phi)+\mathcal{M}(K^{-},p;K^{-})+\mathcal{M}(K^{-},p;K^{*-})+\mathcal{M}(K^{*-},p;K^{-})\\
&+\mathcal{M}(K^{*-},p;K^{*-})+\mathcal{M}(D_s^{-},\Lambda_c^{+};D_s^{-})+\mathcal{M}(D_s^{-},\Lambda_c^{+};D_s^{*-})+\mathcal{M}(D_s^{*-},\Lambda_c^{+};D_s^{-})\\
&+\mathcal{M}(D_s^{*-},\Lambda_c^{+};D_s^{*-})+\mathcal{M}(\bar{K^{0}},n;\bar{K^{0}})+\mathcal{M}(\bar{K^{0}},n;\bar{K^{*0}})+\mathcal{M}(\bar{K^{*0}},n;\bar{K^{0}})\\
&+\mathcal{M}(\bar{K^{*0}},n;\bar{K^{*0}})+\mathcal{M}(\eta,\Lambda;\Lambda)+\mathcal{M}(\phi,\Lambda;\Lambda)+\mathcal{M}(\omega,\Lambda;\Lambda),
\end{aligned}
\end{equation}

\begin{equation}
\hspace{-4em}
\begin{aligned}
\mathcal{A}(\Lambda_{b} \rightarrow \Lambda(1520) f_0(980)) & =\mathcal{S}(\Lambda_{b} \rightarrow \Lambda(1520) f_0(980))+\mathcal{M}(K^{-},p;K^{-})+\mathcal{M}(D_s^{-},\Lambda_c^{+};D_s^{-})\\
&+\mathcal{M}(K^{-},p;p)+\mathcal{M}(K^{*-},p;p)+\mathcal{M}(D_s^{-},\Lambda_c^{+};\Lambda_c^{+})+\mathcal{M}(D_s^{*-},\Lambda_c^{+};\Lambda_c^{+})\\
&+\mathcal{M}(\eta,\Lambda;\eta)+\mathcal{M}(\bar{K^{0}},n;\bar{K^{0}})+\mathcal{M}(\eta,\Lambda;\Lambda)+\mathcal{M}(\phi,\Lambda;\Lambda)+\mathcal{M}(\omega,\Lambda;\Lambda)\\
&+\mathcal{M}(\bar{K^{0}},n;n)+\mathcal{M}(\bar{K^{*0}},n;n),
\end{aligned}
\end{equation}

\begin{equation}
\hspace{-5em}
\begin{aligned}
\mathcal{A}(\Lambda_{b} \rightarrow \Lambda(1520) \rho^{0})& =\mathcal{S}(\Lambda_{b} \rightarrow \Lambda(1520) \rho^{0})+\mathcal{M}(K^{-},p;K^{-})+\mathcal{M}(K^{-},p;K^{*-})+\mathcal{M}(K^{*-},p;K^{-})\\
&+\mathcal{M}(K^{*-},p;K^{*-})+\mathcal{M}(K^{-},p;p)+\mathcal{M}(K^{*-},p;p)+\mathcal{M}(\pi^{0},\Lambda;\omega)\\
&+\mathcal{M}(\bar{K^{0}},n;\bar{K^{0}})+\mathcal{M}(\bar{K^{0}},n;\bar{K^{*0}})+\mathcal{M}(\bar{K^{*0}},n;\bar{K^{0}})+\mathcal{M}(\bar{K^{*0}},n;\bar{K^{*0}})\\
&+\mathcal{M}(\pi^{0},\Lambda;\Sigma^{0})+\mathcal{M}(\rho^{0},\Lambda;\Sigma^{0})+\mathcal{M}(\bar{K^{0}},n;n)+\mathcal{M}(\bar{K^{*0}},n;n),
\end{aligned}
\end{equation}

\begin{equation}
\hspace{-5em}
\begin{aligned}
\mathcal{A}(\Lambda_{b} \rightarrow \Lambda(1520) f_0(500)) & =\mathcal{S}(\Lambda_{b} \rightarrow \Lambda(1520) f_0(500))+\mathcal{M}(K^{-},p;K^{-})+\mathcal{M}(D_s^{-},\Lambda_c^{+};D_s^{-})\\
&+\mathcal{M}(K^{-},p;p)+\mathcal{M}(K^{*-},p;p)+\mathcal{M}(D_s^{-},\Lambda_c^{+};\Lambda_c^{+})+\mathcal{M}(D_s^{*-},\Lambda_c^{+};\Lambda_c^{+})\\
&+\mathcal{M}(\eta,\Lambda;\eta)+\mathcal{M}(\bar{K^{0}},n;\bar{K^{0}})+\mathcal{M}(\eta,\Lambda;\Lambda)+\mathcal{M}(\phi,\Lambda;\Lambda)+\mathcal{M}(\omega,\Lambda;\Lambda)\\
&+\mathcal{M}(\bar{K^{0}},n;n)+\mathcal{M}(\bar{K^{*0}},n;n),
\end{aligned}
\end{equation}

\begin{equation}
\begin{aligned}
\mathcal{A}(\Lambda_{b} \rightarrow \Lambda(1520) K^{*0})& =\mathcal{S}(\Lambda_{b} \rightarrow \Lambda(1520) K^{*0})+\mathcal{M}(\pi^{-},p;K^{-})+\mathcal{M}(\rho^{-},p;K^{-})+\mathcal{M}(\pi^{-},p;K^{*-})\\
&+\mathcal{M}(\rho^{-},p;K^{*-})+\mathcal{M}(D^{-},\Lambda_c^{+};D_{s}^{-})+\mathcal{M}(D^{*-},\Lambda_c^{+};D_{s}^{-})+\mathcal{M}(D^{-},\Lambda_c^{+};D_{s}^{*-})\\
&+\mathcal{M}(D^{*-},\Lambda_c^{+};D_{s}^{*-})+\mathcal{M}(\pi^{-},p;\Sigma^{+})+\mathcal{M}(\rho^{-},p;\Sigma^{+})+\mathcal{M}(D^{-},\Lambda_c^{+};\Xi_{c}^+)\\
&+\mathcal{M}(D^{*-},\Lambda_c^{+};\Xi_{c}^+)+\mathcal{M}(K^0,\Lambda;\eta)+\mathcal{M}(K^0,\Lambda;\phi)+\mathcal{M}(K^0,\Lambda;\omega)\\
&+\mathcal{M}(K^{*0},\Lambda;\eta)+\mathcal{M}(K^{*0},\Lambda;\phi)+\mathcal{M}(K^{*0},\Lambda;\omega)+\mathcal{M}(\pi^{0},n;\bar{K^{0}})+\mathcal{M}(\eta,n;\bar{K^{0}})\\
&+\mathcal{M}(\rho^{0},n;\bar{K^{0}})+\mathcal{M}(\phi,n;\bar{K^{0}})+\mathcal{M}(\omega,n;\bar{K^{0}})+\mathcal{M}(\pi^{0},n;\bar{K^{*0}})+\mathcal{M}(\eta,n;\bar{K^{*0}})\\
&+\mathcal{M}(\rho^{0},n;\bar{K^{*0}})+\mathcal{M}(\phi,n;\bar{K^{*0}})+\mathcal{M}(\omega,n;\bar{K^{*0}})+\mathcal{M}(K^{0},\Lambda;\Xi^{0})+\mathcal{M}(K^{*0},\Lambda;\Xi^{0})\\
&+\mathcal{M}(\eta,n;\Lambda)+\mathcal{M}(\phi,n;\Lambda)+\mathcal{M}(\omega,n;\Lambda)+\mathcal{M}(\pi^{0},n;\Sigma^{0})+\mathcal{M}(\pi^{0},n;\Sigma^{*0})\\
&+\mathcal{M}(\rho^{0},n;\Sigma^{0}),
\end{aligned}
\end{equation}

\begin{equation}
\begin{aligned}
\mathcal{A}(\Lambda_{b} \rightarrow \Lambda(1520) K(700))& =\mathcal{S}(\Lambda_{b} \rightarrow \Lambda(1520) K(700))+\mathcal{M}(\pi^{-},p;K^{-})++\mathcal{M}(D^{-},\Lambda_c^{+};D_{s}^{-})\\
&+\mathcal{M}(\pi^{-},p;\Sigma^{+})+\mathcal{M}(\rho^{-},p;\Sigma^{+})+\mathcal{M}(D^{-},\Lambda_c^{+};\Xi_{c}^+)+\mathcal{M}(D^{*-},\Lambda_c^{+};\Xi_{c}^+)\\
&+\mathcal{M}(K^{0},\Lambda;\eta)+\mathcal{M}(\pi^{0},n;\bar{K^{0}})+\mathcal{M}(\eta,n;\bar{K^{0}})+\mathcal{M}(K^{0},\Lambda;\Xi^0)+\mathcal{M}(K^{*0},\Lambda;\Xi^0)\\
&+\mathcal{M}(\eta,n;\Lambda)+\mathcal{M}(\phi,n;\Lambda)+\mathcal{M}(\omega,n;\Lambda)+\mathcal{M}(\pi^0,n;\Sigma^0)+\mathcal{M}(\rho^0,n;\Sigma^0),
\end{aligned}
\end{equation}

\begin{equation}
\begin{aligned}
\mathcal{A}(\Xi_{b}^- \rightarrow \Lambda(1520) K^{-})& =\mathcal{S}(\Xi_{b}^- \rightarrow \Lambda(1520) K^{-})+\mathcal{M}(D_s^{-},\Xi_c^0;\bar{D^{*0}})+\mathcal{M}(D_s^{*-},\Xi_c^0;\bar{D^{0}})+\mathcal{M}(D_s^{*-},\Xi_c^0;\bar{D^{*0}})\\
&+\mathcal{M}(K^{-},\Sigma^{0};\rho^0)+\mathcal{M}(K^{*-},\Sigma^{0};\pi^0)+\mathcal{M}(K^{*-},\Sigma^{0};\rho^0)+\mathcal{M}(K^{-},\Lambda;\phi)\\
&+\mathcal{M}(K^{-},\Lambda;\omega)+\mathcal{M}(K^{*-},\Lambda;\eta)+\mathcal{M}(K^{*-},\Lambda;\phi)+\mathcal{M}(K^{*-},\Lambda;\omega)+\mathcal{M}(D_s^{-},\Xi_c^0;\Lambda_c^+)\\
&+\mathcal{M}(D_s^{*-},\Xi_c^0;\Lambda_c^+)+\mathcal{M}(K^{-},\Sigma^0;p)+\mathcal{M}(K^{*-},\Sigma^0;p)+\mathcal{M}(K^{-},\Lambda;p)\\
&+\mathcal{M}(K^{*-},\Lambda;p)+\mathcal{M}(\bar{K^{0}},\Sigma^-;\rho^+)+\mathcal{M}(\bar{K^{*0}},\Sigma^-;\pi^+)+\mathcal{M}(\bar{K^{*0}},\Sigma^-;\rho^+)\\
&+\mathcal{M}(\rho^0,\Xi^-;K^{+})+\mathcal{M}(\phi,\Xi^-;K^{+})+\mathcal{M}(\omega,\Xi^-;K^{+})+\mathcal{M}(\pi^0,\Xi^-;K^{*+})\\
&+\mathcal{M}(\eta,\Xi^-;K^{*+})+\mathcal{M}(\rho^0,\Xi^-;K^{*+})+\mathcal{M}(\phi,\Xi^-;K^{*+})+\mathcal{M}(\omega,\Xi^-;K^{*+})\\
&+\mathcal{M}(\bar{K^{0}},\Sigma^-;n)+\mathcal{M}(\bar{K^{*0}},\Sigma^-;n)+\mathcal{M}(\eta,\Xi^-;\Lambda)+\mathcal{M}(\phi,\Xi^-;\Lambda)+\mathcal{M}(\omega,\Xi^-;\Lambda)\\
&+\mathcal{M}(\pi^{0},\Xi^-;\Sigma^{0})+\mathcal{M}(\rho^{0},\Xi^-;\Sigma^{0}).
\end{aligned}
\end{equation}
Here, $\mathcal{S}(\mathcal{B}_b \to f_1f_2)$ represents the direct short-distance weak process. $\mathcal{M}(A,B;C)$ denotes a process in which the initial particle $\mathcal{B}_b$ first undergoes a weak vertex process that produces A and B, then A and B further interact via a strong vertex by exchanging particle C, finally resulting in the final state $f_1f_2$.

\section{Effective Lagrangians}\label{app.B}

All the Effecitve Lagrangians for the hadronic interactions we used are classified and presented as follows~\cite{Cheng:2004ru,Aliev:2006xr,Aliev:2009ei,Yu:2017zst,Riska:2000gd,Erkol:2006eq,Black:1998zc}

\begin{itemize}
    \item The effective Lagrangians for vector meson $V$, pseudoscalar meson octet $P$, and baryon octet $B_8$ :
\begin{equation}
 \begin{aligned}
    \mathcal{L}_{V P P}
    &=\frac{\mathrm{i} g_{\rho\pi\pi}}{\sqrt{2}} \operatorname{Tr}\left[V^\mu\left[P, \partial_\mu P\right]\right],
    \nonumber\\[3mm]
    \mathcal{L}_{V V V}
    &=\frac{i g_{\rho\rho\rho}}{\sqrt{2}} \operatorname{Tr}\left[\left(\partial_\nu V_\mu V^\mu-V^\mu \partial_\nu V_\mu\right) V^\nu\right]=\frac{\mathrm{i} g_{\rho\rho\rho}}{\sqrt{2}} \operatorname{Tr}\left[\left(\partial_\nu V_\mu-\partial_\mu V_\nu\right) V^\mu V^\nu\right],
    \nonumber\\[3mm]
    \mathcal{L}_{V V P}
    &=\frac{4 g_{V V P}}{f_{P}} \varepsilon^{\mu \nu \alpha \beta} \operatorname{Tr}\left(\partial_\mu V_\nu \partial_\alpha V_\beta P\right),
    \nonumber\\[3mm]
    \mathcal{L}_{P B_8 B_8}
    &=\sqrt{2}\Big(D\operatorname{Tr}\left[ \bar{B_8} \{P,B_8\} \right]+F\operatorname{Tr}\left[ \bar{B_8} [P,B_8]\right] \Big),\nonumber\\[3mm]
    \mathcal{L}_{V B_8 B_8}
    &=\sqrt{2}\Big(F \operatorname{Tr} \left[ \bar{B} [V,B_8] \right]+D \operatorname{Tr}\left[ \bar{B_8} \{V,B_8\} \right]+(F-D)\operatorname{Tr}[\bar{B_8} B_8]\;\operatorname{Tr}[V]\Big).\nonumber\\[3mm]
 \end{aligned}
 \end{equation}

\item The Lagrangians involving $D^{(*)}$ meson, pseudoscalar meson octet $P$, and vector meson octet $V$ :
\begin{equation}
 \begin{aligned}
\mathcal{L}_{D^*DP}&=-ig_{D^*DP}\left(D\partial^\mu PD_\mu^{*\dagger}-D_\mu^{*}\partial^\mu PD^{\dagger}\right),\\[3mm]
\mathcal{L}_{{D^{*}D^{*}P}}&=\frac{1}{2}g_{{D^{*}D^{*}P}}\varepsilon_{\mu\nu\alpha\beta}D^{*\mu}\partial^{\nu}P\overleftrightarrow{\partial}^{\alpha}D^{*\beta\dagger},\\[3mm]
\mathcal{L}_{DDV}&=-ig_{DDV}D\overleftrightarrow{\partial}_{\mu}D^{\dagger}\left(V^\mu\right),\\[3mm]
\mathcal{L}_{D^*DV}&=-2f_{D^*DV}\epsilon_{\mu\nu\alpha\beta}\left(\partial^\mu V^\nu\right)\left(D\overleftrightarrow{\partial}^\alpha D^{*\beta\dagger}-D^{*\beta}\overleftrightarrow{\partial}^\alpha D^\dagger\right),\\[3mm]
\mathcal{L}_{{D^{*}D^{*}V}}&=ig_{{D^{*}D^{*}V}}D^{*\nu}\overleftrightarrow{\partial}_{\mu}D_{\nu}^{*\dagger}\left(V^{\mu}\right)+4if_{{D^{*}D^{*}V}}D_{\mu}^{*}\left(\partial^{\mu}V^{\nu}-\partial^{\nu}V^{\mu}\right)D_{\nu}^{*\dagger}\nonumber.\\[3mm]
 \end{aligned}
\end{equation}

 \item The Lagrangians involving charmed baryon sextets $B_6$, anti-triplets $B_{\bar{3}}$, vector meson $V$, and pseudoscalar mesons octet $P$ :
 \begin{equation}
 \begin{aligned}
    \mathcal{L}_{Vhh}
    &=\left\{f_{1V {B}_6 {B}_6} \operatorname{Tr}\left[\bar{{B}}_6 \gamma_\mu V^\mu {B}_6\right]+\frac{f_{2 V {B}_6 {B}_6}}{m_6+m^{\prime}_{6} }\operatorname{Tr}\left[\bar{{B}}_6 \sigma_{\mu \nu} \partial^\mu V^\nu {B}_6\right] \right\}\nonumber\\
    & +\left\{f_{1 V {B}_{\bar{3}} {B}_{\bar{3}}} \operatorname{Tr}\left[\bar{{B}}_{\bar{3}} \gamma_\mu V^\mu {B}_{\bar{3}}\right]+\frac{f_{2 V {B}_{\bar{3}} {B}_{\bar{3}}}}{ m_{\bar{3}}+m_{\bar{3}}^\prime} \operatorname{Tr}\left[\bar{{B}}_{\bar{3}} \sigma_{\mu \nu} \partial^\mu V^\nu {B}_{\bar{3}}\right]\right\}\nonumber \\
    & +\left\{f_{1 V {B}_6 {B}_{\bar{3}}} \operatorname{Tr}\left[\bar{{B}}_6 \gamma_\mu V^\mu {B}_{\bar{3}}\right]+\frac{f_{2 V {B}_6 {B}_{\bar{3}}}}{m_6+m_{\bar{3}}} \operatorname{Tr}\left[\bar{{B}}_6 \sigma_{\mu \nu} \partial^\mu V^\nu {B}_{\bar{3}}\right]+h . c .\right\},\nonumber \\[3mm]
     \mathcal{L}_{Phh}
     &=g_{P {B}_6 {B}_6} \operatorname{Tr}\left[\bar{{B}}_6 i \gamma_5 P {B}_6\right]+g_{P {B}_{\bar{3}} {B}_{\bar{3}}} \operatorname{Tr}\left[\bar{{B}}_{\bar{3}} i \gamma_5 P {B}_{\bar{3}}\right]+\left\{g_{P {B}_6 {B}_{\bar{3}}} \operatorname{Tr}\left[\bar{{B}}_6 i \gamma_5 P {B}_{\bar{3}}\right]+h . c .\right\}.\nonumber\nonumber
 \end{aligned}
 \end{equation}
 \item The Lagrangians for charmed baryon sextets  $B_6$, anti-triplets $B_{\bar{3}}$, baryon octet $B_8$, and $D^{(*)}$-mesons:
  \begin{equation}
 \begin{aligned}
     \mathcal{L}_{\Lambda_{\mathrm{c}} B_8 D}
     &=g_{\Lambda_{\mathrm{c}} B_8 D}\left(\bar{\Lambda}_{\mathrm{c}} i \gamma_5 D B_8+{ h.c. }\right),\nonumber\\[3mm]
     \mathcal{L}_{\Lambda_{\mathrm{c}} B_8 D^*}
     &=f_{1 \Lambda_{\mathrm{c}} B_8 D^*}\left(\bar{\Lambda}_{\mathrm{c}} \gamma_\mu D^{* \mu} B_8+ { h.c. }\right)+\frac{f_{2 \Lambda_{\mathrm{c}} B_8 D^*}}{m_{\Lambda_{\mathrm{c}}}+m_{B_8}}\left(\bar{\Lambda}_{\mathrm{c}} \sigma_{\mu \nu} \partial^\mu D^{* \nu} B_8+{ h.c. }\right),\nonumber\\[3mm]
     \mathcal{L}_{\Sigma_{\mathrm{c}} B_8 D}
     &= g_{\Sigma_{\mathrm{c}} B_8 D}\left(\bar{\Sigma}_{\mathrm{c}} i \gamma_5 D B_8+{ h.c. }\right),\nonumber\\[3mm]
     \mathcal{L}_{\Sigma_{\mathrm{c}} B_8 D^*}
     & =f_{1 \Sigma_{\mathrm{c}} B_8 D^*}\left(\bar{\Sigma}_{\mathrm{c}} \gamma_\mu D^{* \mu} B_8+{ h.c. }\right)+\frac{f_{2 \Sigma_{\mathrm{c}} B_8 D^*}}{m_{\Sigma_{\mathrm{c}}}+m_{B_8}}\left(\bar{\Sigma}_{\mathrm{c}} \sigma_{\mu \nu} \partial^\mu D^{* \nu} B_8+{ h.c. }\right).\nonumber
\end{aligned}
 \end{equation}

 \item The Lagrangians involving scalar meson $S$ :
\begin{equation}
\hspace{-13em} 
\begin{aligned}
    \mathcal{L}_{SPP} &= -\gamma_{SPP}S\partial_{\mu}P\partial^{\mu}P, \\[3mm]
    \mathcal{L}_{SVV} &= \frac{g_{SVV}}{m_{V}}(\partial^{\alpha}V^{\beta}\partial_{\alpha}V_{\beta}-\partial^{\alpha}V^{\beta}\partial_{\beta}V_{\alpha})S, \\[3mm]
    \mathcal{L}_{SB_8B_8} &= -g_{SBB}\bar{B_8}SB_8 \nonumber.\\[3mm]
\end{aligned}
\end{equation}

 \item The Lagrangians involving $\Lambda(1520)$:
\begin{equation}
\begin{aligned}
\mathcal{L}_{PB_8\Lambda}^{1520}&=\frac{f_{PB_8\Lambda}^{1520}}{m_P}\bar{B_8}\gamma_5\partial_\sigma P\Lambda^\sigma+h.c.,\\[3mm]
\mathcal{L}_{VB_8\Lambda}^{1520}&=\frac{ig_{VB_8\Lambda}^{1520}}{m_V^2}\bar{B_8}\sigma_{\mu\nu}\partial^\nu\partial_\sigma V^\mu\Lambda^\sigma+h.c.,\nonumber\\[3mm]
\mathcal{L}_{PB_{10} \Lambda}^{1520} &= \frac{f_{PB_{10} \Lambda}^{1520}}{m_P}\bar{\Lambda}^\sigma\gamma^{\mu}B_{10\sigma}\partial_{\mu}P.\nonumber\\[3mm]
\end{aligned}
\end{equation}

 \end{itemize}
The matrices under SU(3) flavor group representations are given:

  \begin{equation}
 \begin{aligned}
P=\left(\begin{array}{ccc}
\frac{\pi^0}{\sqrt{2}}+\frac{\eta}{\sqrt{6}} & \pi^{+} & \mathrm{K}^{+} \\
\pi^{-} & -\frac{\pi^0}{\sqrt{2}}+\frac{\eta}{\sqrt{6}} & \mathrm{~K}^0 \\
\mathrm{~K}^{-} & \bar{K}^0 & -\sqrt{\frac{2}{3}} \eta
\end{array}\right), \quad {B}_6=\left(\begin{array}{ccc}
\Sigma_{\mathrm{c}}^{++} & \frac{1}{\sqrt{2}} \Sigma_{\mathrm{c}}^{+} & \frac{1}{\sqrt{2}} \Xi_{\mathrm{c}}^{\prime+} \\
\frac{1}{\sqrt{2}} \Sigma_{\mathrm{c}}^{+} & \Sigma_{\mathrm{c}}^0 & \frac{1}{\sqrt{2}} \Xi_{\mathrm{c}}^{\prime 0} \\
\frac{1}{\sqrt{2}} \Xi_{\mathrm{c}}^{\prime+} & \frac{1}{\sqrt{2}} \Xi_{\mathrm{c}}^{\prime 0} & \Omega_{\mathrm{c}}
\end{array}\right),
\end{aligned}
 \end{equation}

   \begin{equation}
 \begin{aligned}
V=\left(\begin{array}{ccc}
\frac{\rho^0}{\sqrt{2}}+\frac{\omega}{\sqrt{2}} & \rho^{+} & \mathrm{K}^{*+} \\
\rho^{-} & -\frac{\rho^0}{\sqrt{2}}+\frac{\omega}{\sqrt{2}} & \mathrm{~K}^{* 0} \\
\mathrm{~K}^{*-} & \bar{\mathrm{K}}^{* 0} & \phi
\end{array}\right), ~~~~~~~~~~~~~\quad {B}_{\bar{3}}=\left(\begin{array}{ccc}
0 & \Lambda_{\mathrm{c}}^{+} & \Xi_{\mathrm{c}}^{+} \\
-\Lambda_{\mathrm{c}}^{+} & 0 & \Xi_{\mathrm{c}}^0 \\
-\Xi_{\mathrm{c}}^{+} & -\Xi_{\mathrm{c}}^0 & 0
\end{array}\right),
\end{aligned}
 \end{equation}
 
  \begin{equation}
 \begin{aligned}
 {B_8}=\left(\begin{array}{ccc}
\frac{\Sigma^0}{\sqrt{2}}+\frac{\Lambda}{\sqrt{6}} & \Sigma^{+} & p \\
\Sigma^{-} & -\frac{\Sigma^0}{\sqrt{2}}+\frac{\Lambda}{\sqrt{6}} & n \\
\Xi^{-} & \Xi^0 & -\frac{2}{\sqrt{6}} \Lambda
\end{array}\right)
 ,~~~~~~~~~~~~~~~\quad D=\left(
\begin{matrix}
    D^0,D^+,D_s^+
\end{matrix}\right).
\end{aligned}
 \end{equation}

\section{The Feynman rules of strong interaction vertex }\label{app.C}
\begin{equation}
	\begin{aligned}
    \langle V(k,\lambda_k)P(p_{2})|i\mathcal{L}_{VPP}|P(p_{1})\rangle&=-ig_{VVP}\varepsilon^{*\mu}(k,\lambda_k)(p_{1}+p_{2})_{\mu},\\
            \end{aligned}
\end{equation}
\begin{equation}
	\begin{aligned}
    \langle V(p_{3},\lambda_{3})V(k,\lambda_{k})|i\mathcal{L}_{VVP}|P(p_{1})\rangle&=-i\frac{g_{VVP}}{f_{p}}\epsilon^{\mu\nu\alpha\beta}p_{3\,\mu}\varepsilon^{*}_{\nu}(\lambda_{3},p_{3})k_{\alpha}\varepsilon^{*}_{\beta}(k,\lambda_{k}),\\
        \end{aligned}
\end{equation}
\begin{equation}
	\begin{aligned}
    \langle V(p_{3},\lambda_{3})V(k,\lambda_{k})|i\mathcal{L}_{VVV}|V(p_{1},\lambda_{1})\rangle
    &= -\frac{ig_{VVV}}{\sqrt{2}}\varepsilon_{\mu}(p_{1},\lambda_{1})\varepsilon^{\mu*}(p_{3},\lambda_{3})\varepsilon^{*}_{\nu}(k,\lambda_{k})\left(p^{\nu}_{3}+p^{\nu}_{1}\right)\\
    & -\frac{ig_{VVV}}{\sqrt{2}}\varepsilon^{*}_{\mu}(k,\lambda_{k})\varepsilon^{\mu}(p_{1},\lambda_{1})\varepsilon^{*}_{\nu}(p_{3},\lambda_{3})\left(-p^{\nu}_{1}-p^{\nu}_{k}\right)\\
    &-\frac{ig_{VVV}}{\sqrt{2}}\varepsilon^{*}_{\mu}(p_{3},\lambda_{3})\varepsilon^{*\mu}(k,\lambda_{k})\varepsilon_{\nu}(p_{1},\lambda_{1})\left(p^{\nu}_{k}-p^{\nu}_{3}\right),
    \end{aligned}
\end{equation}
\begin{equation}
	\begin{aligned}
\langle B(p_{2})P(q)|i\mathcal{L}_{PBB}|B(p_{1})\rangle&=g_{BBP}\bar{u}(p_{2})i\gamma_{5}u(p_{1}),\\
        \end{aligned}
\end{equation}
\begin{equation}
	\begin{aligned}
\langle B(p_{2})V(q,\lambda_q)|i\mathcal{L}_{VBB}|B(p_{1})\rangle&=\bar{u}(p_{2})\left[f_{1}\gamma_{\nu}+f_{2}\frac{i}{m_{1}+m_{2}}\sigma_{\mu\nu}q^{\mu}\right]\varepsilon^{*\nu}(q,\lambda_q)u(p_{1}),\\
    \end{aligned}
\end{equation}
\begin{equation}
	\begin{aligned}
\bra{P(p_{3})D(k,\lambda_{k})}i\mathcal{L}_{D^*DP}\ket{D^{\ast}(p_{1},\lambda_{1})}
    &=ig_{D^{\ast} D P}p^{\mu}_{3}\varepsilon_{\mu}(p_1,\lambda_1),\\
            \end{aligned}
\end{equation}
\begin{equation}
	\begin{aligned}
\bra{P(p_{3})D^*(k,\lambda_{k})}i\mathcal{L}_{D^*D^*P}\ket{D^{\ast}(p_1,\lambda_{1})}
    &= \frac{i}{2}g_{D^*D^*P}\varepsilon_{\mu \nu \alpha \beta}\varepsilon^{*\mu}(k,\lambda_k)\varepsilon^\beta(p_1,\lambda_1)p^\nu_{3} p^\alpha_{1},\\
            \end{aligned}
\end{equation}
\begin{equation}
	\begin{aligned}
\bra{V(p_{3},\lambda_{3})D^*(k,\lambda_{k})}i\mathcal{L}_{D^*D^*V}\ket{D^*(p_{1},\lambda_{1})}
    &=2ig_{D^*D^*V}\varepsilon^{*\nu}(k,\lambda_k)\varepsilon_{\nu}(p_1,\lambda_1)k_\mu \varepsilon^{*\mu}(p_{3},\lambda_3)\\
    &-4if_{D^* D^*V}\varepsilon_{\mu}^*(k,\lambda_k) \Big(p^\mu_{3} \varepsilon^{*\nu}(p_{3},\lambda_3)-p^\nu_{3}\varepsilon^{*\mu}(p_{3},\lambda_3)\Big)\varepsilon_{\nu}(p_{1},\lambda_1),\\
                \end{aligned}
\end{equation}
\begin{equation}
	\begin{aligned}
\bra{V(p_{3},\lambda_{3})D^*(k,\lambda_{k})}i\mathcal{L}_{D^*DV}\ket{D(p_{1})}
   & =2 i f_{D^*DV} \varepsilon_{\mu \nu \alpha \beta} \varepsilon^{*\nu}(p_{3},\lambda_3) \varepsilon^{*\beta}(k,\lambda_k) p^\mu_{3} (k^\alpha +p^\alpha_{1}),\\
               \end{aligned}
\end{equation}
\begin{equation}
	\begin{aligned}
\bra{V(p_{3},\lambda_{3})D(k)}i\mathcal{L}_{DDV}\ket{D(p_{1})}
    &=-i g_{DDV}\varepsilon^{*\mu}(p_{3},\lambda_3)(p_{1,\mu}+k_{\mu}),\\
    \end{aligned}
\end{equation}
\begin{equation}
	\begin{aligned}
\bra{S({p_3})P(k)}i\mathcal{L}_{SPP}\ket{P(p_1)}
    &=-i\gamma_{SPP}p_1^\mu k_\mu,\\
                \end{aligned}
\end{equation}
\begin{equation}
	\begin{aligned}
\bra{S({p_3})B(k)}i\mathcal{L}_{SBB}\ket{B(p_2)}
    &=-ig_{SBB}\bar{u}(p_{2})u(p_{1}),\\
    \end{aligned}
\end{equation}
\begin{equation}
	\begin{aligned}
\bra{\Lambda^*(p_4)}i\mathcal{L}_{PB\Lambda}^{1520}\ket{P(k)B(p_2)}
&=\frac{f_{PB\Lambda}^{1520}}{m_k}\bar{u^\sigma}(p_4)\gamma_5 k_\sigma u(p_2),\\
            \end{aligned}
\end{equation}
\begin{equation}
	\begin{aligned}
\bra{\Lambda^*(p_4)}i\mathcal{L}_{VB\Lambda}^{1520}\ket{V(k,\lambda_k)B(p_2)}
&=-\frac{g_{VB\Lambda}^{1520}}{m_k^2}\bar{u^\sigma}(p_4)\sigma_{\mu\nu}k^\nu k_\sigma\varepsilon^{\mu}(k,\lambda_k)u(p_2).
    \end{aligned}
\end{equation}

\section{Amplitudes of triangle diagram}\label{app.D}

The amplitudes of $\Lambda^{0}_{b}/\Xi_b^- \rightarrow B_{8} P $:
\begin{equation}
	\begin{aligned}
\mathcal{M}[P,B_8;V]&=\displaystyle\int\frac{d^4p}{(2\pi)^4}(+i)\frac{g_{VB\Lambda}}{m_k^2}\frac{g_{PPV}}{\sqrt{2}}\bar{u^{\sigma}}(p_4, s_4)\sigma_{\mu\nu}k^{\nu}k_{\sigma}(\slashed{p_{2}} + m_{2})(A + B\gamma_{5}) u(p,s)(-g^{\mu\alpha} + \frac{k^\mu k^\alpha}{m_{k}^{2}})\\
&(p_1+p_3)_\alpha \mathcal{P}^3\mathcal{F},
    \end{aligned}
\end{equation}

\begin{equation}
	\begin{aligned}
\mathcal{M}[V,B_8;P]&=\displaystyle\int\frac{d^4p}{(2\pi)^4}(-1)\frac{f_{PB\Lambda}}{m_{k}}\frac{g_{PPV}}{\sqrt{2}}\bar{u^{\sigma}}(p_4, s_4)\gamma_{5}k_\sigma (\slashed{p_{2}} + m_{2})(A_{1}\gamma_\alpha \gamma_5+A_{2}\frac{p_{2\alpha}}{M}\gamma_5+B_{1}\gamma_\alpha+B_{2}\frac{p_{2\alpha}}{M})\\
&u(p,s)(p_3-k)_\beta(-g^{\alpha\beta} + \frac{p_1^\alpha p_1^\beta}{m_{1}^{2}})\mathcal{P}^3\mathcal{F},
    \end{aligned}
\end{equation}

\begin{equation}
	\begin{aligned}
\mathcal{M}[V,B_8;V]&=\displaystyle\int\frac{d^4p}{(2\pi)^4}(-1)\frac{g_{VB\Lambda}}{m_k^2}\frac{4g_{VVP}}{f_{3}}\bar{u^{\sigma}}(p_4, s_4)\sigma^{\mu\nu}k_\nu k_\sigma (\slashed{p_{2}} + m_{2}) 
(A_{1}\gamma^\rho\gamma_5+A_{2}\frac{p_{2}^{\rho}}{M}\gamma_5+B_{1}\gamma^{\rho}+B_{2}\frac{p_{2}^{\rho}}{M}) u(p,s)\\   
&\epsilon^{\alpha\beta\gamma\theta}p1_{\alpha}k_{\gamma}(-g_{\mu\theta} + \frac{k_{\mu} k_{\theta}}{m_{k}^{2}})(-g_{\rho\beta} + \frac{p_{1\rho} p_{1\beta}}{m_{1}^{2}})\mathcal{P}^3\mathcal{F},
    \end{aligned}
\end{equation}

\begin{equation}
	\begin{aligned}
\mathcal{M}[D,B_{\bar{3}}/B_6;D^*]&=\displaystyle\int\frac{d^4p}{(2\pi)^4}(+i)\frac{g_{VB\Lambda}}{m_k^2}g_{D^{*}DP}
\bar{u^{\sigma}}(p_4, s_4)\sigma_{\mu\nu}k^{\nu}k_\sigma(\slashed{p_{2}} + m_{2})(A + B\gamma_{5})u(p,s)p_{3\alpha}(-g^{\mu\alpha} + \frac{k^\mu k^\alpha}{m_{k}^{2}})\mathcal{P}^3\mathcal{F},
    \end{aligned}
\end{equation}

\begin{equation}
	\begin{aligned}
\mathcal{M}[D^*,B_{\bar{3}}/B_6;D]&=\displaystyle\int\frac{d^4p}{(2\pi)^4}(+1)\frac{f_{PB\Lambda}}{m_k}g_{D^{*}DP}\bar{u^{\sigma}}(p_4, s_4)\gamma_5k_\sigma(\slashed{p_{2}} + m_{2})
(A_{1}\gamma_\nu \gamma_5+A_{2}\frac{p_{2\nu}}{M}\gamma_5+B_{1}\gamma_\nu +B_{2}\frac{p_{2\nu} }{M})u(p,s)\\
&(-g^{\mu\nu} + \frac{p_1^\mu p_1^\nu}{m_{1}^{2}})
p_{3\mu} \mathcal{P}^3\mathcal{F},
    \end{aligned}
\end{equation}

\begin{equation}
	\begin{aligned}
\mathcal{M}[D^*,B_{\bar{3}}/B_6;D^*]&=\displaystyle\int\frac{d^4p}{(2\pi)^4}(-1)\frac{g_{VB\Lambda}}{m_k^2}\frac{1}{2}g_{D^*D^*P}
\bar{u^{\sigma}}(p_4, s_4)\sigma_{\mu\nu} k^\nu k_\sigma(\slashed{p_{2}} + m_{2})
(A_{1}\gamma_\rho    \gamma_5+A_{2}\frac{p_{2\rho} }{M}\gamma_5+B_{1}\gamma_\rho +B_{2}\frac{p_{2\rho}  }{M})\\
&u(p,s)\epsilon_{\alpha \beta \gamma \delta}p_3^\beta p_1^\gamma(-g^{\alpha\mu} + \frac{k^\alpha k^\mu}{m_{k}^{2}})(-g^{\rho\delta} + \frac{p_1^\rho p_1^\delta}{m_{1}^{2}})
\mathcal{P}^3\mathcal{F},
    \end{aligned}
\end{equation}

\begin{equation}
	\begin{aligned}
\mathcal{M}[P,B_8;B_8]&=\displaystyle\int\frac{d^4p}{(2\pi)^4}(-1)\frac{f_{PB\Lambda}}{m_1}g_{PBB}
\bar{u^{\sigma}}(p_4, s_4)\gamma_5 p_{1\sigma}(\slashed{k} + m_{k})
\gamma_5(\slashed{p_{2}} + m_{2})(A + B\gamma_{5})u(p,s)\mathcal{P}^3\mathcal{F},
    \end{aligned}
\end{equation}

\begin{equation}
	\begin{aligned}
\mathcal{M}[V,B_8;B_8]&=\displaystyle\int\frac{d^4p}{(2\pi)^4}(-i)\frac{g_{VB\Lambda}}{m_1^2}g_{PBB}
\bar{u^{\sigma}}(p_4, s_4)\sigma_{\mu\nu} p_1^\nu p_{1\sigma}(\slashed{k} + m_{k})
\gamma_5(\slashed{p_{2}} + m_{2})(-g^{\mu\rho} + \frac{p_1^\mu p_1^\rho}{m_{1}^{2}})\\
&(A_{1}\gamma_\rho\gamma_5+A_{2}\frac{p_{2\rho} }{M}\gamma_5+B_{1}\gamma_{\rho} +B_{2}\frac{p_{2\rho} }{M})u(p,s)\mathcal{P}^3\mathcal{F}.
    \end{aligned}
\end{equation}

The amplitudes of $\Lambda^{0}_{b} \rightarrow B_{8} V $:

\begin{equation}
	\begin{aligned}
\mathcal{M}[P,B_8;P]&=\displaystyle\int\frac{d^4p}{(2\pi)^4}(-i)\frac{f_{PB\Lambda}}{m_k}\frac{g_{PPV}}{\sqrt{2}}
\bar{u^{\sigma}}(p_4, s_4)\gamma_5 k_\sigma(\slashed{p_{2}} + m_{2})(A + B\gamma_{5})u(p,s)
\varepsilon^{*\mu}(p_3)(p_1+k)_{\mu}\mathcal{P}^3\mathcal{F},
    \end{aligned}
\end{equation}

\begin{equation}
	\begin{aligned}
\mathcal{M}[P,B_8;V]&=\displaystyle\int\frac{d^4p}{(2\pi)^4}(+i)\frac{g_{VB\Lambda}}{m_k^2}\frac{4g_{PVV}}{f_1}
\bar{u^{\sigma}}(p_4, s_4)\sigma _{\mu\nu}  k^{\nu} k_\sigma (\slashed{p_{2}} + m_{2})
(A + B\gamma_{5})u(p,s)
\epsilon_{\alpha\beta\gamma\delta}k^{\alpha}p_3^{\gamma}\varepsilon^{*\delta}(p_3)\\
&(-g^{\mu\beta} + \frac{k^\mu k^\beta}{m_{k}^{2}})\mathcal{P}^3\mathcal{F},
    \end{aligned}
\end{equation}

\begin{equation}
	\begin{aligned}
\mathcal{M}[V,B_8;P]&=\displaystyle\int\frac{d^4p}{(2\pi)^4}(+1)\frac{f_{PB\Lambda}}{m_{k}}\frac{4g_{PVV}}{f_k}
\bar{u^{\sigma}}(p_4, s_4)\gamma_5 k_\sigma(\slashed{p_{2}} + m_{2})
(A_{1}\gamma^\mu\gamma_5+A_{2}\frac{p_{2}^\mu}{M}\gamma_5+B_{1}\gamma^\mu +B_{2}\frac{p_{2}^\mu}{M})u(p,s)\\
&\epsilon^{\alpha\beta\gamma\delta}p_{1\alpha}p_{3\gamma}\varepsilon^{*}_{\delta}(p_3)
(-g_{\mu\beta} + \frac{p_{1\mu }p_{1_\beta}}{m_{1}^{2}})\mathcal{P}^3\mathcal{F},
    \end{aligned}
\end{equation}

\begin{equation}
	\begin{aligned}
\mathcal{M}[V,B_8;V]&=\displaystyle\int\frac{d^4p}{(2\pi)^4}(+1)\frac{g_{VB\Lambda}}{m_k^2}\frac{g_{VVV}}{\sqrt{2}}
\bar{u^{\sigma}}(p_4, s_4)\sigma_{\alpha\beta}k^\beta k_\sigma(\slashed{p_{2}} + m_{2})
(A_{1}\gamma^\rho\gamma_5+A_{2}\frac{p_{2}^\rho}{M}\gamma_5+B_{1}\gamma^\rho +B_{2}\frac{p_{2}^\rho}{M})u(p,s)\\
&[(-g_{\mu\rho} + \frac{p_{1\mu} p_{1\rho}}{m_{1}^{2}})\varepsilon^{*\mu}(p_3)(-g^{\alpha\nu} + \frac{k^\alpha k^\nu}{m_{k}^{2}})(p_3+p_1)_\nu
+(-g^{\alpha\mu} + \frac{k^\alpha k^\mu}{m_{k}^{2}})(-g_{\mu\rho} + \frac{p_{1\mu} p_{1\rho}}{m_{1}^{2}})\varepsilon_\nu^*(p_3)(-p_1-k)^\nu\\
&+\varepsilon^*_\mu(p_3)(-g^{\alpha\mu} + \frac{k^\alpha k^\mu}{m_{k}^{2}})(-g_{\nu\rho} + \frac{p_{1\nu} p_{1\rho}}{m_{1}^{2}})(k-p_3)^\nu]\mathcal{P}^3\mathcal{F},
    \end{aligned}
\end{equation}

\begin{equation}
	\begin{aligned}
\mathcal{M}[D,B_{\bar{3}}/B_6;D]&=\displaystyle\int\frac{d^4p}{(2\pi)^4}(-i)\frac{f_{PB\Lambda}}{m_k}g_{DDV}
\bar{u^{\sigma}}(p_4, s_4)\gamma_5 k_\sigma(\slashed{p_{2}} + m_{2})(A + B\gamma_{5})u(p,s)
(p_1+k)_\mu\varepsilon^{*\mu}(p_3)\mathcal{P}^3\mathcal{F},
    \end{aligned}
\end{equation}

\begin{equation}
	\begin{aligned}
\mathcal{M}[D,B_{\bar{3}}/B_6;D^*]&=\displaystyle\int\frac{d^4p}{(2\pi)^4}(-i)\frac{g_{VB\Lambda}}{m_k^2}2f_{DD^*V}
\bar{u^{\sigma}}(p_4, s_4)\sigma_{\mu\nu}k^\nu k_\sigma(\slashed{p_{2}} + m_{2})(A + B\gamma_{5})u(p,s)
\epsilon_{\alpha\beta\gamma\delta}p_3^\alpha\varepsilon^{*\beta}(p_3)(p_1+k)^\gamma\\
&(-g^{\mu\delta} + \frac{k^\mu k^\delta}{m_{k}^{2}})\mathcal{P}^3\mathcal{F},
    \end{aligned}
\end{equation}

\begin{equation}
	\begin{aligned}
\mathcal{M}[D^*,B_{\bar{3}}/B_6;D]&=\displaystyle\int\frac{d^4p}{(2\pi)^4}(-1)\frac{f_{PB\Lambda}}{m_k}2f_{DD^*V}
\bar{u^{\sigma}}(p_4, s_4)\gamma_5 k_\sigma(\slashed{p_{2}} + m_{2})
(A_{1}\gamma_\rho\gamma_5+A_{2}\frac{p_{2\rho} }{M}\gamma_5+B_{1}\gamma_{\rho} +B_{2}\frac{p_{2\rho} }{M})u(p,s)\\
&\epsilon_{\alpha\beta\gamma\delta}p_3^\alpha\varepsilon^{*\beta}(p_3)(p_1+k)^\gamma(-g^{\rho\delta} + \frac{p_1^\rho p_1^\delta}{m_{1}^{2}})\mathcal{P}^3\mathcal{F},
    \end{aligned}
\end{equation}

\begin{equation}
	\begin{aligned}
\mathcal{M}[D^*,B_{\bar{3}}/B_6;D^*]&=\displaystyle\int\frac{d^4p}{(2\pi)^4}(+1)\frac{g_{VB\Lambda}}{m_k^2}
\bar{u^{\sigma}}(p_4, s_4)\sigma^{\mu\nu} k_\nu k_\sigma(\slashed{p_{2}} + m_{2})
(A_{1}\gamma_\rho\gamma_5+A_{2}\frac{p_{2\rho}    }{M}\gamma_5+B_{1}\gamma_\rho   +B_{2}\frac{p_{2\rho}     }{M})u(p,s)\\
&[-g_{D^*D^*V}(-g^{\rho\alpha} + \frac{p_1^\rho p_1^\alpha}{m_{1}^{2}})(-g_{\mu\alpha} + \frac{k_\mu k_\alpha}{m_{k}^{2}})(k+p_1)_\beta \varepsilon^{*\beta}(p_3)       
-4f_{D^*D^*V}(-g^{\rho\gamma} + \frac{p_1^\rho p_1^\gamma}{m_{1}^{2}})(-g_{\mu\delta} + \frac{k_\mu k_\delta}{m_{k}^{2}})\\
&(p_{3\gamma}\varepsilon^{*\delta}(p_3)-p_3^{\delta}\varepsilon^{*}_{\gamma}(p_3))]
\mathcal{P}^3\mathcal{F},
    \end{aligned}
\end{equation}

\begin{equation}
	\begin{aligned}
\mathcal{M}[P,B_8;B_8]&=\displaystyle\int\frac{d^4p}{(2\pi)^4}(i)\frac{f_{PB\Lambda}}{m_1}
\bar{u^{\sigma}}(p_4, s_4)\gamma_5 p_{1\sigma}(\slashed{k} + m_{k})
[f_{1BBV}\gamma_{\mu}+\frac{if_{2BBV}}{m_k+m_2}\sigma_{\mu\nu}p_3^\mu]
\varepsilon^{*\nu}(p_3)(\slashed{p_{2}} + m_{2})(A + B\gamma_{5})\\
&u(p,s)\mathcal{P}^3\mathcal{F},
    \end{aligned}
\end{equation}

\begin{equation}
	\begin{aligned}
\mathcal{M}[V,B_8;B_8]&=\displaystyle\int\frac{d^4p}{(2\pi)^4}(-1)\frac{g_{VB\Lambda}}{m_1^2}
\bar{u^{\sigma}}(p_4, s_4)\sigma_{\mu\nu}p_1^\nu p_{1\sigma}(\slashed{k} + m_{k})
[f_{1BBV}\gamma_{\alpha}+\frac{if_{2BBV}}{m_k+m_2}\sigma_{\beta\alpha}p_3^\beta]\varepsilon^{*\alpha}(p_3)
(\slashed{p_{2}} + m_{2})\\
&(A_{1}\gamma_\gamma \gamma_5+A_{2}\frac{p_{2\gamma}    }{M}\gamma_5+B_{1}\gamma_\gamma   +B_{2}\frac{p_{2\gamma}     }{M})u(p,s)(-g^{\mu\gamma} + \frac{p_1^\mu p_1^\gamma}{m_{1}^{2}})
\mathcal{P}^3\mathcal{F}.
    \end{aligned}
\end{equation}

The amplitudes of $\Lambda^{0}_{b} \rightarrow B_{8} S $:

\begin{equation}
	\begin{aligned}
\mathcal{M}[P,B_8;P]&=\displaystyle\int\frac{d^4p}{(2\pi)^4}(-i)\frac{f_{PB\Lambda}}{m_k}\gamma_{SPP}
\bar{u^{\sigma}}(p_4, s_4)\gamma_5 k_\sigma(\slashed{p_{2}} + m_{2})(A + B\gamma_{5})u(p,s)
p_{1\mu}k^\mu \mathcal{P}^3\mathcal{F},
    \end{aligned}
\end{equation}

\begin{equation}
	\begin{aligned}
\mathcal{M}[P,B_8;B_8]&=\displaystyle\int\frac{d^4p}{(2\pi)^4}(-i)\frac{f_{PB\Lambda}}{m_1}g_{BBS}
\bar{u^{\sigma}}(p_4, s_4)\gamma_5 p_{1\sigma}(\slashed{k} + m_{k})(\slashed{p_{2}} + m_{2})(A + B\gamma_{5})u(p,s)\mathcal{P}^3\mathcal{F},
    \end{aligned}
\end{equation}

\begin{equation}
	\begin{aligned}
\mathcal{M}[V,B_8;B_8]&=\displaystyle\int\frac{d^4p}{(2\pi)^4}(+1)\frac{g_{VB\Lambda}}{m_1^2}g_{BBS}
\bar{u^{\sigma}}(p_4, s_4)\sigma_{\mu\nu} p_1^\nu p_{1\sigma}(\slashed{k} + m_{k})(\slashed{p_{2}} + m_{2})
(A_{1}\gamma_\alpha\gamma_5+A_{2}\frac{p_{2\alpha} }{M}\gamma_5+B_{1}\gamma_\alpha  +B_{2}\frac{p_{2\alpha} }{M})\\
&u(p,s)
(-g^{\mu\alpha} + \frac{p_1^\mu p_1^\alpha}{m_{1}^{2}})\mathcal{P}^3\mathcal{F}.
    \end{aligned}
\end{equation}
In the above complete derivation, the spinor summation formula is:
         \begin{equation}
 \begin{aligned}
\sum_{s}u(p,s)\bar{u}(p,s) & =\slashed{p}+m\,, 
    \end{aligned}
\end{equation}
the polarization summation for massive vector meson is:
        \begin{equation}
 \begin{aligned}
\sum_{\lambda_{1}}\varepsilon^{*\rho}(p_{1},\lambda_{1})\varepsilon^{\nu}(p_{1},\lambda_{1})&=-g^{\rho\nu}+\frac{p_1^{\rho}p_1^{\nu}}{m^{2}_{1}},
    \end{aligned}
    \end{equation}
and for the convenience of calculation, the Rarita-Schwinger spinors in the amplitude are all selected in the fixed Z-axis direction. Therefore, $u^{\sigma}(p_4, s_4)$ with four different spins are expressed as
\begin{equation}
\begin{aligned}
u^{\sigma}\left(p_{4}, \pm 3 / 2\right)= & \varepsilon^{\sigma}\left(p_{z}, \pm 1\right) u\left(p_{z}, \pm 1 / 2\right), \\
u^{\sigma}\left(p_{4}, \pm 1 / 2\right)= & \frac{1}{\sqrt{3}} \varepsilon^{\sigma}\left(p_{z}, \pm 1\right) u\left(p_{z}, \mp 1 / 2\right)  +\sqrt{\frac{2}{3}} \varepsilon^{\sigma}\left(p_{z}, 0\right) u\left(p_{z}, \pm 1 / 2\right).
\end{aligned}
\end{equation}

\section{$\lambda$ functions for different decay modes}\label{app.E}

In $\Lambda^0_b\rightarrow \Lambda(1520)+\pi^0,$
\begin{equation}
\begin{aligned}
\lambda=\frac{G_F}{\sqrt{2}}f_{\pi^0}^d\left[V_{ub}V_{us}^*(-a_2)-V_{tb}V_{ts}^*\left(\frac{3}{2}a_7-\frac{3}{2}a_9\right)\right],
\end{aligned}
\end{equation}
with $f^d_{\pi^0}=\frac{f_{\pi^0}}{\sqrt{2}}$.
\\
\\
In $\Lambda^0_b\rightarrow \Lambda(1520)+\phi,$
\begin{equation}
\begin{aligned}
\lambda=\frac{G_{F}}{\sqrt{2}}f_{\phi  }\left[-V_{tb}V_{ts}^{*}\left(a_3+a_4+a_5-\frac{1}{2}a_{7}-\frac{1}{2}a_{9}-\frac{1}{2}a_{10}\right)\right].
\end{aligned}
\end{equation}
\\
In $\Lambda^0_b\rightarrow \Lambda(1520)+f_0(980),$
\begin{equation}
\begin{aligned}
\lambda=\frac{G_{F}}{\sqrt{2}}f^{s\bar{s}}_{f_0(980)}R_{f_0(980)}\left[-V_{tb}V_{ts}^{*}\left( a_{6}-\frac{1}{2}a_8\right) \right],
\end{aligned}
\end{equation}
with $R_{f_0(980)}=\frac{q^\mu}{m_b}$.
\\
\\
In $\Lambda^0_b\rightarrow \Lambda(1520)+\rho^0,$
\begin{equation}
\begin{aligned}
\lambda=\frac{G_F}{\sqrt{2}}f_{\rho^0}^d\left[V_{ub}V_{us}^*(-a_2)-V_{tb}V_{ts}^*\left(-\frac{3}{2}a_7-\frac{3}{2}a_9\right)\right],
\end{aligned}
\end{equation}
with $f^d_{\rho^0}=\frac{f_{\rho^0}}{\sqrt{2}}$.
\\
\\
In $\Lambda^0_b\rightarrow \Lambda(1520)+f_0(500),$
\begin{equation}
\begin{aligned}
\lambda=\frac{G_{F}}{\sqrt{2}}f^{s\bar{s}}_{f_0(500)}R_{f_0(500)}\left[-V_{tb}V_{ts}^{*}\left( a_{6}-\frac{1}{2}a_8\right) \right],
\end{aligned}
\end{equation}
with $R_{f_0(500)}=\frac{q^\mu}{m_b}$.
\\
\\
In $\Lambda^0_b\rightarrow \Lambda(1520)+K^{*0},$
\begin{equation}
\begin{aligned}
\frac{G_{F}}{\sqrt{2}}f_{K^*}\left[-V_{tb}V_{td}^{*}\left(a_{4}-\frac12a_{10}\right)\right].
\end{aligned}
\end{equation}
In $\Lambda^0_b\rightarrow \Lambda(1520)+\kappa(700),$
\begin{equation}
\begin{aligned}
\lambda=\frac{G_{F}}{\sqrt{2}}f^{s\bar{s}}_{\kappa(700)}R_{\kappa(700)}\left[-V_{tb}V_{ts}^{*}\left( a_{6}-\frac{1}{2}a_8\right) \right],
\end{aligned}
\end{equation}
with $R_{\kappa(700)}=\frac{q^\mu}{m_b}$.
\\
\\
In $\Xi^-_b\rightarrow \Lambda(1520)+K^-,$\\
A term:
\begin{equation}
\begin{aligned}
 \frac{G_{F}}{\sqrt{2}}f_{K}[V_{ub}V_{us}^{*}a_{1}-V_{tb}V_{ts}^{*}(a_{4}+a_{10})+ R_{K^-}(-V_{tb}V_{ts}^{*}(a_{6}+a_{8}))],
\end{aligned}
\end{equation}
B term:
\begin{equation}
\begin{aligned}
 \frac{G_{F}}{\sqrt{2}}f_{K}[V_{ub}V_{us}^{*}a_{1}-V_{tb}V_{ts}^{*}(a_{4}+a_{10})- R_{K^-}(-V_{tb}V_{ts}^{*}(a_{6}+a_{8}))],
\end{aligned}
\end{equation}
with $R_{K^{-}}=\frac{2 m_{K^{-}}^{2}}{\left(m_{u}+m_{s}\right) m_{b}} .$
\\
The numerical values of the Wilson coefficients are taken from Ref.~\cite{Lu:2009cm} and are collected in Table~\ref{Wilson coef}.

\begin{table}[H]
\caption{Numerical values of the effective Wilson coefficients at $\mu=m_b$, where $m_b$ is taken as 4.8 GeV.}
\centering
\resizebox{\textwidth}{!}
{
\begin{tabular}{cccccccccc}
\toprule
\toprule
$a_1$&$a_2(\times 10^{-2})$&$a_3(\times 10^{-3})$&$a_4(\times 10^{-3})$&$a_5(\times 10^{-3}$)&$a_6(\times 10^{-3})$&$a_7(\times 10^{-4})$&$a_8(\times 10^{-4})$&$a_9(\times 10^{-4})$&$a_{10}(\times 10^{-4})$\\
\midrule
1.03&10.3&3.60&-22.8&-2.29&-29.8&12.2&7.57&-82.2&-8.20\\
\midrule
\midrule
\end{tabular}\label{Wilson coef}
}
\end{table}
\section{SU(4) approach in this work}\label{app.F}
Under SU(4) symmetry, pseudoscalar or vector mesons form a 15-plet. Their matrix representations are structured as follows:
\begin{equation}
\begin{gathered}
P=
\begin{pmatrix}
\frac{\pi^{0}}{\sqrt{2}}+\frac{\eta}{\sqrt{6}}+\frac{\eta_{c}}{\sqrt{12}} & \pi^{+} & K^{+} & \bar{D}^{0} \\
\pi^{-} & -\frac{\pi^{0}}{\sqrt{2}}+\frac{\eta}{\sqrt{6}}+\frac{\eta_{c}}{\sqrt{12}} & K^{0} & D^{-} \\
K^{-} & \bar{K}^{0} & -\sqrt{\frac{2}{3}}\eta+\frac{\eta_{c}}{\sqrt{12}} & D_{s}^{-} \\
D^{0} & D^{+} & D_{s}^{+} & \frac{-3\eta_{c}}{\sqrt{12}}
\end{pmatrix}, \\
\\
V=
\begin{pmatrix}
\frac{\rho_{0}}{\sqrt{2}}+\frac{\omega^{\prime}}{\sqrt{6}}+\frac{J/\psi}{\sqrt{12}} & \rho^{+} & K^{++} & \bar{D}^{*0} \\
\rho^{-} & -\frac{\rho_{0}}{\sqrt{2}}+\frac{\omega^{\prime}}{\sqrt{6}}+\frac{J/\psi}{\sqrt{12}} & K^{*0} & D^{*-} \\
K^{*-} & \bar{K}^{*0} & -\sqrt{\frac{2}{3}}\omega^{\prime}+\frac{J/\psi}{\sqrt{12}} & D_{s}^{*-} \\
D^{*0} & D^{*+} & D_{s}^{*+} & -\frac{3J/\psi}{\sqrt{12}}
\end{pmatrix}.
\end{gathered}
\end{equation}
\\
Analogous to the baryon octet under SU(3) symmetry, the $J^P=\frac12^+$ baryons‌ form a ‌20-plet‌ under SU(4) symmetry, it can be denoted as:
\begin{equation}
\begin{aligned}
p & =\phi_{112}, \quad n=\phi_{221}, \quad \Lambda=\sqrt{\frac{2}{3}}\left(\phi_{321}-\phi_{312}\right), \\
\Sigma^{+} & =\phi_{113}, \quad \Sigma^{0}=\sqrt{2} \phi_{123}, \quad \Sigma^{-}=\phi_{223}, \\
\Xi^{0} & =\phi_{331}, \quad \Xi^{-}=\phi_{332}, \\
\Sigma_{c}^{++} & =\phi_{114}, \quad \Sigma_{c}^{+}=\sqrt{2} \phi_{124}, \quad \Sigma_{c}^{0}=\phi_{224}, \\
\Xi_{c}^{+} & =\sqrt{2} \phi_{134}, \quad \Xi_{c}^{0}=\sqrt{2} \phi_{234}, \\
\Xi_{c}^{+\prime} & =\sqrt{\frac{2}{3}}\left(\phi_{413}-\phi_{431}\right), \quad \Xi_{c}^{0 \prime}=\sqrt{\frac{2}{3}}\left(\phi_{423}-\phi_{432}\right), \\
\Lambda_{c}^{+} & =\sqrt{\frac{2}{3}}\left(\phi_{421}-\phi_{412}\right), \quad \Omega_{c}^{0}=\phi_{334}, \\
\Xi_{c c}^{++} & =\phi_{441}, \quad \Xi_{c c}^{+}=\phi_{442}, \quad \Omega_{c c}^{+}=\phi_{443} ,
\end{aligned}
\end{equation}
\\
where $\phi_{ijk}$ satisfy the following conditions:
\begin{equation}
\begin{aligned}
 & \phi_{ijk}+\phi_{jki}+\phi_{kij}=0, \\
 & \phi_{ijk}=\phi_{jik}.
\end{aligned}
\end{equation}
\\
The SU(4) invariant effictive Lagrangian could be written as:
\begin{equation}
\begin{aligned}
\mathcal{L}_{B_8B_8P}=g_{B_8B_8P}(a\phi^{*ijk}\gamma_{5}P_{i}^{m}\phi_{mjk}+b\phi^{*ijk}\gamma_{5}P_{i}^{m}\phi_{mkj}),\\
\mathcal{L}_{B_8B_8V}=g_{B_8B_8V}(c\phi^{*ijk}\gamma^{\mu}V_{i\mu}^{m}\phi_{mjk}+d\phi^{*ijk}\gamma^{\mu}V_{i\mu}^{m}\phi_{mkj}).
\end{aligned}
\end{equation}
\\
‌All parameters such as $g_{B_8B_8P},a,d$ in ‌the Lagrangian above can be derived by‌ comparing with the $SU(3)$ relations~\cite{Liu:2001yx,Liu:2001ce,Ronchen:2012eg}.We can implement this symmetry into the effective Lagrangian of $\Lambda(1520)$ ‌by modifying only the Lorentz structures‌ while ‌retaining the relations among the $m,i,j,k$ indices.

\bibliographystyle{unsrt}
\bibliography{References}

@article{Sakharov:1967dj,
    author = "Sakharov, A. D.",
    title = "{Violation of CP Invariance, C asymmetry, and baryon asymmetry of the universe}",
    doi = "10.1070/PU1991v034n05ABEH002497",
    journal = "Pisma Zh. Eksp. Teor. Fiz.",
    volume = "5",
    pages = "32--35",
    year = "1967"
}

@article{Cabibbo:1963yz,
    author = "Cabibbo, Nicola",
    title = "{Unitary Symmetry and Leptonic Decays}",
    doi = "10.1103/PhysRevLett.10.531",
    journal = "Phys. Rev. Lett.",
    volume = "10",
    pages = "531--533",
    year = "1963"
}

@article{Kobayashi:1973fv,
    author = "Kobayashi, Makoto and Maskawa, Toshihide",
    title = "{CP Violation in the Renormalizable Theory of Weak Interaction}",
    reportNumber = "KUNS-242",
    doi = "10.1143/PTP.49.652",
    journal = "Prog. Theor. Phys.",
    volume = "49",
    pages = "652--657",
    year = "1973"
}

@article{Muller:1960ph,
    author = "Muller, F. and Birge, R. W. and Fowler, W. B. and Good, R. H. and Hirsch, W. and Matsen, R. P. and Oswald, L. and Powell, W. M. and White, H. S. and Piccioni, O.",
    title = "{Regeneration and Mass Difference of Neutral $K$ Mesons}",
    doi = "10.1103/PhysRevLett.4.418",
    journal = "Phys. Rev. Lett.",
    volume = "4",
    pages = "418--421",
    year = "1960"
}

@article{Christenson:1964fg,
    author = "Christenson, J. H. and Cronin, J. W. and Fitch, V. L. and Turlay, R.",
    title = "{Evidence for the $2\pi$ Decay of the $K_2^0$ Meson}",
    doi = "10.1103/PhysRevLett.13.138",
    journal = "Phys. Rev. Lett.",
    volume = "13",
    pages = "138--140",
    year = "1964"
}

@article{KTeV:1999kad,
    author = "Alavi-Harati, A. and others",
    collaboration = "KTeV",
    title = "{Observation of Direct CP Violation in $K_{S,L} \to \pi \pi$ Decays}",
    eprint = "hep-ex/9905060",
    archivePrefix = "arXiv",
    reportNumber = "EFI-99-25, FERMILAB-PUB-99-150-E",
    doi = "10.1103/PhysRevLett.83.22",
    journal = "Phys. Rev. Lett.",
    volume = "83",
    pages = "22--27",
    year = "1999"
}

@article{BaBar:2001ags,
    author = "Aubert, Bernard and others",
    collaboration = "BaBar",
    title = "{Measurement of CP violating asymmetries in $B^0$ decays to CP eigenstates}",
    eprint = "hep-ex/0102030",
    archivePrefix = "arXiv",
    reportNumber = "SLAC-PUB-8777, BABAR-PUB-01-01, BABAR-PUB-01-001",
    doi = "10.1103/PhysRevLett.86.2515",
    journal = "Phys. Rev. Lett.",
    volume = "86",
    pages = "2515--2522",
    year = "2001"
}

@article{Belle:2001zzw,
    author = "Abe, Kazuo and others",
    collaboration = "Belle",
    title = "{Observation of large CP violation in the neutral $B$ meson system}",
    eprint = "hep-ex/0107061",
    archivePrefix = "arXiv",
    reportNumber = "KEK-PREPRINT-2001-50, BELLE-PREPRINT-2001-10",
    doi = "10.1103/PhysRevLett.87.091802",
    journal = "Phys. Rev. Lett.",
    volume = "87",
    pages = "091802",
    year = "2001"
}

@article{BaBar:2004gyj,
    author = "Aubert, Bernard and others",
    collaboration = "BaBar",
    title = "{Observation of direct CP violation in $B^0 \to K^+ \pi^-$ decays}",
    eprint = "hep-ex/0407057",
    archivePrefix = "arXiv",
    reportNumber = "SLAC-PUB-10582, BABAR-PUB-04-037",
    doi = "10.1103/PhysRevLett.93.131801",
    journal = "Phys. Rev. Lett.",
    volume = "93",
    pages = "131801",
    year = "2004"
}

@article{Belle:2004nch,
    author = "Chao, Y. and others",
    collaboration = "Belle",
    title = "{Evidence for direct CP violation in B0 ---{\ensuremath{>}} K+ pi- decays}",
    eprint = "hep-ex/0408100",
    archivePrefix = "arXiv",
    doi = "10.1103/PhysRevLett.93.191802",
    journal = "Phys. Rev. Lett.",
    volume = "93",
    pages = "191802",
    year = "2004"
}

@article{LHCb:2013syl,
    author = "Aaij, R and others",
    collaboration = "LHCb",
    title = "{First observation of $CP$ violation in the decays of $B^0_s$ mesons}",
    eprint = "1304.6173",
    archivePrefix = "arXiv",
    primaryClass = "hep-ex",
    reportNumber = "CERN-PH-EP-2013-068, LHCB-PAPER-2013-018",
    doi = "10.1103/PhysRevLett.110.221601",
    journal = "Phys. Rev. Lett.",
    volume = "110",
    number = "22",
    pages = "221601",
    year = "2013"
}

@article{LHCb:2019hro,
    author = "Aaij, Roel and others",
    collaboration = "LHCb",
    title = "{Observation of CP Violation in Charm Decays}",
    eprint = "1903.08726",
    archivePrefix = "arXiv",
    primaryClass = "hep-ex",
    reportNumber = "LHCb-PAPER-2019-006, CERN-EP-2019-042",
    doi = "10.1103/PhysRevLett.122.211803",
    journal = "Phys. Rev. Lett.",
    volume = "122",
    number = "21",
    pages = "211803",
    year = "2019"
}

@article{Duan:2024zjv,
    author = "Duan, Zhu-Ding and Wang, Jian-Peng and Li, Run-Hui and Lv, Cai-Dian and Yu, Fu-Sheng",
    title = "{Final-state rescattering mechanism of bottom-baryon decays}",
    eprint = "2412.20458",
    archivePrefix = "arXiv",
    primaryClass = "hep-ph",
    month = "12",
    year = "2024"
}

@article{Buchalla:1995vs,
    author = "Buchalla, Gerhard and Buras, Andrzej J. and Lautenbacher, Markus E.",
    title = "{Weak decays beyond leading logarithms}",
    eprint = "hep-ph/9512380",
    archivePrefix = "arXiv",
    reportNumber = "SLAC-PUB-7009, SLAC-PUB-95-7009, MPI-PH-95-104, TUM-T31-100-95, FERMILAB-PUB-95-305-T",
    doi = "10.1103/RevModPhys.68.1125",
    journal = "Rev. Mod. Phys.",
    volume = "68",
    pages = "1125--1144",
    year = "1996"
}

@article{Cheng:2004ru,
    author = "Cheng, Hai-Yang and Chua, Chun-Khiang and Soni, Amarjit",
    title = "{Final state interactions in hadronic B decays}",
    eprint = "hep-ph/0409317",
    archivePrefix = "arXiv",
    reportNumber = "BNL-HET-04-17",
    doi = "10.1103/PhysRevD.71.014030",
    journal = "Phys. Rev. D",
    volume = "71",
    pages = "014030",
    year = "2005",
    note = {[arXiv:hep-ph/0409317 [hep-ph]]}
}

@article{Cheng:2020ipp,
    author = "Cheng, Hai-Yang and Chua, Chun-Khiang",
    title = "{Branching fractions and $CP$ violation in $B^-\to K^+K^-\pi^-$ and $B^-\to \pi^+\pi^-\pi^-$ decays}",
    eprint = "2007.02558",
    archivePrefix = "arXiv",
    primaryClass = "hep-ph",
    doi = "10.1103/PhysRevD.102.053006",
    journal = "Phys. Rev. D",
    volume = "102",
    number = "5",
    pages = "053006",
    year = "2020"
}

@article{Leibovich:2003tw,
    author = "Leibovich, Adam K. and Ligeti, Zoltan and Stewart, Iain W. and Wise, Mark B.",
    title = "{Predictions for nonleptonic Lambda(b) and Theta(b) decays}",
    eprint = "hep-ph/0312319",
    archivePrefix = "arXiv",
    reportNumber = "LBNL-54153, MIT-CTP-3453, CALT-68-2467",
    doi = "10.1016/j.physletb.2004.02.033",
    journal = "Phys. Lett. B",
    volume = "586",
    pages = "337--344",
    year = "2004",
    note = {[arXiv:hep-ph/0312319 [hep-ph]]}
}

@article{Beneke:2001ev,
    author = "Beneke, M. and Buchalla, G. and Neubert, M. and Sachrajda, Christopher T.",
    title = "{QCD factorization in B ---\ensuremath{>} pi K, pi pi decays and extraction of Wolfenstein parameters}",
    eprint = "hep-ph/0104110",
    archivePrefix = "arXiv",
    reportNumber = "CERN-TH-2001-107, CLNS-01-1728, PITHA-01-01, SHEP-01-11",
    doi = "10.1016/S0550-3213(01)00251-6",
    journal = "Nucl. Phys. B",
    volume = "606",
    pages = "245--321",
    year = "2001",
    note = {[arXiv:hep-ph/0104110 [hep-ph]]}
}

@article{Wolfenstein:1990ks,
    author = "Wolfenstein, Lincoln",
    title = "{Final state interactions and CP violation in weak decays}",
    reportNumber = "CMU-HEP90-14",
    doi = "10.1103/PhysRevD.43.151",
    journal = "Phys. Rev. D",
    volume = "43",
    pages = "151--156",
    year = "1991"
}

@article{Suzuki:1999uc,
    author = "Suzuki, Mahiko and Wolfenstein, Lincoln",
    title = "{Final state interaction phase in B decays}",
    eprint = "hep-ph/9903477",
    archivePrefix = "arXiv",
    reportNumber = "LBNL-43001, UCB-PTH-99-10, LBL-43001",
    doi = "10.1103/PhysRevD.60.074019",
    journal = "Phys. Rev. D",
    volume = "60",
    pages = "074019",
    year = "1999",
    note = {[arXiv:hep-ph/9903477 [hep-ph]]}
}

@article{Mantry:2003uz,
    author = "Mantry, Sonny and Pirjol, Dan and Stewart, Iain W.",
    title = "{Strong phases and factorization for color suppressed decays}",
    eprint = "hep-ph/0306254",
    archivePrefix = "arXiv",
    reportNumber = "MIT-CTP-3370",
    doi = "10.1103/PhysRevD.68.114009",
    journal = "Phys. Rev. D",
    volume = "68",
    pages = "114009",
    year = "2003",
    note = {[arXiv:hep-ph/0306254 [hep-ph]]}
}

@article{Descotes-Genon:2019dbw,
    author = "Descotes-Genon, S. and Novoa-Brunet, Mart{\'\i}n",
    title = "{Angular analysis of the rare decay $\Lambda_b\to \Lambda(1520)(\to NK)\ell^+\ell^-$}",
    eprint = "1903.00448",
    archivePrefix = "arXiv",
    primaryClass = "hep-ph",
    reportNumber = "LPT-Orsay-19-08",
    doi = "10.1007/JHEP06(2019)136",
    journal = "JHEP",
    volume = "06",
    pages = "136",
    year = "2019",
    note = "[Erratum: JHEP 06, 102 (2020)]"
}

@article{Das:2020cpv,
    author = "Das, Diganta and Das, Jaydeb",
    title = "{The $\Lambda_b\to\Lambda^\ast(1520)(\to N\!\bar{K})\ell^+\ell^-$ decay at low-recoil in HQET}",
    eprint = "2003.08366",
    archivePrefix = "arXiv",
    primaryClass = "hep-ph",
    doi = "10.1007/JHEP07(2020)002",
    journal = "JHEP",
    volume = "07",
    pages = "002",
    year = "2020"
}

@article{Huang:2024oik,
    author = "Huang, Ke-Sheng and Jiang, Hua-Yu and Yu, Fu-Sheng",
    title = "{Transition form factors of the $\Lambda _b \rightarrow \Lambda (1520)$ in QCD light-cone sum rules}",
    eprint = "2412.06515",
    archivePrefix = "arXiv",
    primaryClass = "hep-ph",
    doi = "10.1140/epjc/s10052-025-14033-z",
    journal = "Eur. Phys. J. C",
    volume = "85",
    number = "3",
    pages = "351",
    year = "2025"
}

@article{Pakvasa:1990if,
    author = "Pakvasa, S. and Rosen, Simon Peter and Tuan, S. F.",
    title = "{Parity Violation and Flavor Selection Rules in Charmed Baryon Decays}",
    reportNumber = "UH-511-693-90, LA-UR-90-700",
    doi = "10.1103/PhysRevD.42.3746",
    journal = "Phys. Rev. D",
    volume = "42",
    pages = "3746--3754",
    year = "1990"
}

@article{Zhu:2018jet,
    author = "Zhu, Jie and Wei, Zheng-Tao and Ke, Hong-Wei",
    title = "{Semileptonic and nonleptonic weak decays of $\Lambda_b^0$}",
    eprint = "1803.01297",
    archivePrefix = "arXiv",
    primaryClass = "hep-ph",
    doi = "10.1103/PhysRevD.99.054020",
    journal = "Phys. Rev. D",
    volume = "99",
    number = "5",
    pages = "054020",
    year = "2019",
    note = {[arXiv:1803.01297 [hep-ph]]}
}

@article{Faustov:2018ahb,
    author = "Faustov, R. N. and Galkin, V. O.",
    title = "{Relativistic description of the $\Xi_b$ baryon semileptonic decays}",
    eprint = "1810.03388",
    archivePrefix = "arXiv",
    primaryClass = "hep-ph",
    doi = "10.1103/PhysRevD.98.093006",
    journal = "Phys. Rev. D",
    volume = "98",
    number = "9",
    pages = "093006",
    year = "2018"
}

@article{Rui:2025iwa,
    author = "Rui, Zhou and Zou, Zhi-Tian and Li, Ya and Li, Ying",
    title = "{Semileptonic baryon decays {\ensuremath{\Xi}}b{\textrightarrow}{\ensuremath{\Xi}}c{\ensuremath{\ell}}-{\ensuremath{\nu}}{\textasciimacron}{\ensuremath{\ell}} in perturbative QCD}",
    eprint = "2503.23920",
    archivePrefix = "arXiv",
    primaryClass = "hep-ph",
    doi = "10.1103/qncx-7j7b",
    journal = "Phys. Rev. D",
    volume = "111",
    number = "11",
    pages = "113006",
    year = "2025"
}

@article{Jia:2024pyb,
    author = "Jia, Cai-Ping and Jiang, Hua-Yu and Wang, Jian-Peng and Yu, Fu-Sheng",
    title = "{Final-state rescattering mechanism of charmed baryon decays}",
    eprint = "2408.14959",
    archivePrefix = "arXiv",
    primaryClass = "hep-ph",
    doi = "10.1007/JHEP11(2024)072",
    journal = "JHEP",
    volume = "11",
    pages = "072",
    year = "2024"
}

@article{Yue:2024paz,
    author = "Yue, Zi-Li and Guo, Quan-Yun and Chen, Dian-Yong",
    title = "{Strong decays of the \ensuremath{\Lambda}c(2910) and \ensuremath{\Lambda}c(2940) in the ND* molecular frame}",
    eprint = "2402.10594",
    archivePrefix = "arXiv",
    primaryClass = "hep-ph",
    doi = "10.1103/PhysRevD.109.094049",
    journal = "Phys. Rev. D",
    volume = "109",
    number = "9",
    pages = "094049",
    year = "2024",
    note = {[arXiv:2402.10594 [hep-ph]]}
}

@article{ParticleDataGroup:2022pth,
    author = "Workman, R. L. and others",
    collaboration = "Particle Data Group",
    title = "{Review of Particle Physics}",
    doi = "10.1093/ptep/ptac097",
    journal = "PTEP",
    volume = "2022",
    pages = "083C01",
    year = "2022"
}

@article{Qi:2024ddp,
    author = "Qi, Jing-Juan and Zhao, Yu-Jie and Zhang, Zhen-Hua",
    title = "{The forward-backward asymmetry induced $CP$ asymmetry in ${\overline{B}}^{0}\rightarrow K^{-}\pi^{+}\pi^{0}$ in phase space around the resonances ${\overline{K}}^{*}(892)^{0}$ and ${\overline{K}}^{*}_{0}(700)$}",
    eprint = "2410.08539",
    archivePrefix = "arXiv",
    primaryClass = "hep-ph",
    month = "10",
    year = "2024"
}

@article{Liu:2019ymi,
    author = "Liu, Xin and Zou, Zhi-Tian and Li, Ying and Xiao, Zhen-Jun",
    title = "{Phenomenological studies on the $B_{d,s}^0 \to J/\psi f_0(500) [f_0(980)]$ decays}",
    eprint = "1906.02489",
    archivePrefix = "arXiv",
    primaryClass = "hep-ph",
    reportNumber = "JSNU-HEP-2019",
    doi = "10.1103/PhysRevD.100.013006",
    journal = "Phys. Rev. D",
    volume = "100",
    number = "1",
    pages = "013006",
    year = "2019"
}

@article{Aliev:2006xr,
    author = "Aliev, T. M. and Ozpineci, A. and Yakovlev, S. B. and Zamiralov, V.",
    title = "{Meson-octet-baryon couplings using light cone QCD sum rules}",
    doi = "10.1103/PhysRevD.74.116001",
    journal = "Phys. Rev. D",
    volume = "74",
    pages = "116001",
    year = "2006"
}

@article{Aliev:2009ei,
    author = "Aliev, T. M. and Ozpineci, A. and Savci, M. and Zamiralov, V. S.",
    title = "{Vector meson-baryon strong coupling contants in light cone QCD sum rules}",
    eprint = "0905.4664",
    archivePrefix = "arXiv",
    primaryClass = "hep-ph",
    reportNumber = "METU-PHYS-HEP-016-09",
    doi = "10.1103/PhysRevD.80.016010",
    journal = "Phys. Rev. D",
    volume = "80",
    pages = "016010",
    year = "2009",
    note = {[arXiv:0905.4664 [hep-ph]]}
}

@article{Janssen:1996kx,
    author = "Janssen, G. and Holinde, K. and Speth, J.",
    title = "{pi rho correlations in the N N potential}",
    doi = "10.1103/PhysRevC.54.2218",
    journal = "Phys. Rev. C",
    volume = "54",
    pages = "2218--2234",
    year = "1996"
}

@article{Meissner:1987ge,
    author = "Meissner, Ulf G.",
    title = "{Low-Energy Hadron Physics from Effective Chiral Lagrangians with Vector Mesons}",
    reportNumber = "MIT-CTP-1471",
    doi = "10.1016/0370-1573(88)90090-7",
    journal = "Phys. Rept.",
    volume = "161",
    pages = "213",
    year = "1988"
}

@article{Yu:2017zst,
    author = {Yu, Fu-Sheng and Jiang, Hua-Yu and Li, Run-Hui and L\"u, Cai-Dian and Wang, Wei and Zhao, Zhen-Xing},
    title = "{Discovery Potentials of Doubly Charmed Baryons}",
    eprint = "1703.09086",
    archivePrefix = "arXiv",
    primaryClass = "hep-ph",
    doi = "10.1088/1674-1137/42/5/051001",
    journal = "Chin. Phys. C",
    volume = "42",
    number = "5",
    pages = "051001",
    year = "2018",
    note = {[arXiv:1703.09086 [hep-ph]]}
}

@article{Nam:2005uq,
    author = "Nam, Seung-Il and Hosaka, Atsushi and Kim, Hyun-Chul",
    title = "{Lambda(1520,3/2-) photoproduction reaction via gamma N ---{\ensuremath{>}} K Lambda(1520)}",
    eprint = "hep-ph/0503149",
    archivePrefix = "arXiv",
    reportNumber = "PNU-NTG-6-2005",
    doi = "10.1103/PhysRevD.71.114012",
    journal = "Phys. Rev. D",
    volume = "71",
    pages = "114012",
    year = "2005"
}

@article{Riska:2000gd,
    author = "Riska, D. O. and Brown, G. E.",
    title = "{Nucleon resonance transition couplings to vector mesons}",
    eprint = "nucl-th/0005049",
    archivePrefix = "arXiv",
    doi = "10.1016/S0375-9474(00)00362-6",
    journal = "Nucl. Phys. A",
    volume = "679",
    pages = "577--596",
    year = "2001"
}

@article{Ahmed:2020kmp,
    author = "Ahmed, Hiwa A. and Xiao, C. W.",
    title = "{Study the molecular nature of $\sigma$, $f_{0}(980)$, and $a_{0}(980)$ states}",
    eprint = "2001.08141",
    archivePrefix = "arXiv",
    primaryClass = "hep-ph",
    doi = "10.1103/PhysRevD.101.094034",
    journal = "Phys. Rev. D",
    volume = "101",
    number = "9",
    pages = "094034",
    year = "2020"
}

@article{Erkol:2006eq,
    author = "Erkol, G. and Timmermans, R. G. E. and Oka, M. and Rijken, Th. A.",
    title = "{Scalar-meson - baryon coupling constants in QCD sum rules}",
    eprint = "nucl-th/0603058",
    archivePrefix = "arXiv",
    doi = "10.1103/PhysRevC.73.044009",
    journal = "Phys. Rev. C",
    volume = "73",
    pages = "044009",
    year = "2006"
}

@article{LHCb:2025ray,
    author = "Aaij, Roel and others",
    collaboration = "LHCb",
    title = "{Observation of charge{\textendash}parity symmetry breaking in baryon decays}",
    eprint = "2503.16954",
    archivePrefix = "arXiv",
    primaryClass = "hep-ex",
    reportNumber = "LHCb-PAPER-2024-054, CERN-EP-2025-031",
    doi = "10.1038/s41586-025-09119-3",
    journal = "Nature",
    volume = "643",
    number = "8074",
    pages = "1223--1228",
    year = "2025"
}

@article{KLOE:2002kzf,
    author = "Aloisio, A. and others",
    collaboration = "KLOE",
    title = "{Study of the decay $\phi \to \eta \pi^0 \gamma$ with the KLOE detector}",
    eprint = "hep-ex/0204012",
    archivePrefix = "arXiv",
    doi = "10.1016/S0370-2693(02)01821-X",
    journal = "Phys. Lett. B",
    volume = "536",
    pages = "209--216",
    year = "2002"
}

@article{KLOE:2002deh,
    author = "Aloisio, A and others",
    collaboration = "KLOE",
    title = "{Study of the decay $\phi \to \pi^0 \pi^0 \gamma$ with the KLOE detector}",
    eprint = "hep-ex/0204013",
    archivePrefix = "arXiv",
    reportNumber = "LNF-02-003",
    doi = "10.1016/S0370-2693(02)01838-5",
    journal = "Phys. Lett. B",
    volume = "537",
    pages = "21--27",
    year = "2002"
}

@article{Black:1998zc,
    author = "Black, Deirdre and Fariborz, Amir H. and Sannino, Francesco and Schechter, Joseph",
    title = "{Evidence for a scalar kappa(900) resonance in pi K scattering}",
    eprint = "hep-ph/9804273",
    archivePrefix = "arXiv",
    reportNumber = "YCTP-P9-98, SU-4240-678",
    doi = "10.1103/PhysRevD.58.054012",
    journal = "Phys. Rev. D",
    volume = "58",
    pages = "054012",
    year = "1998"
}

@article{Lu:2009cm,
    author = "Lu, Cai-Dian and Wang, Yu-Ming and Zou, Hao and Ali, Ahmed and Kramer, Gustav",
    title = "{Anatomy of the pQCD Approach to the Baryonic Decays Lambda(b) ---{\ensuremath{>}} p pi, p K}",
    eprint = "0906.1479",
    archivePrefix = "arXiv",
    primaryClass = "hep-ph",
    reportNumber = "DESY-09-081",
    doi = "10.1103/PhysRevD.80.034011",
    journal = "Phys. Rev. D",
    volume = "80",
    pages = "034011",
    year = "2009"
}

@inproceedings{Liu:2001yx,
    author = "Liu, W. and Ko, C. M. and Lin, Z. W.",
    title = "{J / psi absorption cross-section by nucleon}",
    eprint = "nucl-th/0107058",
    archivePrefix = "arXiv",
    month = "7",
    year = "2001"
}

@article{Ronchen:2012eg,
    author = "Ronchen, D. and Doring, M. and Huang, F. and Haberzettl, H. and Haidenbauer, J. and Hanhart, C. and Krewald, S. and Meissner, U. -G. and Nakayama, K.",
    title = "{Coupled-channel dynamics in the reactions piN --{\ensuremath{>}} piN, etaN, KLambda, KSigma}",
    eprint = "1211.6998",
    archivePrefix = "arXiv",
    primaryClass = "nucl-th",
    doi = "10.1140/epja/i2013-13044-5",
    journal = "Eur. Phys. J. A",
    volume = "49",
    pages = "44",
    year = "2013"
}

@article{Liu:2001ce,
    author = "Liu, W. and Ko, C. M. and Lin, Z. W.",
    title = "{Cross-section for charmonium absorption by nucleons}",
    doi = "10.1103/PhysRevC.65.015203",
    journal = "Phys. Rev. C",
    volume = "65",
    pages = "015203",
    year = "2002"
}

@article{Ke:2007tg,
    author = "Ke, Hong-Wei and Li, Xue-Qian and Wei, Zheng-Tao",
    title = "{Diquarks and Lambda(b) ---{\ensuremath{>}} Lambda(c) weak decays}",
    eprint = "0710.1927",
    archivePrefix = "arXiv",
    primaryClass = "hep-ph",
    doi = "10.1103/PhysRevD.77.014020",
    journal = "Phys. Rev. D",
    volume = "77",
    pages = "014020",
    year = "2008"
}

@article{Wei:2009np,
    author = "Wei, Zheng-Tao and Ke, Hong-Wei and Li, Xue-Qian",
    title = "{Evaluating decay Rates and Asymmetries of Lambda(b) into Light Baryons in LFQM}",
    eprint = "0909.0100",
    archivePrefix = "arXiv",
    primaryClass = "hep-ph",
    doi = "10.1103/PhysRevD.80.094016",
    journal = "Phys. Rev. D",
    volume = "80",
    pages = "094016",
    year = "2009"
}

@article{Zhu:2016bra,
    author = "Zhu, Jie and Ke, Hong-Wei and Wei, Zheng-Tao",
    title = "{The decay of $\Lambda _b\rightarrow p~K^-$ in QCD factorization approach}",
    eprint = "1603.02800",
    archivePrefix = "arXiv",
    primaryClass = "hep-ph",
    doi = "10.1140/epjc/s10052-016-4134-5",
    journal = "Eur. Phys. J. C",
    volume = "76",
    number = "5",
    pages = "284",
    year = "2016"
}

@article{Feldmann:2011xf,
    author = "Feldmann, Thorsten and Yip, Matthew W. Y.",
    title = "{Form factors for $\Lambda_b \to \Lambda$ transitions in the  soft-collinear effective theory}",
    eprint = "1111.1844",
    archivePrefix = "arXiv",
    primaryClass = "hep-ph",
    reportNumber = "IPPP-11-70, DURHAM-IPPP-11-70, DCPT-11-140",
    doi = "10.1103/PhysRevD.85.014035",
    journal = "Phys. Rev. D",
    volume = "85",
    pages = "014035",
    year = "2012",
    note = "[Erratum: Phys.Rev.D 86, 079901 (2012)]"
}

@article{Khodjamirian:2011jp,
    author = "Khodjamirian, A. and Klein, Ch. and Mannel, Th. and Wang, Y. -M.",
    title = "{Form Factors and Strong Couplings of Heavy Baryons from QCD Light-Cone Sum Rules}",
    eprint = "1108.2971",
    archivePrefix = "arXiv",
    primaryClass = "hep-ph",
    reportNumber = "SI-HEP-2011-05",
    doi = "10.1007/JHEP09(2011)106",
    journal = "JHEP",
    volume = "09",
    pages = "106",
    year = "2011"
}

@article{Azizi:2011mw,
    author = "Azizi, K. and Sarac, Y. and Sundu, H.",
    title = "{Light cone QCD sum rules study of the semileptonic heavy $\Xi_{Q}$ and $\Xi'_{Q}$ transitions to $\Xi$ and $\Sigma $ baryons}",
    eprint = "1107.5925",
    archivePrefix = "arXiv",
    primaryClass = "hep-ph",
    doi = "10.1140/epja/i2012-12002-1",
    journal = "Eur. Phys. J. A",
    volume = "48",
    pages = "2",
    year = "2012"
}

@article{Qi:2025zna,
    author = "Qi, Jing-Juan and Wang, Zhen-Yang and Zhang, Zhen-Hua and Guo, Xin-Heng",
    title = "{Normalization of partial wave CP asymmetries in three-body decays of heavy hadrons}",
    eprint = "2511.12445",
    archivePrefix = "arXiv",
    primaryClass = "hep-ph",
    month = "11",
    year = "2025"
}
\end{document}